\documentclass[twocolumn, times]{aastex631}
\usepackage{CJK}
\usepackage{amsmath}
\usepackage{mathtools}
\usepackage{subfigure}
\usepackage[T1]{fontenc}

\newcommand{\lz}[1]{#1}

\defcitealias{PaperI}{Paper~I}

\def\equationautorefname#1#2\null{(#2)}

\begin{document}
\begin{CJK*}{UTF8}{gbsn}

\title{Radiation GRMHD Models of Accretion onto Stellar-Mass Black Holes: II. Super-Eddington Accretion}

\correspondingauthor{Lizhong Zhang (张力中)}
\email{lizhong4physics@gmail.com}

\author[0000-0003-0232-0879]{Lizhong Zhang (张力中)}
\affiliation{Center for Computational Astrophysics, Flatiron Institute, New York, NY, USA}
\affiliation{School of Natural Sciences, Institute for Advanced Study, Princeton, NJ, USA}

\author[0000-0001-5603-1832]{James M. Stone}
\affiliation{School of Natural Sciences, Institute for Advanced Study, Princeton, NJ, USA}

\author[0000-0001-7448-4253]{Christopher J. White}
\affiliation{Center for Computational Astrophysics, Flatiron Institute, New York, NY, USA}
\affiliation{Department of Astrophysical Sciences, Princeton University, Princeton, NJ, USA}

\author[0000-0001-7488-4468]{Shane W. Davis}
\affiliation{Department of Astronomy, University of Virginia, Charlottesville, VA, USA}
\affiliation{Virginia Institute for Theoretical Astronomy, University of Virginia, Charlottesville, VA, USA}

\author[0000-0002-2624-3399]{Yan-Fei Jiang (姜燕飞)}
\affiliation{Center for Computational Astrophysics, Flatiron Institute, New York, NY, USA}

\author[0000-0003-2131-4634]{Patrick D. Mullen}
\affiliation{Michigan SPARC, Los Alamos National Laboratory, Ann Arbor, MI}
\affiliation{Computational Physics and Methods, Los Alamos National Laboratory, Los Alamos, NM}

\begin{abstract}
We present a comprehensive analysis of super-Eddington black hole accretion simulations that solve the GRMHD equations coupled with angle-discretized radiation transport.  The simulations span a range of accretion rates, two black hole spins, and two magnetic field topologies, and include resolution studies as well as comparisons with non-radiative models.  Super-Eddington accretion flows consistently develop geometrically thick disks supported by radiation pressure, regardless of magnetic field configuration.  Radiation generated in the inner disk drives substantial outflows, forming conical funnel regions that limit photon escape and result in very low radiation efficiency.  The accretion flows are highly turbulent with thermal energy transport dominated by radiation advection rather than diffusion.  Angular momentum is primarily carried outward by Maxwell stress, with turbulent Reynolds stress playing a subdominant role.  Both strong and weak jets are produced.  Strong jets arise from sufficient net vertical magnetic flux and rapid black hole spin and can effectively evacuate the funnel, enabling radiation to escape through strong geometric beaming.  In contrast, weak jets fail to clear the funnel, which becomes obscured by radiation-driven outflows and leads to distinct observational signatures.  Spiral structures are observed in the plunging region, behaving like density waves.  These super-Eddington models are applicable to a variety of astronomical systems, including ultraluminous X-ray sources, little red dots, and black hole transients.  
\end{abstract}

\keywords{\uat{Radiative magnetohydrodynamics}{2009} --- \uat{General relativity}{641} --- \uat{Black hole physics}{159} --- \uat{Accretion}{14} --- \uat{Jets}{870} --- \uat{Ultraluminous X-ray sources}{2164}}

\section{Introduction}
\label{sec:introduction}

Accretion onto a compact object is regulated by radiation feedback, where radiation produced by the accretion process can further inhibit inflow by exerting outward pressure.  The Eddington limit is defined as the threshold at which the outward radiation force on isotropically infalling material balances the inward gravitational pull.  While this classical limit applies to spherically symmetric accretion and sets an upper bound on the accretion rate, theoretical studies have shown that accretion beyond this limit (i.e., super-Eddington accretion) is possible through anisotropic flows (see review by \citealt{Jiang2024}, and references therein), in which geometric collimation and powerful equatorial outflows redirect energy and allow super-Eddington mass inflow to persist \citep{Shakura1973}.  

Super-Eddington accretion is increasingly recognized as a fundamental process underlying many high-energy astrophysical phenomena.  Observational evidence suggests that such accretion flows can power ultraluminous X-ray sources (ULXs; \citealt{Kaaret2017, King2023}), little red dots (LRDs; \citealt{Greene2024,Liu2025}), tidal disruption events (TDEs; \citealt{Gezari2021}), and stellar-mass black hole binaries (BHBs; \citealt{Stone2017,Rodriguez-Ramirez2025}), and may also contribute to the rapid growth of both stellar-mass and supermassive black holes \citep[e.g., ][]{Madau2014,Volonteri2021}.  Both black holes and neutron stars have been identified as central engines in these systems \citep{Sutton2013,Cseh2014,Bachetti2014}, often accompanied by relativistic jets or optically thick winds. 

Due to their distinct characteristics, super-Eddington accretion flows are often described using the slim disk models \citep{Abramowicz1988, Beloborodov1998, Sadowski2009, Abramowicz2013}, which accounts for radial energy advection -- a key component in regulating the energy balance within optically thick inflows near black holes.  However, the theoretical understanding of super-Eddington accretion remains incomplete.  Critical aspects such as energy transport, radiation collimation, angular momentum redistribution, and jet formation require detailed modeling across a range of plausible physical conditions. Although significant progress has been made through radiation magnetohydrodynamic simulations \citep[see, e.g.,][]{Jiang2014a,Sadowski2014,McKinney2014,McKinney2015,Sadowski2016b,McKinney2017,Narayan2017,Jiang2019b,Utsumi2022,Curd2023,Fragile2025}, a robust numerical framework, systematic parameter exploration, and comprehensive analysis that connects simulations, analytical models, and observations are still under development.  

Establishing a coherent physical picture across different parameter regimes is one of the primary goals of this project.  In \citetalias{PaperI}, we presented an overview of simulation results covering a broad range of mass accretion rates.  This paper focuses specifically on the super-Eddington models and introduces the analysis routine that will be systematically applied throughout the remainder of the series (i.e., Papers~III and IV).  

This paper is organized as follows.  In \autoref{sec:numerical_methods}, we briefly review the numerical methods and define auxiliary operators used throughout the paper.  In \autoref{sec:results}, we present a detailed analysis of the super-Eddington models, including time evolution, steady-state structure, angular momentum and energy transport, outflows and jet properties, and the plunging region.  In \autoref{sec:discussion}, we assess the quality of our numerical results, explore observational applications, and compare our findings with non-radiative models, slim disk theory, and previous simulation works.  Finally, in \autoref{sec:conclusions}, we summarize the main results of this work.

\section{Numerical Methods}
\label{sec:numerical_methods}

Simulations are performed using \textsc{AthenaK} \citep{Stone2024} via solving the radiation GRMHD equations \citep{White2023} in Cartesian Kerr-Schild coordinates. Radiation intensity is frequency-integrated and computed in an angle-dependent manner.  The full set of governing equations including radiation source terms, as well as the initial and boundary conditions, are detailed in Section~2 of \citetalias{PaperI}. 

We adopt both intermediate- and low-resolution mesh configurations for our super-Eddington models, as described in Section~2 of \citetalias{PaperI}.  In addition to the details provided there, the intermediate-resolution setup consists of 2048 cells in both the $x$- and $y$-directions and 1152 cells in the $z$-direction passing through the origin.  For comparison, the low-resolution setup contains 1152 cells in the $x$- and $y$-directions and 640 cells in the $z$-direction.  

Both grid configurations are used for all single-loop models, with the low-resolution runs designated for the resolution study.  For the double-loop and non-radiative models, only low-resolution setups are used, as their intrinsically thick disk structures reduce resolution requirements.  In addition to spatial resolution, we also increase the radiation angular resolution in one representative single-loop model to examine its effect. 

We present a suite of 12 simulations to study black hole accretion in the super-Eddington regime, with 7 main models summarized in \autoref{tab:sim_overall}.  The remaining 5 are dedicated to resolution studies and discussed in detail in \autoref{sec:resolution_effect}.  The parameter space spans two black hole spins (0.3 and 0.9375), two magnetic field topologies, two resolution levels, both radiative and non-radiative cases, and a range of accretion rates.  The accretion rate is set by the density unit $\rho_0$, as the radiative models are not scale-free. 

The model naming convention follows \citetalias{PaperI} and reflects multiple parameters: normalized accretion rate (`E'), black hole spin (`a'), magnetic field topology (single-loop by default and ‘DL’ for double-loop), and resolution level (`LR' for low-resolution models).  The non-radiative runs use single-loop magnetic fields and are scale-free, labeled as `NoRad' followed by the black hole spin (`a').  
  
In the low-density funnel region where the jet forms, the fluid density approaches the numerical floor, and the high-density unit can artificially enhance the Compton process, thereby overcooling the funnel.  To mitigate this issue, we apply a density reduction to the radiation-gas coupling terms when the local density approaches the floor value.  Further details of this numerical treatment are provided in \autoref{appendix:density_rescale}. 

For convenience, we define the following notations used in the subsequent sections to denote time, azimuthal, and combined time-azimuthal averages: 
\begin{subequations}    
\begin{align}
    &\left<\cdot\right>_t = \frac{1}{\Delta t}\int\cdot~dt
    \ , \qquad\qquad
    \left<\cdot\right>_{\phi} = \frac{1}{2\pi}\int\cdot~d\phi
    \ , 
    \\
    &\left<\cdot\right>_{\phi, t} = \frac{1}{2\pi\Delta t}\int\cdot~d\phi dt
    \ .
\end{align}
\end{subequations}

\begin{deluxetable*}{l c c c c c c c c}
\tablecaption{Comparison of time- and azimuthally averaged properties of the super-Eddington models \label{tab:sim_overall}}
\tablehead{
\colhead{Name} & 
\colhead{$\rho_0$} & 
\colhead{$\Delta z_{\mathrm{disk},30}$} & 
\colhead{$r_{\mathrm{eq}}$} & 
\colhead{$r_{\mathrm{tr}}$} & 
\colhead{$\tilde{P}_{\mathrm{therm}}^{(\mathrm{disk})}$} & 
\colhead{$\tilde{P}_{\mathrm{m}}^{(\mathrm{disk})}$} & 
\colhead{$\tilde{\alpha}_{\mathrm{Max}}^{r~(\mathrm{disk})}$} & 
\colhead{$\tilde{\alpha}_{\mathrm{Rey,turb}}^{r~(\mathrm{disk})}$}
\\
& \colhead{$\left(\mathrm{g/cm^3}\right)$} &  
\colhead{$(r_g)$} & \colhead{$(r_g)$} & \colhead{$(r_g)$} & 
\colhead{$(10^{15}~\mathrm{dyn/cm^{2}})$} & \colhead{$(10^{15}~\mathrm{dyn/cm^{2}})$} & 
\colhead{$(10^{-2})$} & \colhead{$(10^{-2})$}
\\
\quad\;\;\;(1) & (2) & (3) & (4) & (5) & (6) & (7) & (8) & (9)
}
\startdata
    E150-a9    & 1e-1 & 111 & 38 & 50 & \;\;$24.04~r_{10}^{-1.54}$    & \quad$0.57~r_{10}^{-1.60}$    & $3.52~r_{10}^{+0.01}$ & $1.04~r_{10}^{-0.01}$ \\ 
    E88-a3     & 3e-2 & 89 & 46 & 63 & \;\;$12.91~r_{10}^{-1.30}$    & \quad$0.42~r_{10}^{-1.81}$    & $3.96~r_{10}^{-0.30}$ & $1.35~r_{10}^{-0.42}$ \\
    E31-a3-DL  & 8e-4 & 71 & 30 & 87 & \;\;\;\:$3.55~r_{10}^{-1.33}$ & \quad$0.17~r_{10}^{-1.69}$    & $4.03~r_{10}^{-0.65}$ & $1.66~r_{10}^{-0.55}$ \\
    E15-a9     & 1e-2 & 80 & 49 & 25 & \;\;\;\:$2.77~r_{10}^{-1.51}$ & \quad$0.08~r_{10}^{-1.43}$    & $2.92~r_{10}^{-0.05}$ & $0.85~r_{10}^{+0.02}$ \\
    E9-a3      & 3e-3 & 65 & 32 & 17 & \;\;\;\:$1.64~r_{10}^{-1.04}$ & \quad$0.04~r_{10}^{-1.64}$    & $3.72~r_{10}^{-0.55}$ & $1.08~r_{10}^{-0.42}$ \\
    NoRad-a9   & -    & 178 & 49 & -  & $\propto 0.95~r_{10}^{-1.89}$ & $\propto 0.05~r_{10}^{-1.97}$ & $5.78~r_{10}^{+0.10}$ & $1.05~r_{10}^{-0.21}$ \\
    NoRad-a3   & -    & 134 & 59 & -  & $\propto 0.94~r_{10}^{-1.64}$ & $\propto 0.06~r_{10}^{-1.91}$ & $7.22~r_{10}^{-0.40}$ & $2.22~r_{10}^{-0.59}$ \\
    \hline
\enddata
\tablecomments{
    Recall that non-radiative models are dimensionless; thus, their pressure measurements are normalized by the total pressure (indicated by $\propto$).  \lz{Additional quantities for the radiative models are summarized in Table 1 of \citetalias{PaperI}. }
    {\bf Columns (from left to right):} 
    (1) Model name \lz{(`E': accretion rate; `a': black hole spin)}; 
    (2) Simulation density unit; 
    (3) Disk vertical thickness at $r=30r_g$; 
    (4) Maximum inflow equilibrium radius; 
    (5) Photon trapping radius; 
    (6) Power-law fitted, disk-averaged thermal pressure (gas + radiation); 
    (7) Power-law fitted, disk-averaged magnetic pressure; 
    (8) Power-law fitted, disk-averaged radial Maxwell component of the angular momentum flux, normalized by the total pressure (see equation~\ref{eq:viscosity_approx}); 
    (9) Power-law fitted, disk-averaged radial turbulent Reynolds component of the angular momentum flux, normalized by the total pressure (see equation~\ref{eq:viscosity_approx}). 
}
\end{deluxetable*}

\begin{figure*}
    \centering
    \subfigure[\centering Gas density with magnetic streamlines]{{\includegraphics[width=\columnwidth]{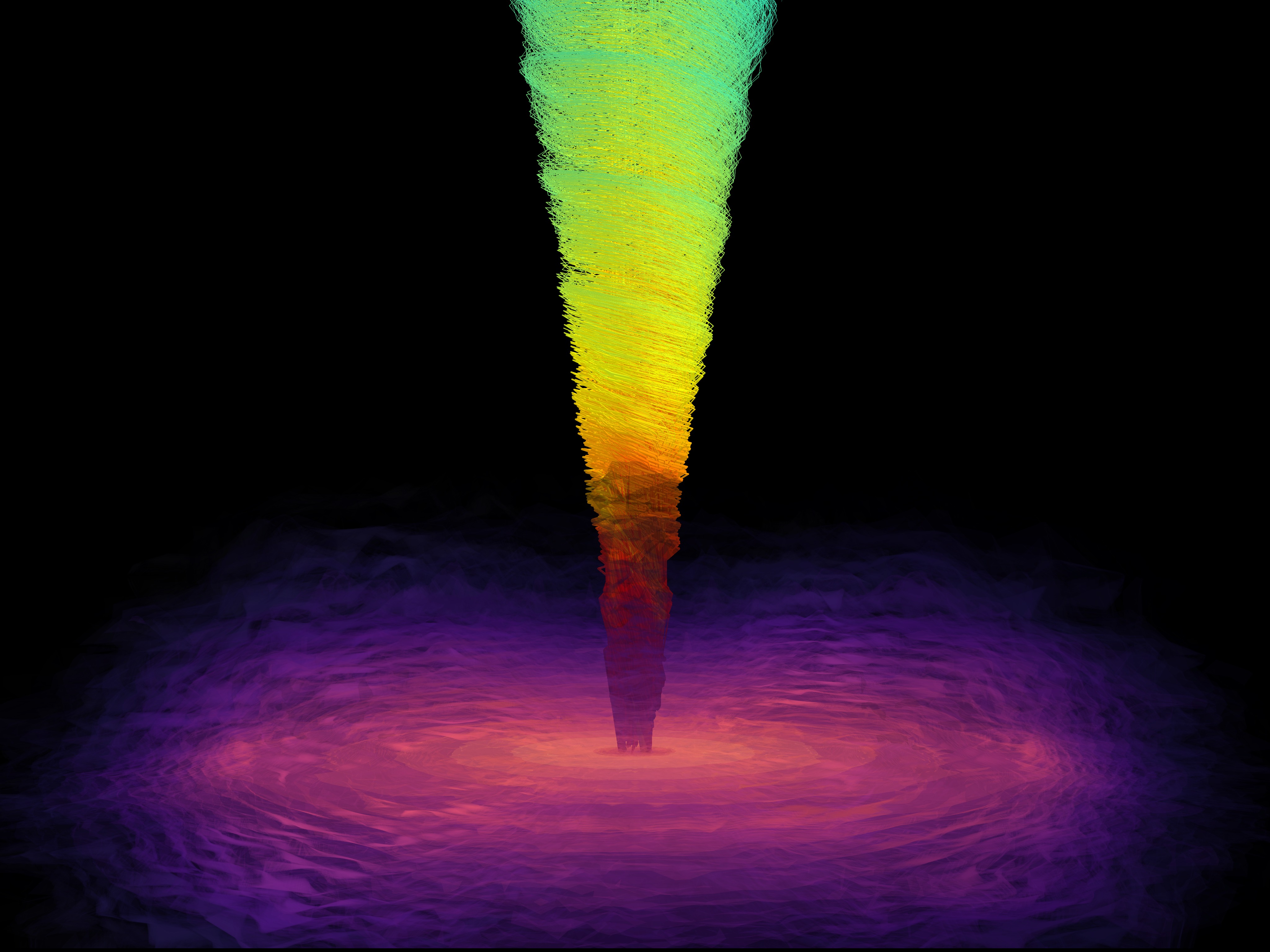} }}%
    \quad
    \subfigure[\centering Radiation energy density in the fluid frame]{{\includegraphics[width=\columnwidth]{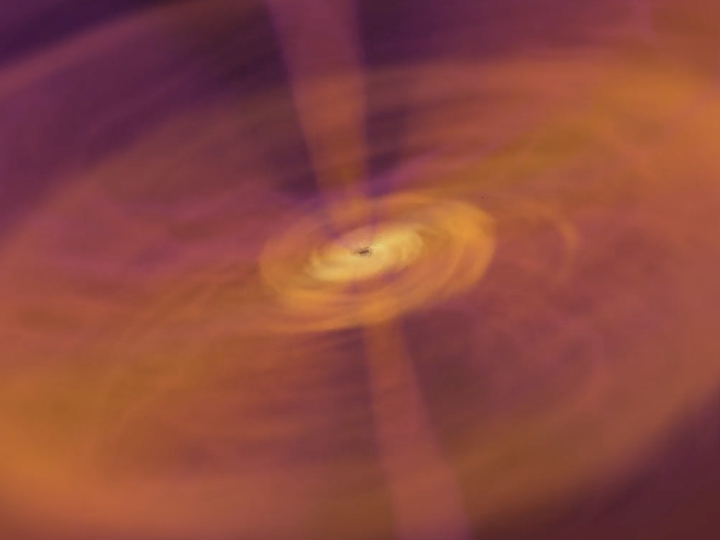} }}%
    \caption{
    3D rendering of model E88-a3-LR. 
    {\bf Left panel:} Gas density with magnetic streamlines highlighting the jet region at $t=50000r_g/c$.  The accretion disk is geometrically thick, with density increasing toward the midplane.  The jet is launched near the spinning black hole and is threaded by helical magnetic fields that extend to large distances.  Streamline colors indicate magnetic field strength, transitioning from green/yellow (weaker) to orange/red (stronger).  
    {\bf Right panel:} \lz{(Image credit: Joseph Insley)} Radiation energy density in the fluid frame at $t=32700r_g/c$.  Most of the radiation originates from the accretion disk, and is concentrated in the inner regions.  The vertical structures above and below the central black hole represent the radiation field within the funnel, which is significantly energized by the relativistic jet through scattering processes.  
    }
    \label{fig:3d_render}
\end{figure*}

When analyzing disk properties, we define the vertical average over the disk region as: 
\begin{subequations}    
\begin{equation}
    \left<\cdot\right>_{\mathrm{disk}} = \frac{1}{\Delta z_{\mathrm{disk}}}\int_{\mathrm{disk}} \left<\cdot\right>_{\phi,t} dz
    \ ,
\end{equation}
where $\Delta z_{\mathrm{disk}}$ is the vertical thickness of the disk (defined as the gravitationally bound region) measured at a given cylindrical radius.  The disk thickness at $r=30r_g$ for all models is summarized in column~(3) of \autoref{tab:sim_overall}. 

When analyzing wind and jet properties at a given radius, we define the region-specific spherical average as: 
\begin{equation}
    \left<\cdot\right>_{\mathrm{wind/jet}} = \frac{\displaystyle \int_{\mathrm{wind/jet}} \left<\cdot\right>_{\phi,t}\sqrt{-g} d\theta d\phi}{\displaystyle\int_{\mathrm{wind/jet}} \sqrt{-g} d\theta d\phi}
    \ . 
\end{equation} 
\end{subequations}

\section{Results} 
\label{sec:results}

In this section, we review the evolution history and describe the physical picture of the super-Eddington accreting system.  We then conduct a detailed steady-state analysis of the disk, wind, and jet regions by examining representative profiles and the conservation of momentum and energy. Additionally, we investigate the accretion flow in the plunging region, with particular focus on the physical properties of the spiral structures. 

All of our super-Eddington accreting systems approximately reach steady states by $15000r_g/c$ and are further \lz{run} to $60000r_g/c$.  For comparison, the non-radiative runs are evolved to $34000r_g/c$, with steady state similarly established by around $20000r_g/c$.  As an example, the morphological structure of a super-Eddington accretion system is demonstrated in \autoref{fig:3d_render}, using snapshots of E88-a3-LR.  In the left panel, the 3D-rendered density profile is color-coded in purple and overlaid with magnetic streamlines that highlight the relativistic jet.  The right panel shows the radiation energy density in the fluid frame, where the vertically conical distribution effectively traces the jet region, resulting from the scattering process by the relativistic jet.

\subsection{Time Evolution}
\label{sec:time_evolution}

\begin{figure*}
    \centering
    \includegraphics[width=\textwidth]{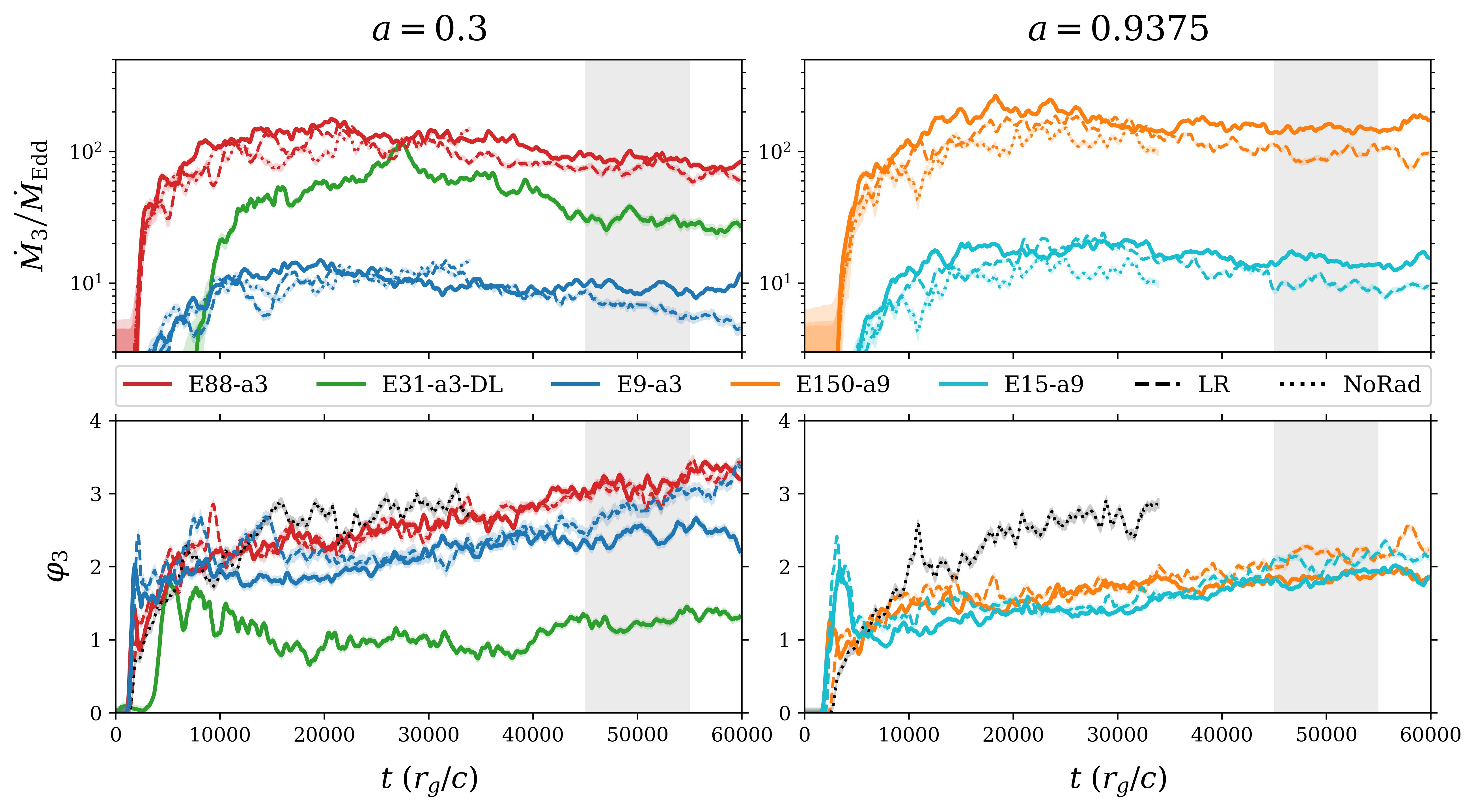}
    \caption{
    Time evolution of the mass accretion rate $\dot{M}_3$ and magnetic flux $\varphi_3$ measured at $3r_g$.  Left panels show low-spin cases, while right panels present high-spin cases.  Each color represents a different accretion rate, as indicated in the legend.  Dashed lines denote low-resolution simulations used for resolution studies, and dotted lines indicate non-radiative models.  The gray-shaded region marks the time window used for the time-averaged analysis of radiative models.  All systems evolve into steady states and maintain super-Eddington accretion. 
    }
    \label{fig:hst_compare}
\end{figure*}

\autoref{fig:hst_compare} presents the evolution of the mass accretion rate $\dot{M}$ and magnetic flux $\varphi$ near the horizon at $3r_g$, with low-spin black hole runs in the left column and high-spin runs in the right.  These quantities are defined by equation~(5) in \citetalias{PaperI}, and the subscript `3' denotes measurements taken at $3r_g$.  The mass accretion rate is normalized by the Eddington mass accretion rate $\dot{M}_{\mathrm{Edd}}=4\pi G M_{\mathrm{BH}}/(0.1\kappa_Tc)$, assuming 0.1 radiation efficiency.  Solid colored lines represent five simulations with varying accretion rates and black hole spins at the highest applicable resolution.  Dashed lines in the same colors indicate their low-resolution counterparts, included for resolution studies.  Since the non-radiative runs are scale-free, their mass accretion rates are rescaled using the corresponding density units from the radiative runs and shown as dotted lines in matching colors. The dimensionless magnetic fluxes of the non-radiative runs are shown as black dotted lines.  Each curve is smoothed using a moving boxcar with a time window of $400r_g/c$, and the color-shaded region along each curve represents the 1$\sigma$ variation within the local time window.  The gray-shaded region marks the $10000r_g/c$ interval used for the time-averaged analysis of radiative runs in subsequent sections.  For the non-radiative runs, the final $10000r_g/c$ of the simulation is used for time-averaging. 

\begin{figure*}
    \centering
    \includegraphics[width=\textwidth]{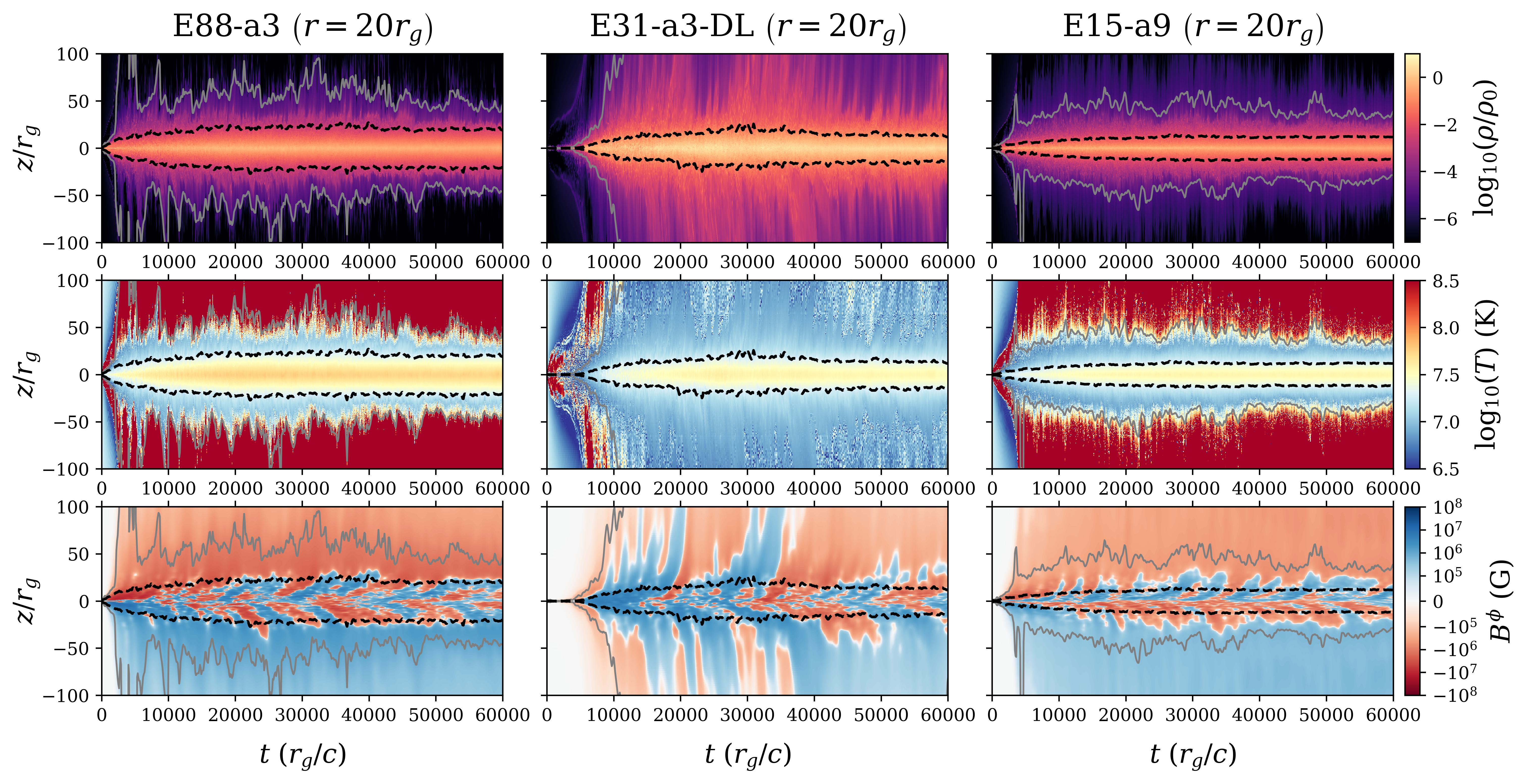}
    \caption{
    Space-time diagrams of gas density, gas temperature, and the toroidal component of the 3-magnetic field measured at \lz{a spherical Kerr-Schild radius} $r=20r_g$ \lz{on the midplane}.  The black dashed lines mark the effective photosphere, and the gray solid lines denote the scattering photosphere. 
    }
    \label{fig:timestream}
\end{figure*}

As shown in \autoref{fig:hst_compare}, all radiative simulations remain in the SANE state with $\varphi<4$ and achieve the super-Eddington accretion regime after reaching steady state.  
The system initialized with a double-loop magnetic configuration shows significantly lower magnetic flux compared to the single-loop runs at the same black hole spin.  In single-loop simulations, the magnetic flux weakly depends on accretion rate in low-spin cases, with slightly higher magnetic flux observed at much higher accretion rates; in contrast, there is almost no difference in high-spin cases.  Between different spins, low-spin systems generally maintain higher magnetic flux at comparable accretion rates.  The evolution of accretion rate and magnetic flux is remarkably consistent between non-radiative and radiative runs, except in high-spin cases, where radiative runs exhibit higher accretion rates and a corresponding reduction in magnetic flux by approximately a factor of 1.5.  Low-resolution runs yield slightly lower accretion rates compared to their intermediate-resolution counterparts.  A detailed discussion of the resolution effect is provided in \autoref{sec:resolution_effect}. 

\autoref{fig:timestream} presents space-time diagrams of gas density, gas temperature, and the toroidal magnetic field at a \lz{spherical Kerr-Schild} radius $r=20r_g$ \lz{on the midplane (corresponding to a cylindrical radius $\sqrt{r^2+a^2}$)}.  This radius lies within the mass inflow equilibrium region and represents the main body of the disk.  The black dashed lines and gray solid lines denote the effective and scattering photospheres, respectively.  The photosphere is defined as the locations where the vertically integrated optical depth, computed from the domain boundaries toward the midplane, equals unity.  The scattering photosphere is determined using the Thomson scattering opacity $\kappa_T$, while the effective photosphere is based on the effective opacity, defined as $\sqrt{\kappa_T\kappa_P}$.  The vertical structure of the disk stabilizes after approximately $15000r_g/c$, with gas density and temperature peaking near the midplane.  The evolution of the toroidal magnetic field exhibits butterfly patterns near the midplane, driven by magnetic buoyancy and the dynamo process associated with MRI \citep{Brandenburg1995, Oliver2010, Blackman2012}, a behavior also observed in a wide range of global \citep{ONeill2011, Jiang2014a, Jiang2019b, Jiang2019a} and shearing-box \citep{Stone1996, Miller2000, Davis2010, Shi2010, Simon2012, Jiang2013, Jiang2014b, Salvesen2016a, Salvesen2016b} simulations.

\subsection{Steady State}
\label{sec:steady_state}

\begin{figure}
    \centering
    \includegraphics[width=\columnwidth]{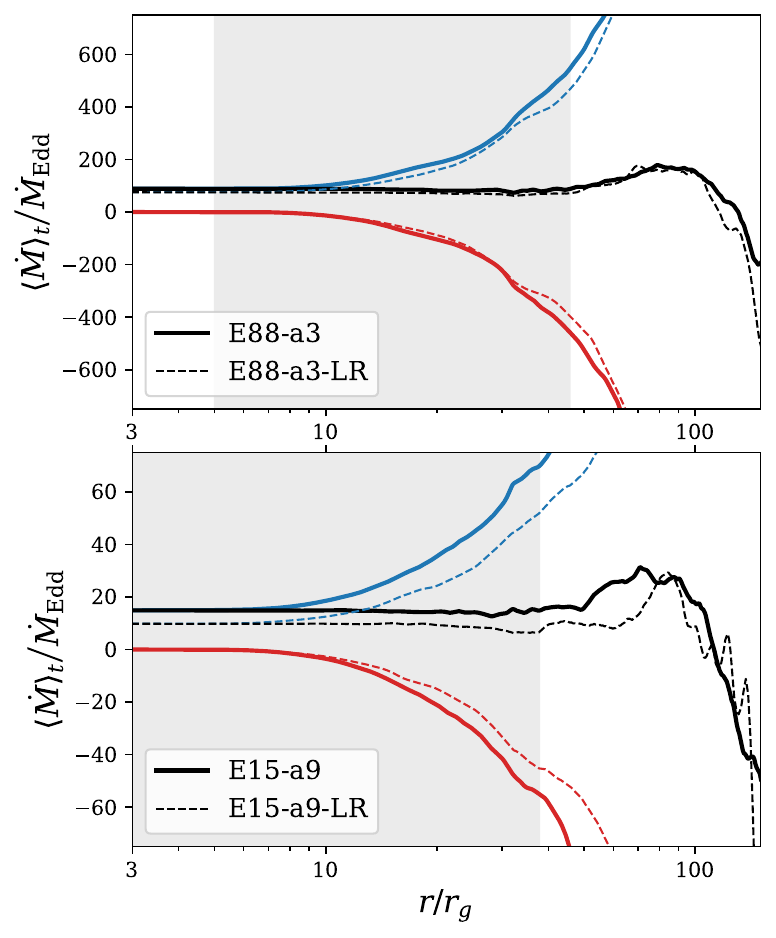}
    \caption{
    Time-averaged mass accretion rate as a function of radius.  The upper and lower panels present two representative models with different black hole spins, each including both intermediate- and low-resolution runs for comparison.  Model names are indicated in the legend.  Black lines show the net mass accretion rates, while blue and red lines represent the spherically integrated mass fluxes, accounting for only inflow and outflow, respectively. 
    }
    \label{fig:mdot_compare}
\end{figure}

\begin{figure*}
    \centering
    \includegraphics[width=\textwidth]{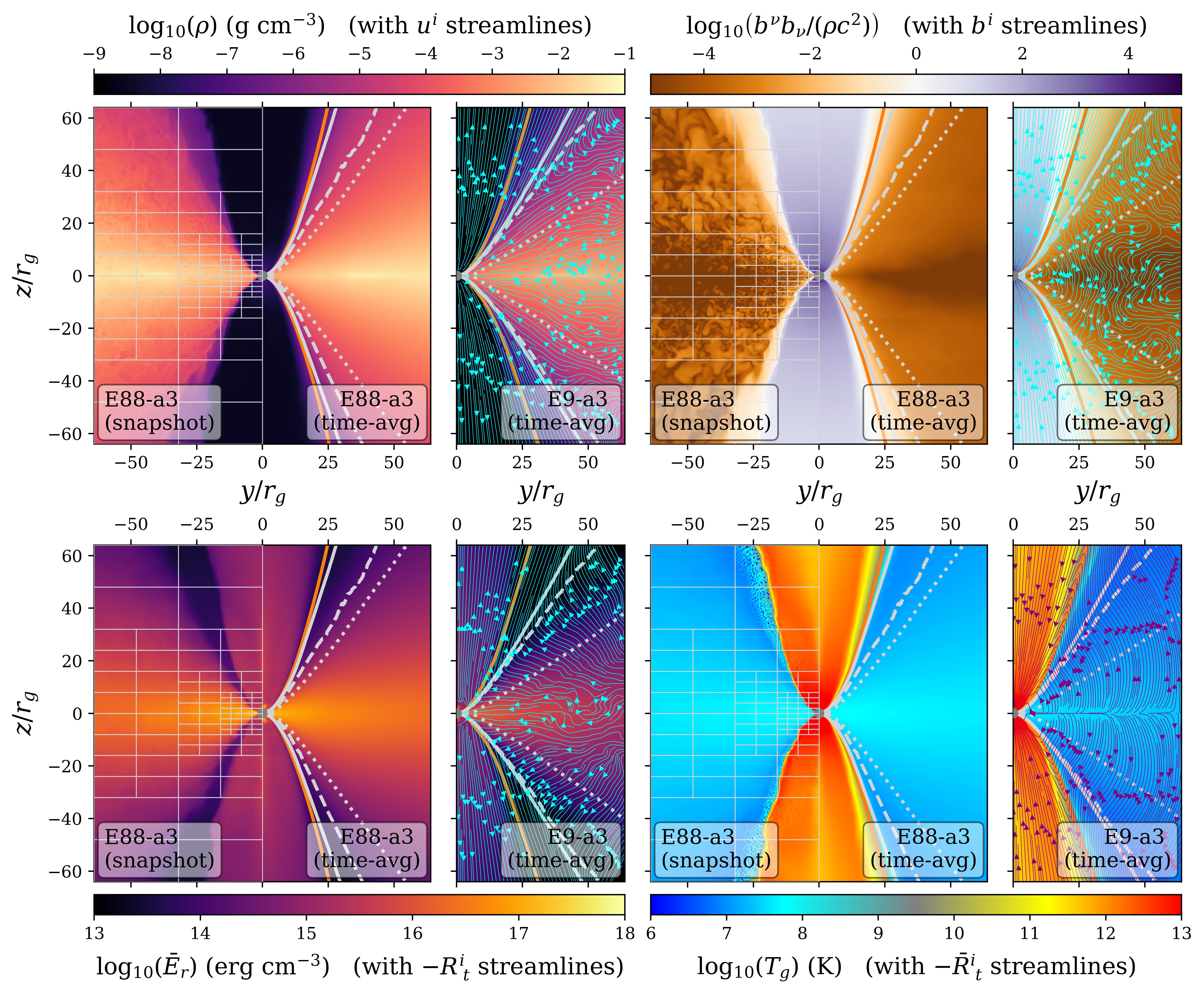}
    \caption{
    2D steady-state profiles of gas density (upper left), magnetization (upper right), fluid-frame radiation energy density (lower left), and gas temperature (lower right) for representative super-Eddington models.  Each panel presents a side-by-side comparison between a snapshot and its time average of the highly super-Eddington model (E88-a3), followed by the time-averaged profile of a lower accretion rate model (E9-a3).  Different \lz{time-averaged} streamlines are overlaid: velocity in the density panel, magnetic field in the magnetization panel, coordinate-frame radiation flux in the radiation energy density panel, and fluid-frame radiation flux in the gas temperature panel.  Auxiliary lines are overplotted on the 2D profiles to delineate different regions: orange lines indicate jet boundaries, solid gray lines mark the scattering photosphere, gray dashed lines denote disk boundaries, and gray dotted lines represent the effective photosphere. 
    }
    \label{fig:profile2d}
\end{figure*}

After steady state is established, inflow equilibrium is achieved in all simulations within approximately $30$ -- $50r_g$.  \autoref{fig:mdot_compare} shows the time-averaged mass accretion rates as a function of radius for selected models with different black hole spins.  The gray-shaded regions indicate the radial extent from the innermost stable circular orbit (ISCO) to the outer boundary of the inflow equilibrium.  Blue and red lines represent the spherically integrated net mass inflow and outflow rates, respectively, with simulation names labeled in the legend.  The outermost radii of the inflow equilibrium region $r_{\mathrm{eq}}$ for all models are summarized in column~(4) of \autoref{tab:sim_overall}. 

As demonstrated in Figure~4 of \citetalias{PaperI}, the accretion system can be partitioned into three regions: disk, wind, and jet.  The disk region is identified as gravitationally bound, where the Bernoulli parameter $\mathrm{Be}$ is negative.  The Bernoulli parameter is defined to include radiation enthalpy within the scattering photosphere and to exclude it outside the scattering photosphere as:
\begin{equation}    
    \mathrm{Be} = \begin{dcases}
    -\frac{w_{\mathrm{disk}} u_t}{\rho} - 1
    \ \ (\text{within scattering photosphere})
    \\
    -\frac{w u_t}{\rho} - 1
    \qquad(\text{outside scattering photosphere})
    \end{dcases}
    \ ,
\end{equation}
where the total enthalpy per unit volume $w_{\mathrm{disk}} = w + 4\bar{E}_r/3$ is also extensively used in the subsequent analysis of the disk region.  \lz{The physical motivation for this piecewise definition lies in the thermal and dynamical coupling between gas and radiation. Within the scattering photosphere where gas and radiation are well coupled, the Bernoulli parameter must account for the radiation contribution to the total enthalpy and inertia. Similar definitions have been adopted in previous studies (e.g., \citealt{Curd2023}).}  The jet region is identified by tracing the outermost velocity streamlines within the strongly magnetized funnel located outside the disk, where $b^{\nu}b_{\nu}/(\rho c^2)>1$.  A detailed description of the jet identification is provided in \autoref{appendix:jet_id}. The wind region then lies between the disk and jet regions.  

\autoref{fig:profile2d} shows the 2D profiles of gas density (upper left), magnetization (upper right), radiation energy density (lower left), and gas temperature (lower right) in the steady state.  Since the super-Eddington accretion systems exhibit broadly similar structures, we present only two representative models (E88-a3 and E9-a3) for clarity.  As labeled in the figure, each panel presents a side-by-side comparison between a snapshot and its time-averaged profile of the highly super-Eddington model (E88-a3), followed by the time-averaged profile for comparison at a lower accretion rate (E9-a3).  Note that the time-averaged profiles are also azimuthally averaged, and the snapshot is taken at $t=50000r_g/c$, corresponding to the midpoint of the averaging interval.  The mesh block structure is overplotted in gray solid lines on the snapshot to illustrate the static mesh refinement.  In the time-averaged profiles, auxiliary lines are included to distinguish different regions: solid orange lines indicate the jet boundary, gray dashed lines mark the disk surface, and gray solid and dotted lines represent the scattering and effective photospheres, respectively.  Additionally, streamlines are overlaid in the rightmost 2D profile of each panel to visualize the corresponding vector fields: velocity ($u^i$) in the density panel, magnetic field ($b^i$) in the magnetization panel, coordinate-frame radiation flux ($R^i_{\ t}$) in the radiation energy density panel, and fluid-frame radiation flux ($\bar{R}^i_{\ t}$) in the temperature panel. 

In the super-Eddington accretion regime, the accretion disk is thermally expanded, forming a narrow conical funnel along the polar axis.  The disk body is highly turbulent due to MRI, and radiation is strongly beamed within the funnel as a result of geometric confinement.  As shown in the upper right panel, magnetization increases from the disk to the jet regions: the disk region (within dashed lines) is weakly magnetized, the wind region (between the solid orange and gray lines) is mildly magnetized, and the jet region (from the pole to the solid orange line) is strongly magnetized.  The initial magnetic field topology (a single loop in this case), is well preserved, as traced by the field streamlines, where the jet outflows open the dipole loop, leaving net poloidal fields threading through the black hole.  

The accretion disk (within gray dashed lines) is primarily supported by radiation pressure, with the local radiative force indicated by the streamlines of the fluid-frame radiation flux (lower right panel).  At small radii, radiation is largely trapped by the optically thick accretion flow, as shown by the coordinate-frame radiation flux, which follows the inflow into the black hole within approximately $15r_g$ in the E9-a3 case (lower left panel).  Since thermal transport near the black hole is dominated by advection, the trapping radius is more accurately defined based on radiation advection (see Figure~5 in \citetalias{PaperI}).  The trapping radii $r_{\mathrm{tr}}$ for all super-Eddington models are summarized in column~(5) of \autoref{tab:sim_overall}.  Note that there exist gravitationally bound outflows within the disk body, as indicated by the outward fluid velocity just below the disk surface (upper left panel).  The fraction of these bound outflows further increases with accretion rate (e.g., see the upper panels of Figure~4 in \citetalias{PaperI} for E88-a3). 

The outflowing wind (between the solid orange and gray dashed lines) is launched near the disk surface, as shown by the velocity streamlines in the upper left panel, and is primarily driven by excess local radiation pressure.  The wind region is generally narrow and consists of a hotter, optically thin component (between the solid gray and orange lines) and a cooler, optically thick component (between the gray dashed and solid lines).  The optically thick wind facilitates outward radiation transport via advection and becomes increasingly dominant at higher accretion rates.  For example, in E88-a3, the majority of the wind is optically thick, in contrast to E9-a3.  \lz{The transition between the optically thick and thin wind components produces a narrow region of reduced fluid-frame radiation energy density along the jet boundary, where radiation and gas become progressively decoupled.}

The jet region (between the polar axis and the solid orange line) is typically optically thin and characterized by extremely low density.  In the single-loop magnetic configuration, the jet is highly relativistic and powered by the Blandford-Znajek process \citep{Blandford1977}.  In this case, the jet is strong enough to evacuate a well-defined conical funnel, where photons originating from the disk exert a drag on the jet outflow, as evidenced by the negative fluid-frame radiation flux in the lower right panel.  However, for the double-loop case (E31-a3-DL), the MHD jet is intermittent and weak, unable to clear the funnel.  Instead, a mildly relativistic outflow forms along the polar axis, driven by radiation pressure at larger radii.  Further details are provided in \autoref{sec:jet_properties}.

\subsection{Radial Structure}
\label{sec:radial_structure}

\begin{figure*}
    \centering
    \includegraphics[width=0.95\textwidth]{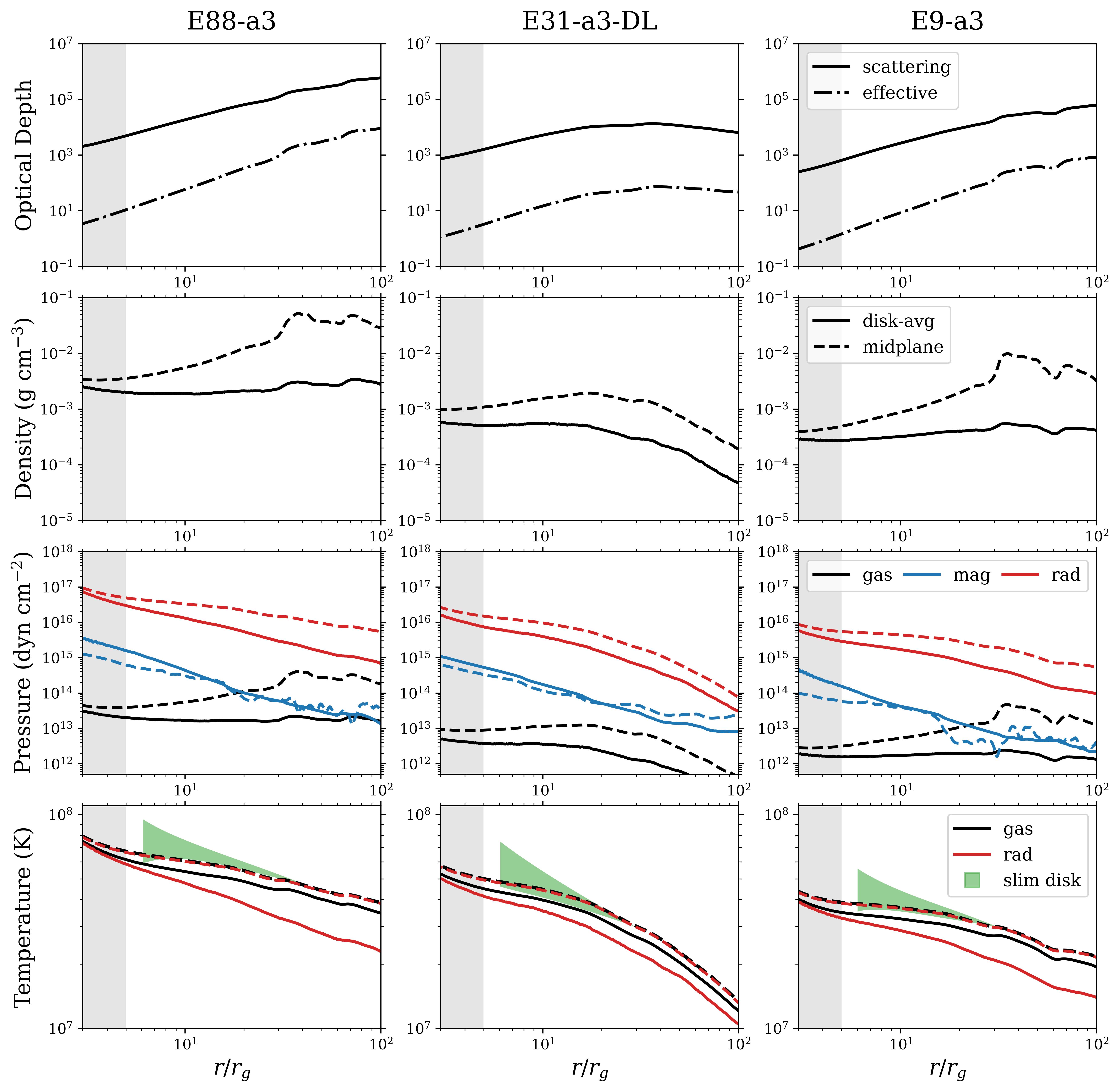}
    \caption{
    Radial profiles of accretion disks in time- and azimuthal average for representative models.  The first row shows two types of disk optical depths: scattering (solid lines) and effective (dot-dashed lines).  The remaining panels display density, pressure, and temperature, measured as disk averages (solid lines) and at the midplane (dashed lines).  Gas, magnetic, and radiation quantities are shown in black, blue, and red, respectively.  For comparison, temperature profiles inferred from the slim disk model over a range of \lz{values for the} $\alpha$ \lz{parameter} are shown in green shading.  Details on how these quantities are computed are provided in \autoref{sec:radial_structure}.  
    }
    \label{fig:hori_compare}
\end{figure*}

The disk body is defined as the gravitationally bound region, within which we compute disk-averaged quantities, including gas density, gas/magnetic/radiation pressure, and gas/radiation temperature, as follows: 
\begin{subequations}    
\begin{align}
    &\rho^{(\mathrm{disk})} = \left<\rho\right>_{\mathrm{disk}}
    \ , 
    && P_m^{(\mathrm{disk})} = \frac{1}{2}\left<b^{\nu}b_{\nu}\right>_{\mathrm{disk}}
    \ , 
    \nonumber
    \\
    & P_g^{(\mathrm{disk})} = \left<P_g\right>_{\mathrm{disk}}  
    \ , 
    && \bar{P}_r^{(\mathrm{disk})} = \frac{1}{3}\left<\bar{E}_r\right>_{\mathrm{disk}}
    \ , 
    \nonumber
    \\
    & T_g^{\mathrm{(disk)}} = \frac{\mu m_p}{k_B} \frac{\left<P_g\right>_{\mathrm{disk}}}{\left<\rho\right>_{\mathrm{disk}}}
    \ , 
    && T_r^{\mathrm{(disk)}} = \left(\frac{\left<\bar{E}_r\right>_{\mathrm{disk}}}{a_r}\right)^{\frac{1}{4}}
    \ ,
    \nonumber
\end{align}
\end{subequations}
where $k_B$ is the Boltzmann constant, and $m_p$ is the proton mass.  The mean molecular weight $\mu=0.5$, is adopted for ionized hydrogen.  The quantity $a_r$ refers to the radiation density constant.  

\autoref{fig:hori_compare} presents 1D profiles of representative models, showing optical depth, gas density, pressure, and temperature.  In the following, we define these quantities and discuss their trends across the simulations. 

The first row of \autoref{fig:hori_compare} shows two types of disk optical depths: the scattering optical depth (solid lines) and the effective optical depth (dot-dashed lines), defined as 
\begin{subequations}    
\begin{align}
    \tau_s^{\mathrm{(disk)}} &= \int_{\mathrm{disk}} \left<\rho\kappa_T\right>_{\phi,t} dz
    \ ,
    \\
    \tau_{\mathrm{eff}}^{\mathrm{(disk)}} &= \int_{\mathrm{disk}} \left<\rho\sqrt{\kappa_T\kappa_P}\right>_{\phi,t} dz
    \ .    
\end{align}
\end{subequations}
Since the scattering opacity is constant, $\tau_s^{\mathrm{(disk)}}$ effectively traces the disk surface density.  The effective optical depth is always smaller than the scattering optical depth at a given radius, as the system resides in a high-temperature regime where the Planck mean opacity is much lower than the Thomson opacity.  Within the inflow equilibrium region, both optical depths increase with radius, reflecting the wedge-like geometry of the disk.  

The levels of density, pressure, and temperature generally increase with higher accretion rate. 

The second row of \autoref{fig:hori_compare} presents gas density profiles, measured both in disk average (solid line) and at the midplane (dashed line).  The higher midplane density relative to the disk average suggests a concentration of mass towards the midplane, which is further examined in \autoref{sec:vertical_structure}.  Additionally, the density profiles do not follow a clear power-law trend with radius, reflecting the radiation-dominated nature of the disk. 

Pressure profiles, shown in the third row, are also measured as disk averages (solid lines) and at the midplane (dashed lines), with gas, magnetic, and radiation pressures plotted in black, blue, and red, respectively.  All systems are dominated by radiation pressure, while magnetic pressure remains subdominant. Both radiation and magnetic pressures follow a clear power-law dependence on radius.  Gas pressure, on the other hand, mirrors the behavior of the density profiles and shows almost no power-law dependence.  The power-law fitted thermal (gas + radiation) and magnetic pressures for all models are listed in columns~(6) and (7) of \autoref{tab:sim_overall}. 

The fourth row presents the temperatures of both radiation and gas, measured as disk-averaged values and at the midplane (using the same line styles and colors as before).  Both temperatures exhibit a power-law dependence on radius, with the fitted gas temperatures for all models provided in Table~1 of \citetalias{PaperI}.  While gas and radiation temperatures are in thermal equilibrium near the midplane, the disk-averaged gas temperature is slightly higher than the disk-averaged radiation temperature.  This difference arises from contributions near the disk surface, where gas and radiation become less thermally decoupled. 

In the temperature panels, the green shading represents temperature profiles inferred from the slim disk model over a range of constant $\alpha$, with the minimum and maximum values corresponding to the lower and upper bounds of the shaded region, respectively.  The slim disk solutions are computed by integrating from the outer edge of the inflow equilibrium region ($r_{\mathrm{eq}}$) down to $6r_g$, using the midplane profiles as initial conditions.  From left to right, we select $\alpha$ ranges of $[0.0165, 0.02]$, $[0.04, 0.08]$, and $[0.025, 0.035]$, respectively, to approximately match the midplane temperature profiles.  Note that the green shaded region remains relatively broad despite the narrow range of $\alpha$, indicating the slim disk solution is highly sensitive to the choice of $\alpha$.

\subsection{Vertical Structure}
\label{sec:vertical_structure}

\begin{figure*}
    \centering
    \includegraphics[width=0.95\textwidth]{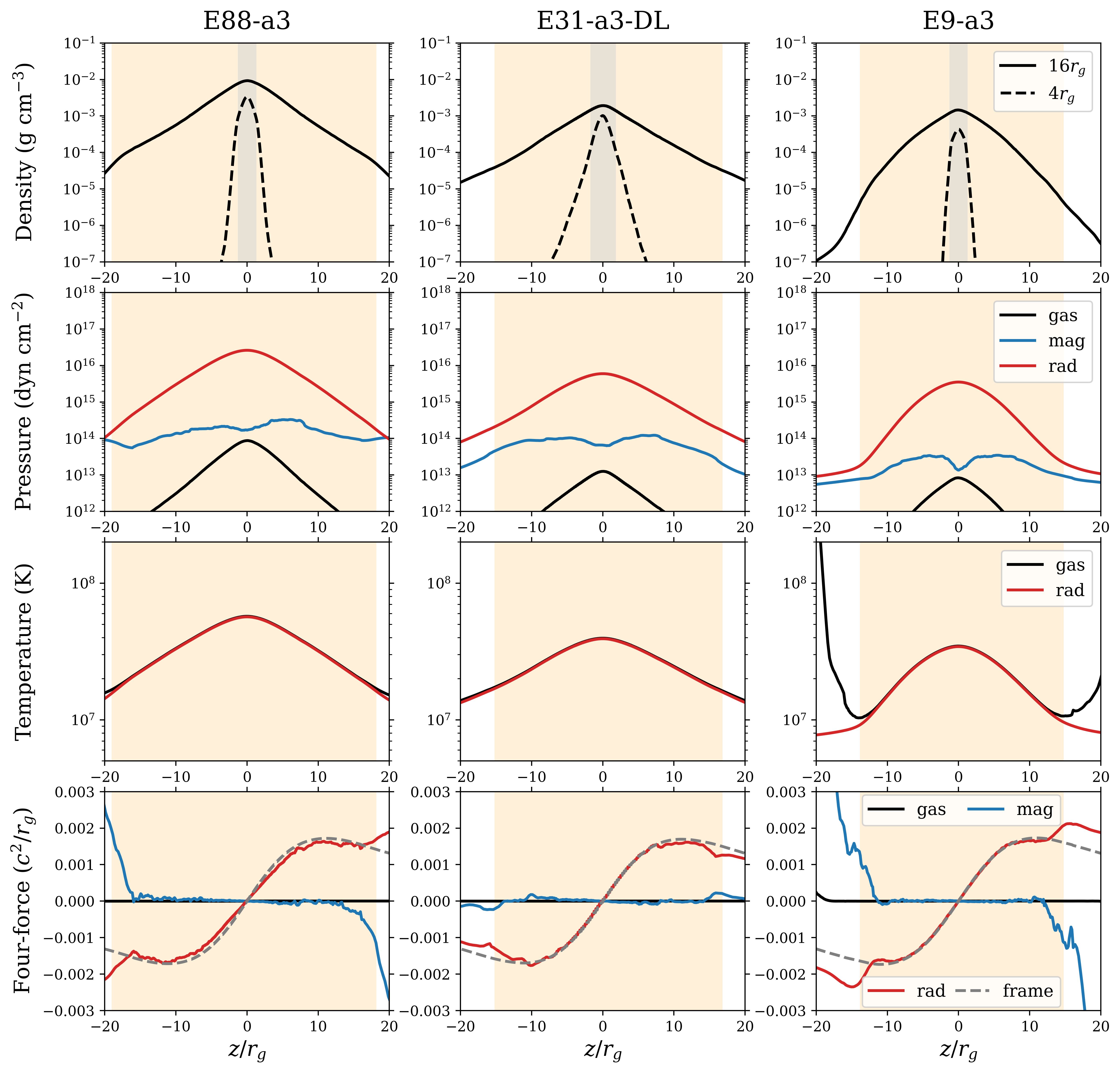}
    \caption{
    Vertical profiles of accretion disks in time- and azimuthal average for representative models.  From top to bottom, the rows show density, pressure, temperature, and four-force profiles measured at a horizontal distance of $16r_g$, with the disk region highlighted in yellow.  Gas, magnetic, and radiation components are shown in black, blue, and red, respectively.  In the first row, the density profile at $4r_g$ is also included (dashed lines), with the corresponding disk region shaded in gray.  In the last row, the \lz{negative} frame force (primarily gravity) is indicated by dashed gray lines.  Details on these measurements are provided in \autoref{sec:vertical_structure}.  
    }
    \label{fig:vert_compare}
\end{figure*}

\autoref{fig:vert_compare} presents the vertical profiles of density, pressure, temperature, and four-force components at selected radii.  A horizontal distance of $16r_g$ is used, with the disk body highlighted in yellow.  For comparison, the density profile at $4r_g$ is also included, with the corresponding disk region shaded in gray.

The first row shows the vertical density profiles, where measurements at $16r_g$ and $4r_g$ are represented by solid and dashed lines, respectively.  Gas density peaks near the midplane, consistent with the radial profiles in \autoref{fig:hori_compare}, where the midplane density exceeds the disk average.  The double-loop model (E31-a3-DL) exhibits a slightly lower degree of midplane density concentration compared to the single-loop models.  

The second row shows the vertical profiles of gas (black), magnetic (blue), and radiation (red) pressures at $16r_g$.  Both radiation and gas pressures increase toward the midplane, providing vertical support against gravity, with radiation pressure being dominant.  In contrast, magnetic pressure exhibits a local minimum near the midplane, which compresses the disk.  This inverse magnetic energy distribution is likely a result of magnetic buoyancy, as evidenced by the butterfly patterns in \autoref{fig:timestream}.  However, the relatively flat vertical distribution of magnetic pressure results in a small gradient, making its dynamical contribution negligible.

The third row displays the vertical profiles of gas and radiation temperatures at $16r_g$, with radiation temperature in red and gas temperature in black.  The overall temperature regime increases with higher accretion rate.  Local thermal equilibrium is largely maintained within the disk region, although gas and radiation temperatures begin to decouple near the disk surface.  

The last row presents the vertical components of the gas pressure force (black), total magnetic force (blue), and radiation force (red) within the disk body at horizontal distance of $16r_g$.  The frame force, dominated by gravity and computed using the four-velocity $u^{\mu}=(1,0,0,0)$, is multiplied by -1 and shown as gray dashed lines.  Gravity is primarily balanced by the radiation force, with gas pressure and magnetic forces being negligible.  Note that near the disk surface, the vertical radiation force slightly exceeds gravity, enabling the launch of a radiation-driven outflow.

The four-forces plotted in the last row originate from the geodesic equation, where they appear as source terms.  These forces are obtained by projecting the stress-energy equation onto the space-like directions orthogonal to the local four-velocity \citep{Moller2015,White2020,Utsumi2022}.  The detailed derivation is provided in \autoref{appendix:four_force}.  Here, we summarize the geodesic equation along with the corresponding force components:
\begin{equation}
    \frac{du^{\alpha}}{d\tau} = f^{\alpha}_{\mathrm{frame}} + f^{\alpha}_{\mathrm{gas}} + f^{\alpha}_{\mathrm{pmag}} + f^{\alpha}_{\mathrm{tmag}} + f^{\alpha}_{\mathrm{rad}}
    \ ,
\end{equation}
where $f^{\alpha}_{\mathrm{frame}}$ is the frame force arising from the spacetime geometry, $f^{\alpha}_{\mathrm{gas}}$ is the gas pressure force, $f^{\alpha}_{\mathrm{pmag}}$ is the magnetic pressure force, $f^{\alpha}_{\mathrm{tmag}}$ is the magnetic tension force, and $f^{\alpha}_{\mathrm{rad}}$ is the radiation force.  These forces are defined as: 
\begin{subequations}
\begin{align}
    f^{\alpha}_{\mathrm{frame}} &= -\Gamma^{\alpha}_{\mu\nu}u^{\mu}u^{\nu}
    \ ,
    \\
    f^{\alpha}_{\mathrm{gas}} &= -\frac{1}{w}\mathcal{P}^{\alpha\nu}\nabla_{\nu}P_g
    \ ,
    \\
    f^{\alpha}_{\mathrm{pmag}} &= -\frac{1}{w}\mathcal{P}^{\alpha\nu} \nabla_{\nu}P_m
    \ ,
    \\
    f^{\alpha}_{\mathrm{tmag}} &= \frac{1}{w}\mathcal{P}^{\alpha\nu} \nabla_{\mu}b^{\mu}b_{\nu}
    \ ,
    \\
    f^{\alpha}_{\mathrm{rad}} &= \frac{1}{w}\mathcal{P}^{\alpha\nu}G_{\nu}
    \ . 
\end{align}
\label{eq:four_forces}
\end{subequations}
Here, $\Gamma^{\alpha}_{\mu\nu}$ denotes the connection coefficients, and $\mathcal{P}^{\alpha\nu}=g^{\alpha\nu}+u^{\alpha}u^{\nu}$ is the space-like projection tensor.  Since this decomposition is directly derived from the stress-energy equation solved in the simulations,  it provides valuable insight into how different physical processes drive fluid motion.  Moreover, these definitions reduce to their Newtonian counterparts in the classical limit, ensuring consistency with familiar physical intuition.  

When the system reaches force balance, the four-acceleration becomes negligible, allowing each force component to be evaluated in terms of its contribution to maintaining the disk structure.  However, in our force analysis, the evaluation of time derivatives is limited by the coarser time intervals of the full 3D data dumps, which are constrained by I/O limitations.  To overcome this, we utilize 2D sliced simulation data at $x=0$, which provides a finer time resolution of approximately $0.5$~ms ($10r_g/c$) for computing the time derivatives required in the four-force calculations, as well as shorter intervals for time averaging.   This approach, however, restricts the analysis to azimuthal averages within the 2D plane defined by $x=0$, divided into the regions $y>0$ and $y<0$.

\subsection{Angular Momentum Transport}
\label{sec:angmom_transport}

The accretion process is driven by the outward transport of angular momentum.  In this section, we first partition the angular momentum flux to identify the dominant transport mechanisms. We then compute the effective viscosity in a co-rotating frame to cross-check its consistency with the flux analysis.  Finally, we summarize and compare the results across different models to provide a comprehensive overview.

\subsubsection{Angular Momentum Flux}
\label{sec:angular_momentum_flux}

\begin{figure*}
    \centering
    \includegraphics[width=\textwidth]{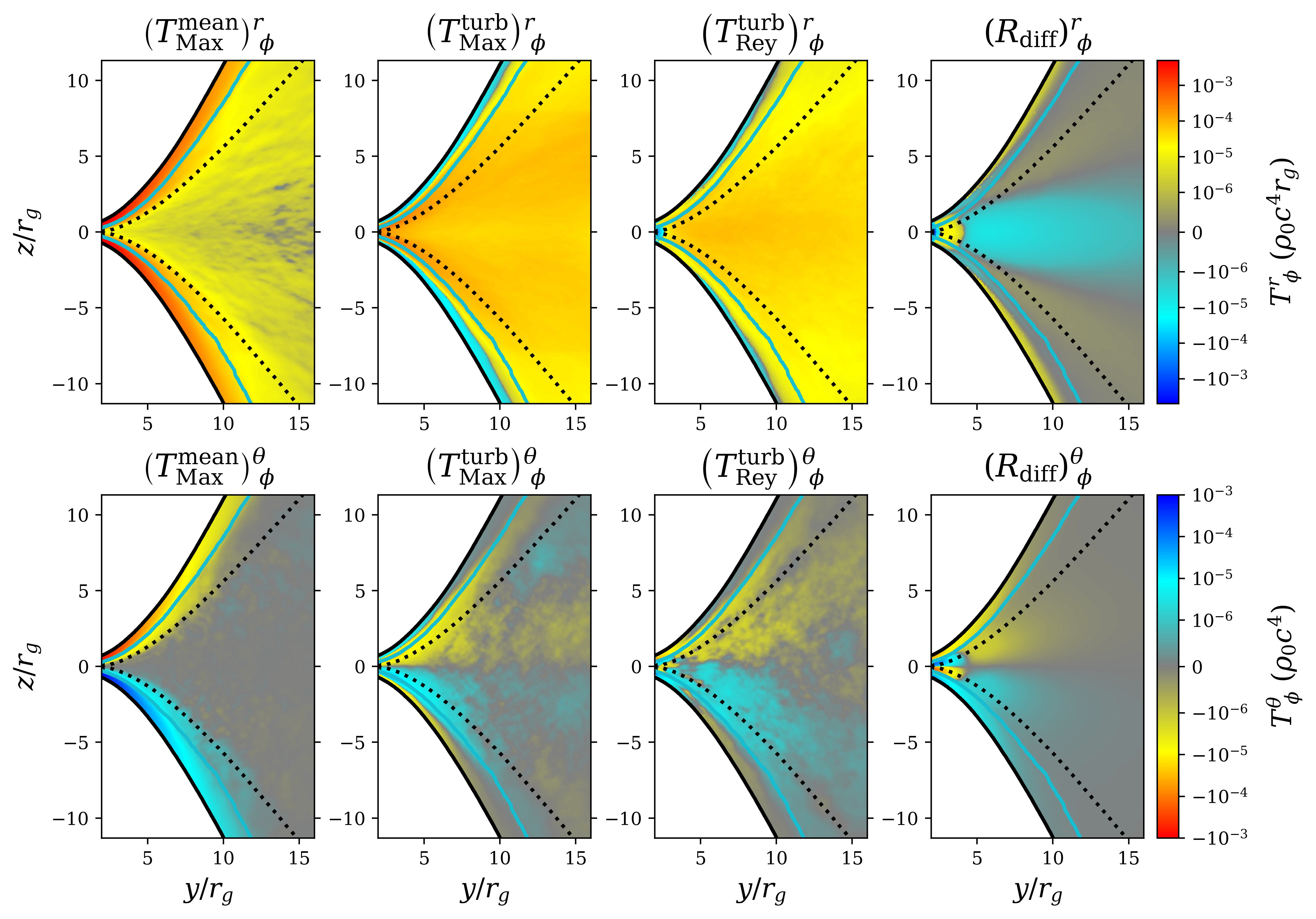}
    \caption{
    2D angular momentum flux components of model E88-a3, averaged temporally and azimuthally.  The first and second rows show the radial and polar components, respectively, of the mean-field Maxwell stress, turbulent Maxwell stress, turbulent Reynolds stress, and radiation diffusion stress.  Auxiliary lines mark the scattering photosphere (black solid), disk surface (cyan), and effective photosphere (dotted).  Detailed definitions of each component are provided in \autoref{sec:angmom_transport}.
    }
    \label{fig:angmom2d}
\end{figure*}

To identify the primary mechanism facilitating accretion, we first decompose the total angular momentum flux as
\begin{equation}
    \left(T^i_{\ \phi} + R^i_{\ \phi}\right) = \left(T_{\mathrm{Rey}}\right)^{i}_{\ \phi} + \left(T_{\mathrm{Max}}\right)^{i}_{\ \phi} + \left(R_{\mathrm{diff}}\right)^{i}_{\ \phi}
    \ ,
\end{equation}
where the Reynolds ($T_{\mathrm{Rey}}$), Maxwell ($T_{\mathrm{Max}}$), and radiation diffusion ($R_{\mathrm{diff}}$) components are defined as follows: 
\begin{subequations}    
\begin{align}
    \left(T_{\mathrm{Rey}}\right)^{i}_{\ \phi} &= w_{\mathrm{disk}} u^i u_{\phi}
    \ ,
\end{align}
\begin{align}
    \left(T_{\mathrm{Max}}\right)^{i}_{\ \phi} &= -b^i b_{\phi}
    \ ,
    \\
    \left(R_{\mathrm{diff}}\right)^{i}_{\ \phi} &= R^{i}_{\ \phi} - \frac{4}{3}\bar{E}_r u^i u_{\phi}
    \ . 
\end{align}
\end{subequations}
The Reynold stress is weighted by the total enthalpy, including radiation, since gas and radiation are well coupled within the disk region.  Accordingly, the radiation advection term is subtracted from the total radiation stress, isolating the remaining term that primarily represents radiation diffusion.  

To separate mean and turbulent contributions, we further partition the Reynolds and Maxwell stresses as follows:
\begin{subequations}    
\begin{align}
    \left(T_{\mathrm{Rey}}^{\mathrm{mean}}\right)^{i}_{\ \phi} &= \frac{\left<w_{\mathrm{disk}}u^i\right>_{\phi} \left<w_{\mathrm{disk}}u_{\phi}\right>_{\phi}}{\left<w_{\mathrm{disk}}\right>_{\phi}}
    \ ,
    \\
    \left(T_{\mathrm{Rey}}^{\mathrm{turb}}\right)^{i}_{\ \phi} &= \left(T_{\mathrm{Rey}}\right)^{i}_{\ \phi} - \left(T_{\mathrm{Rey}}^{\mathrm{mean}}\right)^{i}_{\ \phi}  
    \ ,
    \\
    \left(T_{\mathrm{Max}}^{\mathrm{mean}}\right)^{i}_{\ \phi} &= -\left<b^i\right>_{\phi} \left<b_{\phi}\right>_{\phi}
    \ ,
    \\
    \left(T_{\mathrm{Max}}^{\mathrm{turb}}\right)^{i}_{\ \phi} &= \left(T_{\mathrm{Max}}\right)^{i}_{\ \phi} - \left(T_{\mathrm{Max}}^{\mathrm{mean}}\right)^{i}_{\ \phi}
    \ ,
\end{align}
\label{eq:angmom_partition}
\end{subequations}
where $T_{\mathrm{Rey}}^{\mathrm{mean}}$ represents the mean-flow Reynolds stress, and the turbulent component $T_{\mathrm{Rey}}^{\mathrm{turb}}$ is defined as the residual beyond the mean flow, i.e., the difference between the total and mean-flow Reynolds stresses.  Similarly, $T_{\mathrm{Max}}^{\mathrm{mean}}$ denotes the mean-field Maxwell stress, while $T_{\mathrm{Max}}^{\mathrm{turb}}$ captures the turbulent field component.  Note that we define the mean flow and mean field using azimuthal averages, which treat all non-axisymmetric components as turbulent.  Given the axisymmetric nature of the disk and the alignment between the disk rotation and the black hole spin, this assumption is reasonable, provided the time average is sufficiently long.  In our analysis, we apply a time average of $10000r_g/c$, corresponding to approximately 200 and 25 orbital periods at $4r_g$ and $16r_g$, respectively.  

In \autoref{fig:angmom2d}, we use model E88-a3 as an example to show the time- and azimuthally averaged 2D profiles of angular momentum transport, decomposed into flux components in both the radial and polar directions. Auxiliary lines mark the scattering photosphere (black solid lines), disk region (cyan lines), and effective photosphere (dotted lines).  Since the mean flow tracks the accretion process, the mean-flow Reynolds stress consistently transports angular momentum inward.  Therefore, we focus on the turbulent Reynolds component, Maxwell stress, and radiation contribution, which are responsible for outward angular momentum transport. 

As shown in the first row of \autoref{fig:angmom2d}, outward angular momentum transport is dominated by the Maxwell stress: the mean-field component is strongest near the disk surface, while the turbulent component dominates within the main disk body.  The turbulent Reynolds stress is subdominant, and radiation diffusion has negligible impact.  The second row shows vertical angular momentum transport, where turbulence and radiation diffusion carry angular momentum away from the midplane, thereby facilitating accretion within the disk body. 

\subsubsection{Effective Viscosity}
\label{sec:eff_viscosity}

Similar to the angular momentum flux, we compute the effective viscosity to evaluate local angular momentum transport and relate it to the classical $\alpha$-prescription.  While both diagnostics characterize angular momentum transfer, they differ in a subtle way: the flux analysis uses conservative quantities in the coordinate frame, whereas effective viscosity is defined in a co-rotating frame.  Below, we present the formulation and demonstrate through a consistency check that both approaches yield nearly identical results with negligible differences.  

To compute the effective viscosity in the general relativistic framework, we boost into the co-rotating tetrad frame, where the mean flow is defined by azimuthal averages as in \autoref{sec:angular_momentum_flux}.  This approach follows \citet{Krolik2005} (see also \citealt{Penna2013, White2019}) and is summarized in \autoref{appendix:viscosity}.  The expressions for the different viscosity components are given below:

\begin{subequations}    
\begin{align}
    \alpha_{\mathrm{Rey}}^i &= \left<
    \frac{w_{\mathrm{disk}}u^{\tilde{i}}u^{\tilde{\phi}}}{P_{\mathrm{tot}}}
    \right>_{\phi}
    \ , 
    \\
    \alpha_{\mathrm{Max}}^i &= \left<
    \frac{-b^{\tilde{i}}b^{\tilde{\phi}}}{P_{\mathrm{tot}}} 
    \right>_{\phi}
    \ , 
    \\
    \alpha_{\mathrm{diff}}^i &= \left< 
    \frac{R^{\tilde{i}\tilde{\phi}}}{P_{\mathrm{tot}}}
    - \frac{4\bar{E}_r u^{\tilde{i}}u^{\tilde{\phi}}}{3P_{\mathrm{tot}}} 
    \right>_{\phi}
    \ , 
\end{align}
\end{subequations}
where the tilde denotes quantities measured in the co-rotating tetrad frame, and $P_{\mathrm{tot}} = P_g + b^{\nu}b_{\nu}/2 + \bar{E}_r/3$ is the total pressure.  Similar to the angular momentum flux, we further decompose the Reynolds component to isolate the turbulent contribution: 
\begin{subequations}
\begin{align}
    \alpha_{\mathrm{Rey, mean}}^i &= \left< \frac{\left<w_{\mathrm{disk}}u^{\tilde{i}}\right>_{\phi}\left<w_{\mathrm{disk}}u^{\tilde{\phi}}\right>_{\phi}}{\left<w_{\mathrm{disk}}\right>_{\phi}P_{\mathrm{tot}}}
    \right>_{\phi}
    \ , 
    \\
    \alpha_{\mathrm{Rey, turb}}^i &= \alpha_{\mathrm{Rey}}^i - \alpha_{\mathrm{Rey, mean}}^i
    \ . 
\end{align}
\end{subequations}

An alternative estimate of the effective viscosity is to rescale the coordinate-frame angular momentum flux by the total pressure: 
\begin{subequations}    
\begin{align}
    \widehat{\alpha}^{r} &= \frac{1}{\sqrt{g_{\phi\phi}}}\frac{T^r_{\ \phi}}{P_{\mathrm{tot}}}
    \ ,
    \\
    \widehat{\alpha}^{\theta} &= \sqrt{\frac{g_{\theta\theta}}{g_{\phi\phi}}}\frac{T^{\theta}_{\ \phi}}{P_{\mathrm{tot}}}
    \ , 
\end{align}
\label{eq:viscosity_approx}
\end{subequations}
where the wide hat denotes the effective viscosity estimated by this approximation.  Across all simulations, this estimate in the main disk body closely matches the viscosity measured in the co-rotating tetrad frame.  For example, \autoref{fig:alpha2d} in \autoref{appendix:viscosity} presents a side-by-side comparison for model E88-a3, demonstrating the consistency between the two approaches.  The last two columns of \autoref{tab:sim_overall} presents power-law fits to the Maxwell and turbulent Reynolds components of the pressure-scaled radial angular momentum flux for all models.  Interestingly, both high-spin radiative models exhibit an approximately constant value of $\alpha \approx 0.04$.

\subsubsection{Comparison across Simulations}

\begin{figure*}
    \centering
    \includegraphics[width=\textwidth]{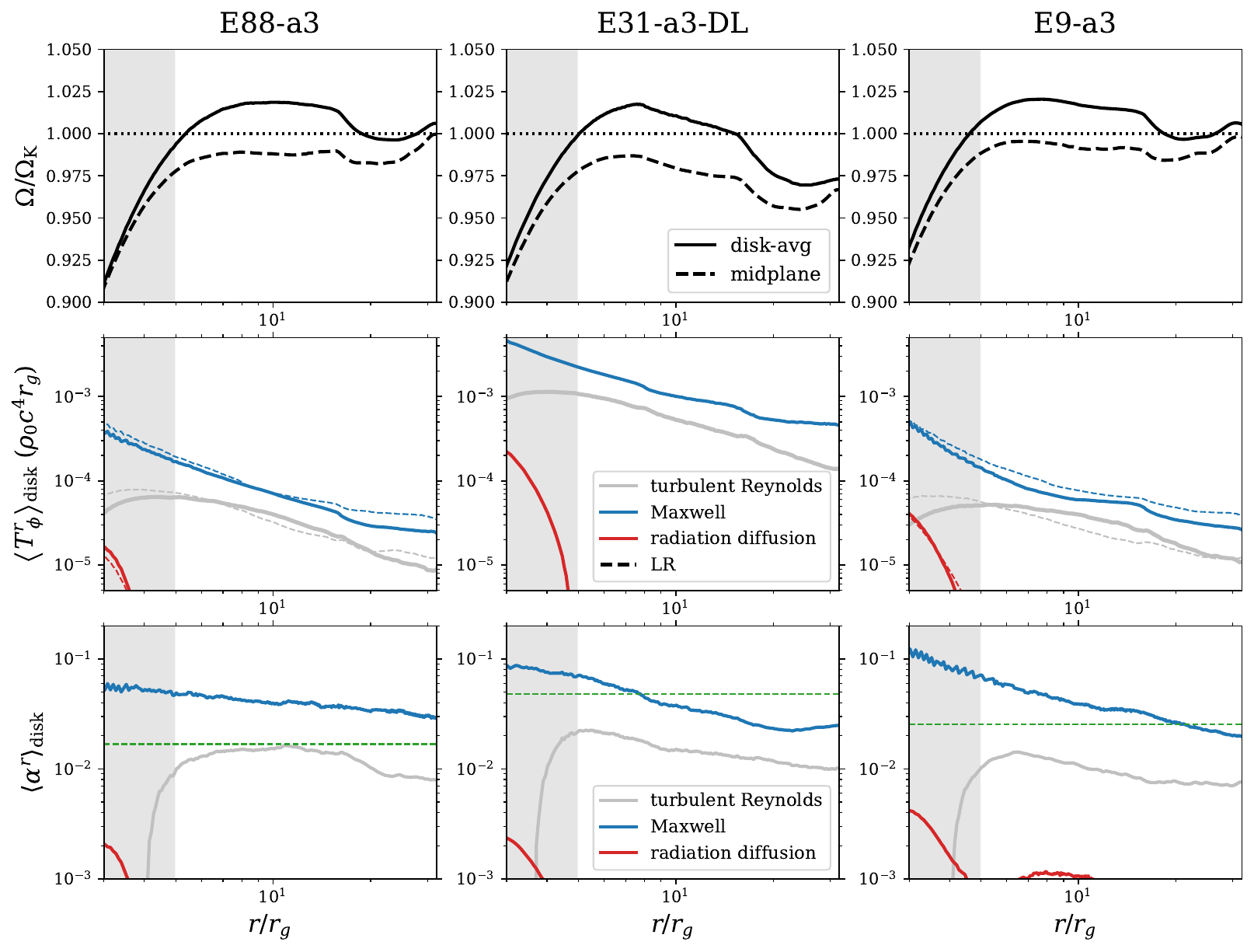}
    \caption{
    1D profiles of angular velocity, radial angular momentum flux, and radial effective viscosity for selected models.  In the first row, angular velocity is normalized by the Keplerian speed and shown as disk-averaged (solid) and midplane (dashed) profiles.  The second and third rows show angular momentum flux and effective viscosity, respectively, averaged over the disk body.  Each includes stress components from turbulent Reynolds (gray), total Maxwell (blue), and radiation diffusion (red).  In the second row, dashed lines represent measurements from the corresponding low-resolution models.  In the third row, horizontal green dashed lines denote the best-fit $\alpha$ values inferred from the slim disk model.  Details on these measurements are provided in \autoref{sec:angmom_transport}.  
    }
    \label{fig:angmom1d}
\end{figure*}

\autoref{fig:angmom1d} shows the radial profiles of angular velocity, angular momentum flux, and effective viscosity for three selected models.  In the first row, angular velocity is normalized by the Keplerian speed, defined as: 
\begin{equation}
    \Omega_{\mathrm{K}} = \left( \sqrt{\frac{r^3}{M}} + a \right)^{-1}
    \ .
\end{equation}
Disk-averaged and midplane angular velocities are shown as solid and dashed lines, respectively.  The disk-averaged angular velocity is computed as 
\begin{equation}
    \Omega^{(\mathrm{disk})} = \frac{\left<w u^{\phi}\right>_{\mathrm{disk}}}{\left<w u^t\right>_{\mathrm{disk}}}
    \ .
\end{equation}
Most of the disk remains close to Keplerian rotation, while the inner region becomes increasingly sub-Keplerian.  The midplane exhibits a more sub-Keplerian profile than the disk average, indicating that accretion predominantly proceeds through inflow near the midplane. 

The second row shows the stress components responsible for outward angular momentum transport, with dashed lines representing the corresponding low-resolution results, where applicable.  All quantities are computed as disk averages from time- and azimuthally averaged data.  In the super-Eddington regime, Maxwell stress (blue) dominates, confirming that accretion remains primarily MRI-driven in radiation-dominated disks.  Turbulent Reynolds stress (gray) contributes subdominantly to outward angular momentum transport, while radiation enthalpy accounts for up to 10\% through advection.  In contrast, radiation diffusion (red) has a negligible effect in all cases.  The agreement in Maxwell stress between intermediate- and low-resolution runs suggests a degree of numerical convergence, as further discussed in \autoref{sec:resolution_effect}. 

The last row presents the radial profiles of effective viscosity measured in the co-rotating tetrad frame, computed as disk averages from time- and azimuthally averaged data.  The results align with the angular momentum flux analysis: Maxwell stress dominates, turbulent Reynolds stress is subdominant, and radiation diffusion has a negligible impact on outward angular momentum transport.  Both the Maxwell and turbulent Reynolds components follow power-law trends, although the latter declines sharply in the plunging region, likely due to geodesic stretching of the turbulent flow near ISCO \citep{Mummery2024a,Rule2025}.  \lz{The persistence of power-law stress extending to the ISCO implies a nonzero torque at the ISCO \citep{Gammie1999,Krolik1999}.}  For each model, the best-fit $\alpha$ value is selected from the range applied in the slim disk model to reproduce the temperature profile, as indicated by the green shading in \autoref{fig:hori_compare}, and is shown as a horizontal green dashed line.   This value agrees well with the numerical result in model E31-a3-DL, but appears moderately underestimated in models E88-a3 and E9-a3.  However, since the disk profile can be highly sensitive to the choice of $\alpha$, such discrepancies are likely to lead to significant deviations from the outcomes of our numerical models. 

\subsection{Outflows and Jet}
\label{sec:outflow_and_jet}

\subsubsection{Wind Properties}
\label{sec:wind_properties}

\begin{figure}
    \centering
    \includegraphics[width=\columnwidth]{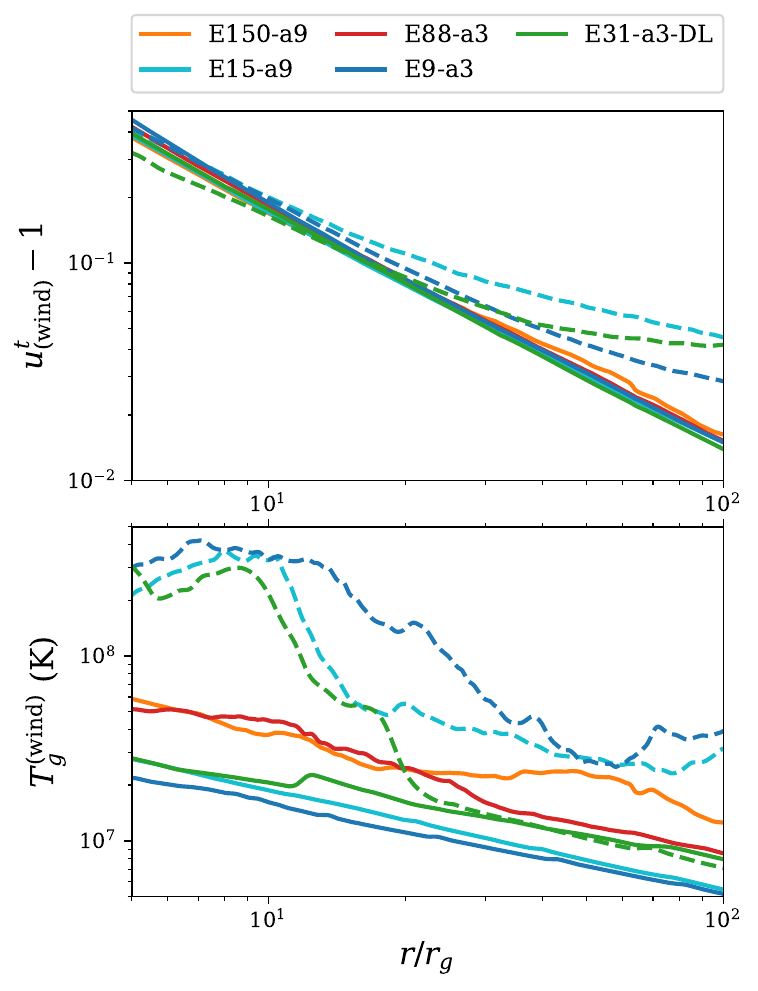}
    \caption{
    1D profiles of wind speed (upper) and temperature (lower), averaged over the optically thick and thin wind regions using temporally and azimuthally averaged data.  Solid lines represent measurements in the optically thick wind, while dashed lines correspond to the optically thin wind.  For model E150-a9 and E88-a3, the wind is predominantly optically thick, so measurements for the optically thin component are omitted. 
    }
    \label{fig:wind_profile}
\end{figure}

The radiation-driven outflow originates near the inner disk surface and can propagate to large radii. High-altitude components escape as unbound winds, gradually cooling via radiation, while low-altitude components near the disk surface (failed wind) decelerate due to gravity and magnetic drag before eventually rejoining the accretion flow.  The origin of the wind is evident from: (1) the force balance analysis (last row of \autoref{fig:vert_compare}), where the radiation force exceeds gravity near the disk surface, and (2) the velocity streamlines (upper left panel of \autoref{fig:profile2d} or Figure~4 in \citetalias{PaperI}), which show that the wind is primarily launched from the inner disk, as revealed by backward tracing of the fluid trajectories.

In the super-Eddington regime, the radiation-driven wind can become optically thick.  As shown in \autoref{fig:profile2d}, the wind region lies between the orange solid and gray dashed lines.  The scattering photosphere, marked by the gray solid line, separates the optically thick wind below from the optically thin wind above.  In model E88-a3, the wind is predominantly opaque, whereas in E9-a3, it comprises both optically thick and thin components. 

\autoref{fig:wind_profile} presents wind-averaged 1D profiles of the time component of the four-velocity and the gas temperature. Solid and dashed lines represent measurements in the optically thick and thin parts of the wind, respectively. These quantities are computed from temporally and azimuthally averaged data as follows:
\begin{subequations}    
\begin{align}
    u^t_{(\mathrm{wind})} &= \frac{\left<wu^t\right>_{\mathrm{wind}}}{\left<w\right>_{\mathrm{wind}}}
    \ ,
    \\
    T_g^{(\mathrm{wind})} &= \frac{\mu m_p}{k_B} \frac{\left<P_g\right>_{\mathrm{wind}}}{\left<\rho\right>_{\mathrm{wind}}}
    \ ,
\end{align}
\end{subequations}
where the subscript `wind' refers to either the optically thin or thick portion of the wind.  In both panels of \autoref{fig:wind_profile}, we only plot the optically thin components for models in the moderately super-Eddington regime (i.e., E9-a3, E15-a9, and E31-a3-DL), as the wind regions in the highly super-Eddington regime (i.e., E150-a9 and E88-a3) are nearly entirely optically thick.  

In the upper panel of \autoref{fig:wind_profile}, the wind in the super-Eddington regime is mildly relativistic, with velocities exceeding 15\% of the speed of light.  The wind speed gradually decreases with radius \lz{as} it propagates outward.  In the lower panel, the gas temperature also declines with increasing radius.  Within the optically thick wind, the temperature follows a power-law decline and remains comparable to the disk temperature, whereas the optically thin component is generally hotter, indicating the presence of a hot corona.

\subsubsection{Jet Properties}
\label{sec:jet_properties}

\begin{figure}
    \centering
    \includegraphics[width=\columnwidth]{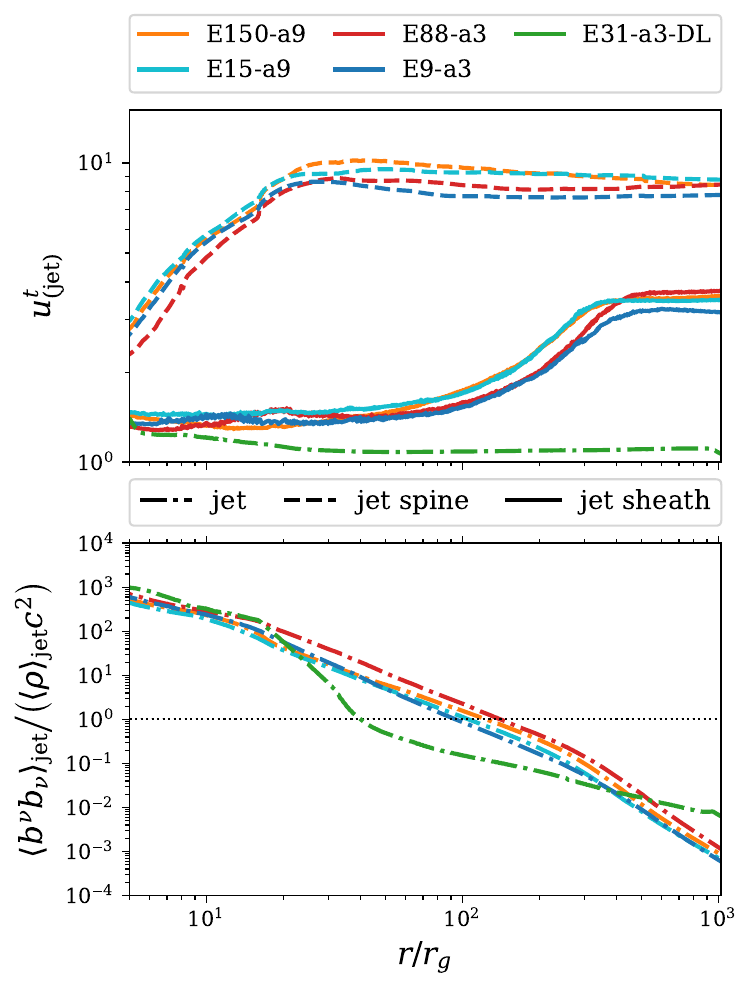}
    \caption{
    1D profiles of jet speed and magnetization, averaged over different jet regions using time- and azimuthally averaged data.  Each model is shown in a different color.     
    {\bf Upper panel:} Time component of jet four-velocity.  The strong jet is averaged separately over the jet sheath (solid) and jet spine (dashed), while the weak jet is averaged over the entire jet region (dot-dashed). 
    {\bf Lower panel:} Magnetization averaged over the full jet region (dot-dashed), with the unity value indicated by a dotted line. 
    }
    \label{fig:jet_profile}
\end{figure}

Similar to the wind analysis, we divide the relativistic jet, characterized by an opening angle $\theta^{(\mathrm{jet})}$ \lz{(see also Section~3.4 of \citetalias{PaperI})}, into two regions: the jet spine and the jet sheath.  The jet spine, which focuses on the central region, is defined within an opening angle of $\min(\theta^{(\mathrm{jet})}, r_g/r)$, while the jet sheath encompasses the remaining outer portion.  \lz{In practice, for most radii where $\theta^{(\mathrm{jet})}\ge r_g/r$, the jet spine corresponds to a cylinder of radius $\sim 1\ r_g$, while the sheath occupies the surrounding funnel region.}

\autoref{fig:jet_profile} shows 1D profiles of jet speed and magnetization, averaged over the northern and southern jets.  The jet speed is represented by the time component of the four-velocity $u^t_{\mathrm{(jet)}}$, computed similarly to $u^t_{\mathrm{(wind)}}$ but integrated over the jet spine (dashed lines) and sheath (solid lines).  The fluid magnetization, defined as the ratio of magnetic to rest-mass energy, is averaged over the entire jet region (dot-dashed lines).  For model E31-a3-DL, initialized with a double-loop magnetic configuration, only a weak jet develops; thus, we do not separate jet components and instead integrate over the entire jet region. 

All single-loop models produce strong jets launched near the horizon that accelerate to relativistic speeds, remaining well below the numerical ceiling ($\sqrt{-1/g^{00}}u^t\le 20$).  As shown in the upper panel of \autoref{fig:jet_profile}, the jet spine reaches peak velocity around $25r_g$ and then remains nearly constant.  The jet sheath, though generally slower, begins accelerating beyond $\sim 100r_g$, where the jet loses strong magnetic confinement and transitions into a freely streaming state.  This transition is evident in the lower panel, where the fluid magnetization declines with radius following a power law and drops below unity around $100r_g$.  In contrast, the double-loop model E31-a3-DL produces only a weak jet with significantly lower speed.  Further details are provided in \autoref{sec:weak_jet}.  

The gas temperature at the jet base above the photosphere can reach $\sim 10^{13}$~K for a strong jet and $\sim 10^{8}$~K for a weak jet.  However, these funnel-region measurements should be interpreted qualitatively for physical insight rather than precise quantification, due to: (1) the density approaching the floor value and (2) incomplete thermal physics.  \lz{In particular, our models assume a single-temperature plasma, which may not be appropriate in the highly magnetized, optically thin funnel region where electrons and ions can decouple. In addition,} the Compton approximation (i.e., the last term in equation~4a of \citetalias{PaperI}) breaks down in the funnel region where the radiation field becomes highly anisotropic, and pair creation is not included in the current models.  In future work, we plan to incorporate a more accurate treatment of Compton scattering using \lz{multiple frequency groups, and including pair production}. 

\subsubsection{Strong Jet}
\label{sec:strong_jet}

\begin{figure}
    \centering
    \includegraphics[width=\columnwidth]{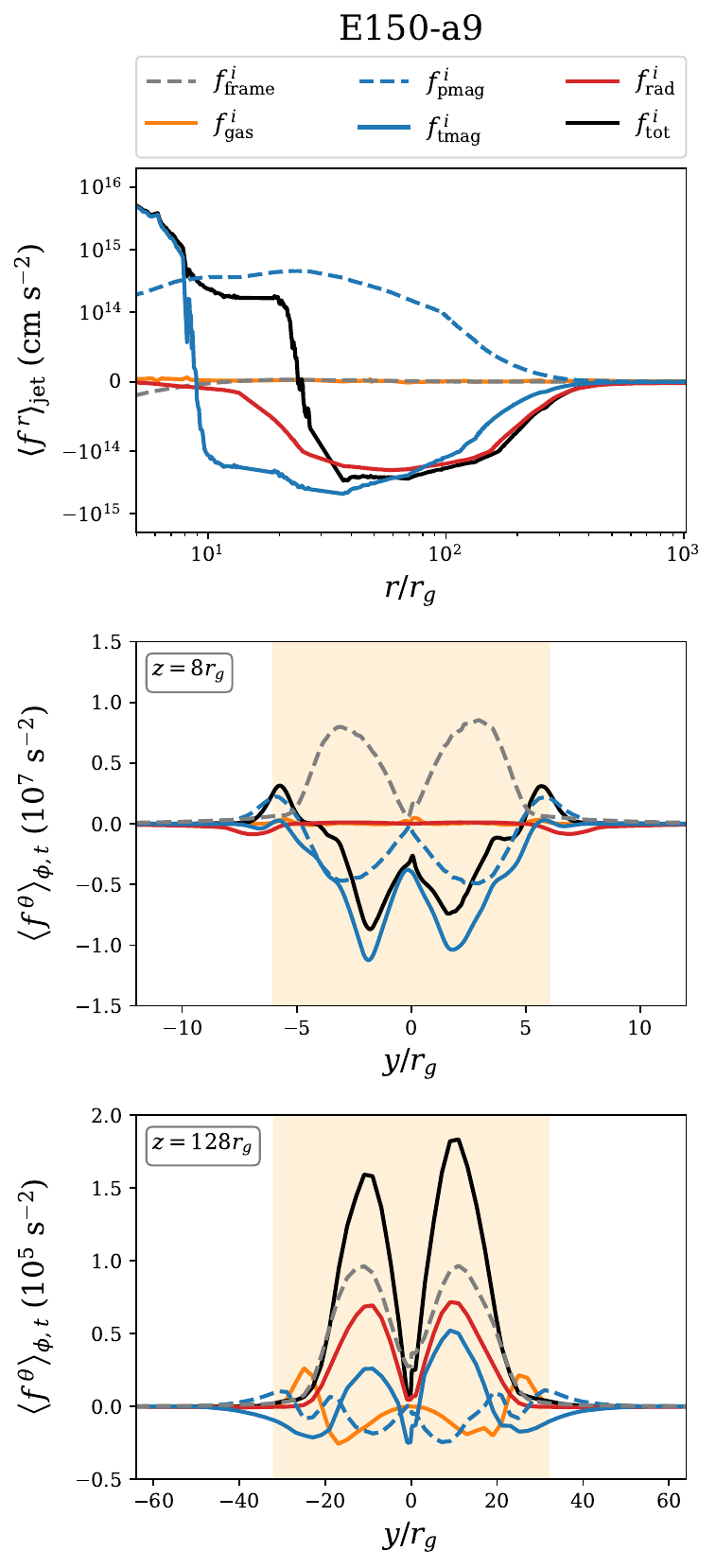}
    \caption{
    1D profiles of jet four-forces for model E150-a9, which features the strongest jet, based on time- and azimuthally averaged data.  The four-forces are defined by equation~\autoref{eq:four_forces}, including the frame force (gray dashed), gas pressure force (orange solid), magnetic pressure force (blue dashed), magnetic tension force (blue solid), and radiation force (red).  The total force is shown as a black solid line.  The upper panel presents the radial forces averaged over the jet.  The middle and lower panels show the lateral forces near and far from the jet launching region, measured at radii of $8r_g$ and $128r_g$, respectively.  The jet region is highlighted in yellow. 
    }
    \label{fig:jet_force}
\end{figure}

We apply a four-force analysis (see \autoref{appendix:four_force} for details) to understand the formation of strong jets in single-loop models.  As an example, \autoref{fig:jet_force} presents results from model E150-a9, which produces the highest jet power, illustrating the mechanisms responsible for jet propulsion and collimation.  

The upper panel shows the radial forces averaged over the jet region.  The frame (gray) and gas pressure (orange) forces have negligible influence on the jet dynamics.  Instead, the jet is primarily driven by magnetic (blue) and radiation (red) forces.  The radiation force is consistently negative and acts as a drag, whereas the magnetic pressure force (blue dashed) is always positive and provides continuous outward acceleration.  The magnetic tension force (blue solid) exhibits a dual behavior: it accelerate the jet near the black hole but transitions into a braking force at larger radii.  

The total force (black) provides an overview of jet formation, tracing its launch near the black hole to larger radii where it transitions into a freely streaming state.  The jet is launched within a compact region ($\lesssim 10r_g$), primarily driven by magnetic tension.  As it propagates outward, the magnetic field becomes `overstretched', turning the tension force negative and exerting a drag on the flow.  At this stage, magnetic pressure takes over, continuing to accelerate the jet.  Beyond $\sim 25r_g$, magnetic tension force once again dominates, gradually braking the flow.  Meanwhile, radiation consistently exerts a drag, which becomes increasingly important at larger radii.  Eventually, the total force diminishes, and the jet transitions into free streaming.  This trend is consistent with the 1D velocity profiles in \autoref{fig:jet_profile}, where the jet spine reaches peak velocity near $25r_g$, then decelerates, and ultimately streams freely. 

The middle and lower panels of \autoref{fig:jet_force} demonstrate jet collimation through lateral forces, using the same color and line styles as in the upper panel.  The jet region is highlighted in yellow.  In the middle panel, forces are measured at $8r_g$ within the jet launching region, showing that the jet is magnetically dominated and primarily confined laterally by magnetic tension.  The lower panel presents force measurements at $128r_g$, where the jet begins transitioning into the free-streaming regime and loses its lateral confinement.  At this stage, gravity, radiation and magnetic pressure forces all act to expand the jet opening angle.  This behavior further explains the velocity evolution in the jet sheath: as the jet enters the free-streaming regime and loses lateral confinement beyond $\sim 100r_g$, the high-velocity flow from the jet spine expands laterally, merges into the jet sheath, and boosts its overall velocity. 

\subsubsection{Weak Jet}
\label{sec:weak_jet}

\begin{figure}
    \centering
    \includegraphics[width=\columnwidth]{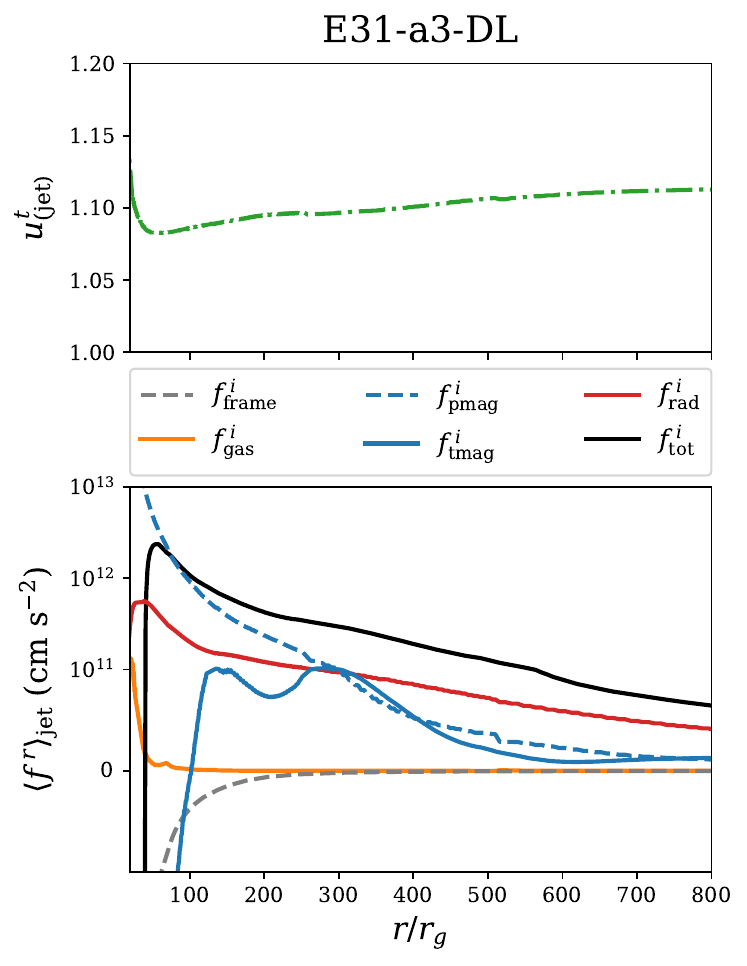}
    \caption{
    1D radial profiles of jet speed and four-forces for model E31-a3-DL, which features a weak jet, based on time- and azimuthally averaged data.  Both quantities are averaged over the entire jet region, with the four-forces defined by equation~\autoref{eq:four_forces} and each component labeled in the legend. 
    }
    \label{fig:jet_weak_profile}
\end{figure}

Unlike the strong, steady jets produced in all single-loop models, the double-loop model E31-a3-DL forms a weak and intermittent jet.  The jet becomes more persistent only after sufficient net vertical magnetic flux accumulates near the black hole, as radiation-driven outflows stochastically advect magnetic fields and break the initial magnetic symmetry.  Even then, the jet remains too weak to fully evacuate the funnel region, resulting in jet properties that differ significantly from those of strong MHD jets.  In this weak jet, radiation plays a more prominent role in driving the outflow at large radii. 

\begin{figure*}
    \centering
    \includegraphics[width=\textwidth]{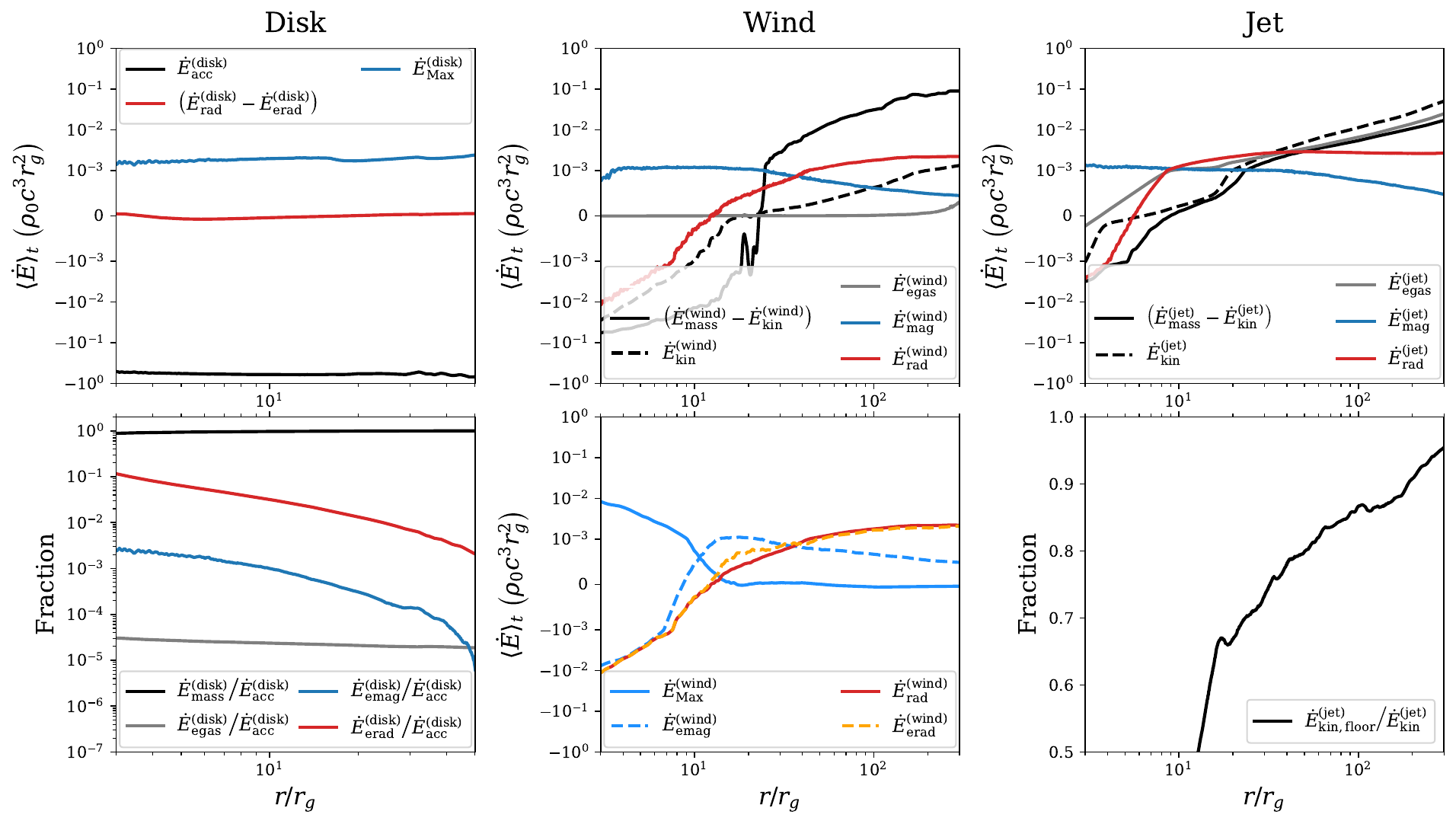}
    \caption{
    Energetic analysis of E88-a3 based on time-averaged simulation data.  The top row present the energy partitioning of heating and cooling processes across different regions.  The bottom row further breaks down the heating contributions within the disk, analyzes the cooling contributions in the wind region, and evaluates the reliability of the kinetic jet power. 
    }
    \label{fig:energy1d}
\end{figure*}

\autoref{fig:jet_weak_profile} presents 1D profiles of velocity and forces, with radius scaled linearly for clarity.  As shown in the upper panel, the jet gradually accelerates after experiencing a sharp deceleration near $20r_g$, caused by the drag of magnetic tension, similar to that observed in strong MHD jets.  The lower panel shows the radial forces averaged over the weak jet.  The total force remains positive at large radii, with magnetic forces dominating the acceleration up to $\sim 300r_g$, and radiation taking over beyond that radius.  Unlike strong MHD jets, this weak jet is powered by a combination of magnetic fields and radiation.

\subsection{Energetics}
\label{sec:energetics}

In this section, we analyze the energetics of the accretion system by separately examining the disk, wind, and jet regions, as well as the radiation luminosity emerging from the photosphere.  We then summarize the energetics of all super-Eddington models in a table. 

\subsubsection{Energy Transport}
\label{sec:energy_transport}

In the super-Eddington regime, all simulations reach a steady state with broadly similar energetics.  As an example, \autoref{fig:energy1d} presents the energy analysis of model E88-a3, based on time and azimuthal averages.  The top row shows the individual energy components within each region of the system, while the bottom panels delve into the detailed heating and cooling mechanisms.  The energy transport decompositions in each region, along with their corresponding definitions, are provided in \autoref{appendix:energy_decomposition}. 

In the disk region (upper-left panel), the accretion process ($\dot{E}_{\mathrm{acc}}^{(\mathrm{disk})}$, black) provides the only energy input, while magnetic convection ($\dot{E}_{\mathrm{Max}}^{(\mathrm{disk})}$, blue) carries away about 0.1\% of the accretion power.  Based on the decomposition of the accretion power (lower-left panel), fluid rest-mass inward advection ($\dot{E}_{\mathrm{mass}}^{(\mathrm{disk})}$, black) dominates, radiation advection ($\dot{E}_{\mathrm{rad}}^{(\mathrm{disk})}$, red) contributes up to 10\%, and the remaining components (i.e., gas thermal energy $\dot{E}_{\mathrm{egas}}^{(\mathrm{disk})}$ and magnetic advection $\dot{E}_{\mathrm{emag}}^{(\mathrm{disk})}$) are insignificant ($<1\%$).  

In the wind region (upper-middle panel), the Poynting flux ($\dot{E}_{\mathrm{mag}}^{(\mathrm{wind})}$, blue) dominates cooling in the inner region but decreases with radius, while radiative cooling ($\dot{E}_{\mathrm{rad}}^{(\mathrm{wind})}$, red) and kinetic outflow ($\dot{E}_{\mathrm{kin}}^{(\mathrm{wind})}$, black dashed) become increasingly important at larger radii.  The black solid line, which accounts for both rest-mass and gravitational binding energy, primarily reflects the mass loss.  The lower-middle panel provides a detailed breakdown of magnetic and radiative cooling.  Near the black hole, magnetized fluid and trapped radiation are primarily advected inward by the accretion process. \lz{In the time- and azimuthally averaged wind region, fluctuations of the disk surface can mix accreting material into the wind mask, and the resulting inward advection can produce negative kinetic and radiative fluxes; this inward motion is further enhanced by strong relativistic gravity near and within the plunging region}. Nonetheless, magnetic convection ($\dot{E}_{\mathrm{Max}}^{(\mathrm{wind})}$, blue solid) remains the dominant cooling mechanism over magnetic advection ($\dot{E}_{\mathrm{emag}}^{(\mathrm{wind})}$, blue dashed), resulting in an outward Poynting flux.  At larger radii, radiation-driven outflows develop, and outward advection becomes the major cooling mechanism, mostly through radiation advection ($\dot{E}_{\mathrm{erad}}^{(\mathrm{wind})}$, orange dashed). 

In the jet region (upper-right panel), the Poynting flux ($\dot{E}_{\mathrm{mag}}^{(\mathrm{jet})}$, blue) dominates the energy outflow at small radii but gradually declines, while the radiative cooling ($\dot{E}_{\mathrm{rad}}^{(\mathrm{jet})}$, red) and gas outflow (black and gray) become increasingly significant.  Radiative cooling stabilizes beyond approximately $10r_g$, and gas outflow continues to increase at larger radii, where the funnel region is evacuated and approaches the density floor.  The energy carried by the gas outflow is primarily in the form of kinetic energy ($\dot{E}_{\mathrm{kin}}^{(\mathrm{jet})}$, black dashed), with smaller contributions from gas internal energy ($\dot{E}_{\mathrm{egas}}^{(\mathrm{jet})}$, gray) and rest-mass energy ($\dot{E}_{\mathrm{mass}}^{(\mathrm{jet})}$).  In the lower-right panel, we quantify the density floor contribution to the kinetic outflow as a function of radius, defined as:
\begin{equation}
    \dot{E}^{(\mathrm{jet})}_{\mathrm{kin,\:flr}} = \displaystyle\int_{\mathrm{jet}} \left< -\rho_{\mathrm{flr}} u^r(u_t+\sqrt{-g_{tt}})\right>_{t} \sqrt{-g} d\theta d\phi
    \ ,
\end{equation}
where $\rho_{\mathrm{flr}}$ denotes the numerical density floor.  Although the jet power may be overestimated when the funnel density approaches this floor value, the fraction of jet power attributed to the density floor offers a useful metric for quantifying the reliability of the jet power estimate.  We therefore define the density-floor contribution to the kinetic energy outflow, expressed as a percentage as
\begin{equation}
    \epsilon_{\mathrm{floor}}^{(\mathrm{jet})} = \left. \dot{E}^{(\mathrm{jet})}_{\mathrm{kin,\:flr}} \right/ \dot{E}^{(\mathrm{jet})}_{\mathrm{kin}}
    \ . 
\end{equation}

\begin{deluxetable*}{l c c c c c c c c c}
\tablecaption{Comparison of cooling efficiencies and contributions from different cooling mechanisms \label{tab:energy_eql}}
\tablehead{
\colhead{Name} & 
\colhead{$\eta_{50}^{(\mathrm{rad})}$} & 
\colhead{$\eta_{50}^{(\mathrm{wind})}$} & 
\colhead{$\eta_{50}^{(\mathrm{jet})}$} & 
\colhead{$\eta_{200}^{(\mathrm{wind})}$} &
\colhead{$\epsilon_{\mathrm{kin, 50}}^{(\mathrm{wind})}$} & 
\colhead{$\epsilon_{\mathrm{mag, 50}}^{(\mathrm{wind})}$} & 
\colhead{$\epsilon_{\mathrm{kin, 50}}^{(\mathrm{jet})}$} & 
\colhead{$\epsilon_{\mathrm{mag, 50}}^{(\mathrm{jet})}$} & 
\colhead{$\epsilon_{\mathrm{floor, 50}}^{(\mathrm{jet})}$} 
\\
& \colhead{(\%)} & \colhead{(\%)}& \colhead{(\%)}& \colhead{(\%)}& \colhead{(\%)}
& \colhead{(\%)} & \colhead{(\%)}& \colhead{(\%)}& \colhead{(\%)}
\\
\quad\;\;\;(1) & (2) & (3) & (4) & (5) & (6) & (7) & (8) & (9) & (10)
}
\startdata
    E150-a9    & 0.54 & 0.21 & 2.59 & 0.32 & 33 & 67 & 50 & 32    & 78  \\ 
    E88-a3     & 0.63 & 0.18 & 1.48 & 0.25 & 28 & 72 & 53 & 10    & 80  \\
    E31-a3-DL  & 0.60 & 0.04 & 0.01 & 0.23 & 21 & 79 & 42 & 57    & 0.4 \\
    E15-a9     & 1.95 & 0.33 & 3.15 & 0.59 & 49 & 51 & 54 & 28    & 82  \\
    E9-a3      & 2.44 & 0.20 & 1.44 & 0.43 & 54 & 46 & 58 & \:\:7 & 82  \\
    \hline
\enddata
\tablecomments{
    The subscript indicates the radius at which the measurement is taken.
    {\bf Columns (from left to right):}  
    (1) Model name; 
    (2) Radiation efficiency; 
    (3) Wind efficiency; 
    (4) Jet efficiency;
    (5) Wind efficiency measured at a larger radius (where the wind power stabilizes);
    (6) Kinetic contribution to wind cooling;
    (7) Magnetic contribution to wind cooling;
    (8) Kinetic contribution to jet cooling;
    (9) Magnetic contribution to jet cooling;
    (10) Fraction of jet kinetic energy attributed to the density floor. 
}
\end{deluxetable*}

\subsubsection{Radiation Luminosity}
\label{sec:rad_luminosity}

\begin{figure}
    \centering
    \includegraphics[width=\columnwidth]{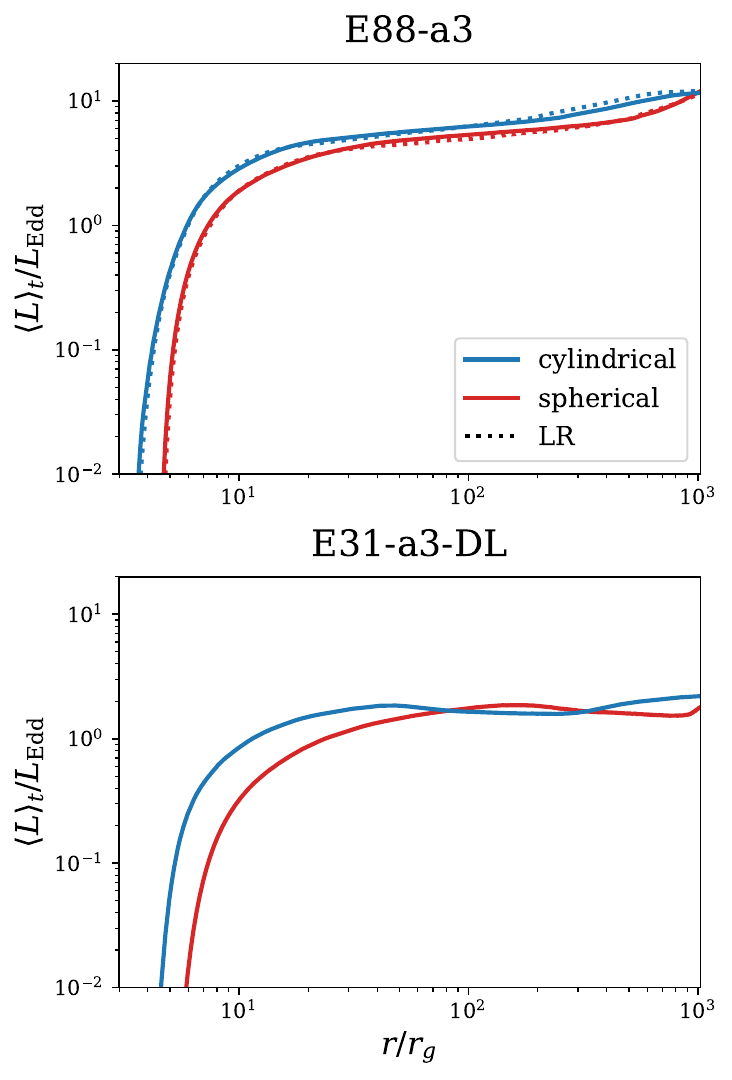}
    \caption{
    Time-averaged radiation luminosity above the scattering photosphere for models E88-a3 and E31-a3-DL.  The luminosity is integrated cylindrically (blue) and spherically (red) for comparison.  When applicable, low-resolution counterparts are overplotted with dotted lines.  In both models, most of the radiation is generated within the region in inflow equilibrium. 
    }
    \label{fig:lum_compare}
\end{figure}

We evaluate the radiation luminosity using both cylindrical and spherical surface integrations above the scattering photosphere, defined as:
\begin{subequations}
\begin{align}
    L_{\mathrm{cyl}} &= -\int_{\mathrm{ph}} \left<R^z_{\ t}ds + R^s_{\ t}dz\right>_{t} sd\phi
    \ ,
    \\
    L_{\mathrm{sph}} &= -\int_{\mathrm{ph}} \left<R^r_{\ t}\right>_{t} \sqrt{-g} d\theta d\phi
    \ ,
\end{align}
\end{subequations}
where $s=\sqrt{x^2+y^2}$ is the cylindrical radius.  The cylinder height at each radius is set by the local height of the scattering photosphere.  For radii beyond the location where the photosphere reaches its maximum height, this maximum value is adopted as the cylinder height.  

\autoref{fig:lum_compare} presents the radiation luminosities for models E88-a3 and E31-a3-DL, computed using spherical (red) and cylindrical (blue) integrations.  When applicable, their lower-resolution counterparts are shown as dotted lines for comparison.  The luminosity is normalized by the Eddington luminosity, defined as $L_{\mathrm{Edd}}=4\pi GM_{\mathrm{BH}}c/\kappa_T$.  

Since radiation escapes the accretion system primarily in the vertical direction (see coordinate-frame radiation flux streamlines in \autoref{fig:profile2d} or Figure~6 of \citetalias{PaperI}), the cylindrically integrated luminosity more accurately captures the intrinsic radiation output.  Spherical integration introduces a projection effect that reduces the measured luminosity.  However, this discrepancy diminishes at larger radii, where both integration methods converge to similar values.  In all models, more than 50\% of the radiation is produced within $50r_g$, indicating that the majority of the output luminosity originates from the steady-state accretion flow.

\subsubsection{Comparison across Simulations}

\autoref{tab:energy_eql} provides a detailed summary of the cooling processes due to radiation, wind, and jet, with measurements taken at $50r_g$ to capture the steady-state accretion flow.  Note we wind efficiency is also evaluated at $200r_g$, where it generally stabilizes beyond this radius. 

The radiative cooling efficiency is defined as 
\begin{equation}
    \eta^{(\mathrm{rad})} = \left. L_{\mathrm{cyl}} \right/ (\dot{M}c^2)
    \ ,
\end{equation}
where the cylindrically integrated luminosity is adopted, as it more accurately captures the radiation that vertically escapes from the funnel region at smaller radii. 

The cooling efficiencies in the wind and jet zones are defined by excluding radiative contributions and considering only the outflow of gas thermal energy, kinetic energy, and magnetic energy, as follows:
\begin{equation}
    \eta^{(\mathrm{zone})} = \frac{\dot{E}^{\mathrm{(zone)}}_{\mathrm{egas}} + \dot{E}^{\mathrm{(zone)}}_{\mathrm{kin}} + \dot{E}^{\mathrm{(zone)}}_{\mathrm{mag}}}{\dot{M}c^2}
    \ ,
\end{equation}
where we further define the kinetic and magnetic cooling fractions in the wind and jet zones as
\begin{subequations}
\begin{align}
    \epsilon^{(\mathrm{zone})}_{\mathrm{kin}} &= \frac{\dot{E}^{\mathrm{(zone)}}_{\mathrm{kin}}}{\dot{E}^{\mathrm{(zone)}}_{\mathrm{egas}} + \dot{E}^{\mathrm{(zone)}}_{\mathrm{kin}} + \dot{E}^{\mathrm{(zone)}}_{\mathrm{mag}}}
    \ ,
    \\
    \epsilon^{(\mathrm{zone})}_{\mathrm{mag}} &= \frac{\dot{E}^{\mathrm{(zone)}}_{\mathrm{mag}}}{\dot{E}^{\mathrm{(zone)}}_{\mathrm{egas}} + \dot{E}^{\mathrm{(zone)}}_{\mathrm{kin}} + \dot{E}^{\mathrm{(zone)}}_{\mathrm{mag}}}
    \ . 
\end{align}
\end{subequations}

As shown in \autoref{tab:energy_eql}, super-Eddington accretion systems generally exhibit low radiation efficiency, which further decreases with increasing accretion rate, primarily due to the reduced angular size of the funnel region.  Radiation-driven winds contribute even less to cooling, with their outflows predominantly carried in the form of kinetic and magnetic energy.  

The jet can significantly cool the system, with its power dominated by Poynting flux at small radii and by kinetic energy outflow at larger radii.  In the absence of sufficient net vertical magnetic flux (as in the double-loop model), the jet remains weak and produces negligible energy output.  When a strong jet forms (as in the single-loop model), it can dominate the cooling process, with higher black hole spin significantly enhancing the jet power.  In such cases, the jet power can exceed or at least become comparable to the radiation luminosity.  However, the kinetic component of the jet power may be overestimated in the simulations when the funnel density approaches the numerical floor value.

\subsection{Plunging Region}
\label{sec:plunging_region}

\begin{figure}
    \centering
    \includegraphics[width=0.98\columnwidth]{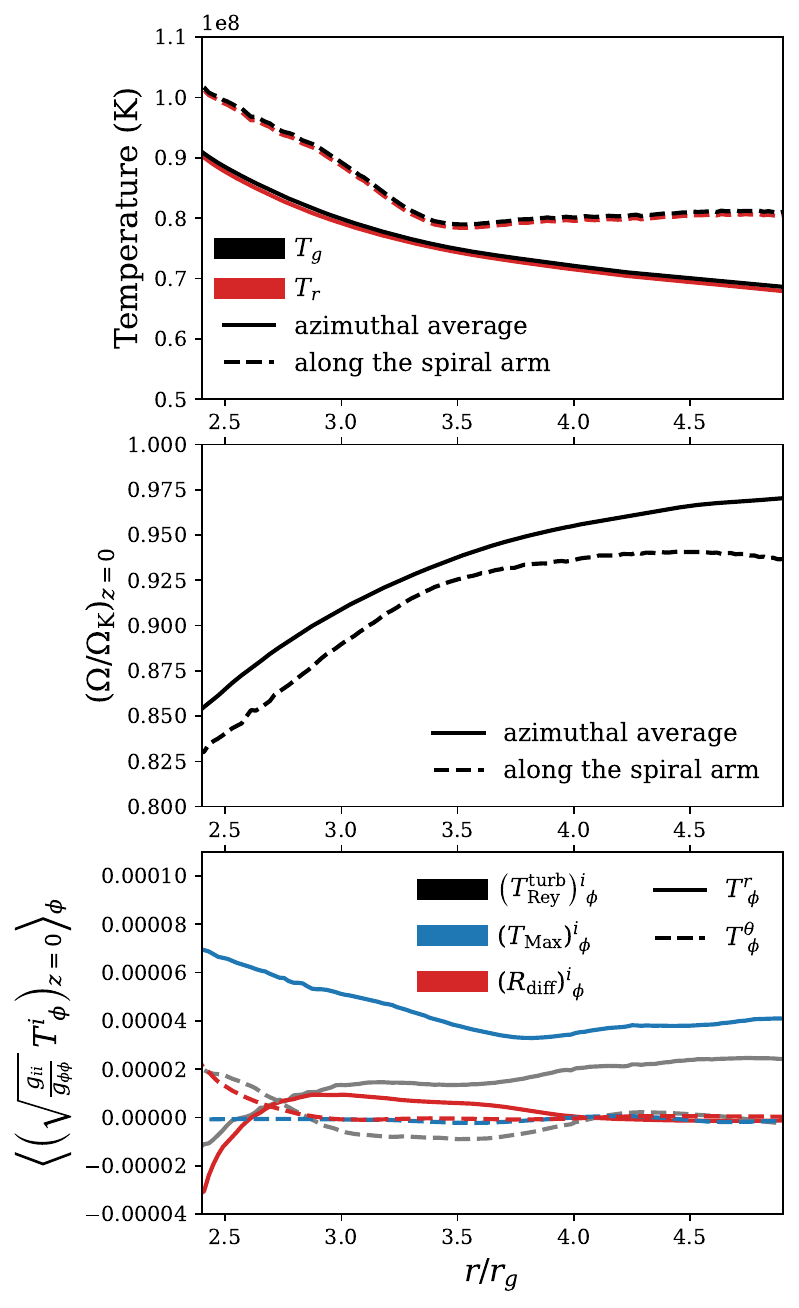}
    \caption{
    Selected fluid properties in the plunging region for model E88-a3 at the midplane ($z=0$) in the final snapshot ($t=60000r_g/c$).  
    {\bf Upper panel:} Gas (black) and radiation (red) temperatures, measured in azimuthal average (solid lines) and along a spiral arm (dashed lines).  
    {\bf Middle panel:} Angular velocity normalized by the Keplerian speed, measured in azimuthal average (solid lines) and along a spiral arm (dashed lines).  
    {\bf Lower panel:} Contributions to the outward angular momentum flux from various components, indicated by different colors (see equation~\ref{eq:angmom_partition} for detailed definitions).  Radial fluxes are shown as solid lines and poloidal fluxes as dashed lines, all normalized to a consistent unit for comparison. 
    }
    \label{fig:plunging_1d}
\end{figure}

\begin{figure}
    \centering
    \includegraphics[width=\columnwidth]{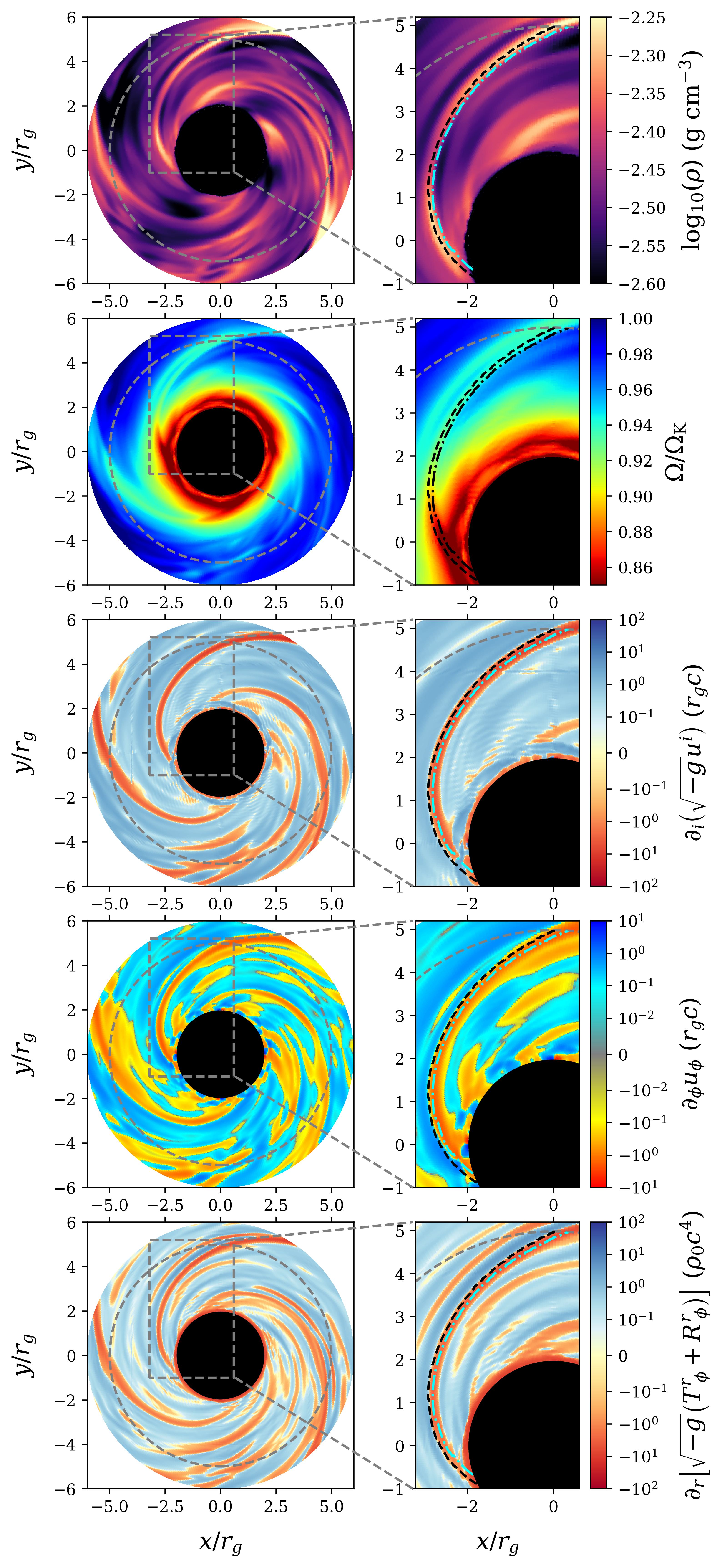}
    \caption{
    2D profiles of density, angular velocity, velocity divergence, azimuthal velocity derivative, and divergence of angular momentum flux for model E88-a3 at the midplane ($z=0$) from the final snapshot ($t=60000r_g/c$).  The left panels show the entire plunging region, outlined by gray dashed circles, with a selected spiral arm highlighted by a gray dashed box.  The right panels provide a zoomed-in view of this spiral arm.  Dashed and dot-dashed lines in the right panels trace the density maximum and minimum along the spiral arm, respectively. These auxiliary lines are shown in different colors across panels solely for visual clarity.  
    }
    \label{fig:plunging_2d}
\end{figure}

In this section, we examine the fluid properties of the plunging region using azimuthal averages.  Within this region, spiral arms form near the midplane and extend outward.  Inspired by \citet{Mummery2024b}, we also analyze the physical properties of a selected spiral arm.  Since the dynamical behavior is qualitatively similar across all simulations, we use model E88-a3 for demonstration due to its relatively large plunging region associated with low black hole spin.

We begin by examining the azimuthally averaged profiles at the midplane ($z=0$) from the last snapshot ($t=60000r_g/c$).  The solid lines in the upper panel of \autoref{fig:plunging_1d} show the azimuthally averaged gas and radiation temperatures, which are calculated as: 
\begin{subequations}    
\begin{align}
    T_g^{(\phi)} &= \frac{\mu m_p}{k_B} \frac{\left<P_g\right>_{\phi}}{\left<\rho\right>_{\phi}}
    \ , 
    \\
    T_r^{(\phi)} &= \left(\frac{\left<\bar{E}_r\right>_{\phi}}{a_r}\right)^{1/4}
    \ .
\end{align}
\end{subequations}
The gas and radiation are nearly in thermal equilibrium, with the gas temperature slightly exceeding the radiation temperature.  Both temperatures increase toward the event horizon. 

The middle panel shows the azimuthally averaged angular velocity, normalized by the Keplerian value.  The solid line represents this angular velocity, computed using the enthalpy-weighted four-velocity as
\begin{equation}
    \Omega^{(\phi)} = \frac{\left<w u^{\phi}\right>_{\phi}}{\left<w u^t\right>_{\phi}}
    \ .
\end{equation}
The angular velocity decreases gradually from near-Keplerian at the ISCO to sub-Keplerian as the radius approaches the event horizon. 

The lower panel of \autoref{fig:plunging_1d} shows the azimuthally averaged outward angular momentum transport at the midplane.  Solid and dashed lines represent the radial and poloidal flux components, respectively.  All fluxes are normalized to the same unit using the diagonal metric components.  Different contributions to the angular momentum flux are color-coded as indicated in the legend, with definitions provided in \autoref{sec:angular_momentum_flux}.  In the plunging region, angular momentum is primarily extracted by the radial Maxwell stress, which becomes increasingly dominant as the flow approaches the event horizon.  This behavior aligns with the decrease in angular velocity shown in the middle panel.  

We select a spiral arm for detailed analysis, as indicated by the gray dashed box in the left column of \autoref{fig:plunging_2d}.  The plunging region is within the gray dashed circle.  The density maximum and minimum along this spiral arm are traced by the dashed and dot-dashed lines, respectively, in the right column of \autoref{fig:plunging_2d}.  Along this structure, we track its gas and radiation temperatures, as well as the normalized angular velocity, shown by the dashed lines in upper and middle panels of \autoref{fig:plunging_1d}.  Both temperatures are higher and the angular velocity is lower compared to the azimuthal averages, indicating compression along the spiral arm.  This compression arises from velocity differences, suggesting that the spiral wave is phase-related and analogous to a density wave, unlike the characteristic curves proposed in \citet{Mummery2024b}.  In the third row of \autoref{fig:plunging_2d}, the density maximum (minimum) of the spiral arm aligns with the minimum (maximum) in velocity divergence, corresponding to a 90-degree phase shift.  The negative velocity divergence (dominated by the radial component) indicates that the outer edge of the spiral arm drifts inward faster than the inner edge, leading to fluid compression within the arm. 

By examining the azimuthal gradient of $u_{\phi}$ in the forth row of \autoref{fig:plunging_2d}, we find that the negative gradient indicates a decrease in angular velocity as the fluid passes through the spiral arm.  This suggests that fluid parcels are losing angular momentum and consequently drifting inward.  In the bottom panel, we compute the divergence of the total angular momentum flux.  The total flux is negative due to the inward accretion flow, and the negative divergence implies that less angular momentum carried by the accretion flow is being transported inward.  This indicates that angular momentum is being extracted within the spiral arm, primarily by the Maxwell stress as indicted in the last panel of \autoref{fig:plunging_1d}.  

\begin{deluxetable*}{l c c c c c c c c c c c c}
\tablecaption{Simulation quality check \label{tab:quality}}
\tablehead{
    \colhead{Name} 
    & \colhead{$\dfrac{\big<\dot{M}_{10}\big>_{t}^{(\mathrm{MR})}}{\big<\dot{M}_{10}\big>_{t}^{(\mathrm{LR})}}$} 
    & \colhead{$\dfrac{\big<(T_{\mathrm{Rey}}^{\mathrm{turb}})^r_{\ \phi, 10}\big>_{\mathrm{disk}}^{(\mathrm{MR})}}{\big<(T_{\mathrm{Rey}}^{\mathrm{turb}})^r_{\ \phi, 10}\big>_{\mathrm{disk}}^{(\mathrm{LR})}}$}
    & \colhead{$\dfrac{\big<(T_{\mathrm{Max}})^r_{\ \phi, 10}\big>_{\mathrm{disk}}^{(\mathrm{MR})}}{\big<(T_{\mathrm{Max}})^r_{\ \phi, 10}\big>_{\mathrm{disk}}^{(\mathrm{LR})}}$}
    & \colhead{$\big<Q_{\mathrm{MRI,10}}^{\phi}\big>_{\mathrm{disk}}$}
    & \colhead{$\big<Q_{\mathrm{MRI,10}}^z\big>_{\mathrm{disk}}$}
    & \colhead{$\big<Q_{\mathrm{therm,10}}\big>_{\mathrm{disk}}$}
    \\
    \quad\;\;\;(1) & (2) & (3) & (4) & (5) & (6) & (7)
}
\startdata
    E150-a9   & 1.52 & 1.51 & 1.23 & \:\:94 (52) & 30 (12)         & 60 (31)         &  \\ 
    E88-a3    & 1.18 & 1.24 & 0.98 & 116    (69) & 30 (18)         & 60 (30)         &  \\
    E15-a9    & 1.52 & 1.58 & 1.27 & 101    (62) & 31 (16)         & 57 (29)         &  \\
    E9-a3     & 1.38 & 1.46 & 0.78 & 124    (99) & 35 (31)         & 49 (26)         &  \\
    E31-a3-DL & -    & -    & -    & 67\quad\:\: & 14\quad\:\:\:\: & 32\quad\:\:\:\: &  \\
    \hline
\enddata
\tablecomments{
    The subscript `10' indicates measurements at $10r_g$. Superscripts `(MR)' and `(LR)' denote intermediate- and low-resolution models, respectively.  Quality factors in parenthesis are measured from low-resolution models.  
    {\bf Columns (from left to right):}
    (1) Model name; 
    (2) Ratio of mass accretion rates (between intermediate- and low-resolution models); 
    (3) Ratio of turbulent Reynolds stress; 
    (4) Ratio of Maxwell stress; 
    (5) Azimuthal MRI quality factor; 
    (6) Vertical MRI quality factor; 
    (7) Thermal quality factor. 
    }
\end{deluxetable*} 

\section{Discussion}
\label{sec:discussion}

\subsection{Quality Check}
\label{sec:quality_check}

We conduct a quality check on our simulations by computing several diagnostic factors to evaluate how well the MRI and thermal scale height are resolved.  Following \citet{Porth2019}, we define the MRI quality factors in the vertical and azimuthal directions as: 
\begin{subequations}    
\begin{align}
    Q_{\mathrm{MRI}}^z &= \frac{2\pi}{\Omega_{\mathrm{K}}}\frac{b^{z} + \beta^{z} b^{t}}{\Delta z \sqrt{w}}
    \ , 
    \\
    Q_{\mathrm{MRI}}^{\phi} &= \frac{2\pi}{\Omega_{\mathrm{K}}}\frac{g_{r\phi}(b^r + \beta^r b^t) + g_{\phi\phi}b^{\phi}}{\Delta l \sqrt{g_{\phi\phi}w}}
    \ , 
\end{align}
\end{subequations}
where $\beta^i=-g^{ti}/g^{tt}$ is the shift vector in the standard 3+1 formalism.  The quantity $\Delta z$ denotes the vertical cell size, and the azimuthal resolution is given by $\Delta l=\sqrt{\Delta x^2 + \Delta y^2}$, with $\Delta x$ and $\Delta y$ representing the cell sizes in the $x$ and $y$ directions, respectively.  We also define a thermal quality factor in a similar manner: 
\begin{equation}
    Q_{\mathrm{therm}} = \frac{2\pi c_s}{\Omega_{\mathrm{K}}\Delta z}
    \ ,
\end{equation}
where the thermal sound speed of the gas-radiation coupled fluid is computed following \citet{Tao2011} as: 
\begin{subequations}    
\begin{align}
    & c_s = \sqrt{\gamma_{\mathrm{disk}} (P_g + \bar{E}_r/3)/\rho}
    \ ,
    \\
    & \gamma_{\mathrm{disk}} = \frac{16\bar{E}_r^2 + 60P_g\bar{E}_r + 9\gamma P_g^2/(\gamma-1)}{9\left[P_g/(\gamma-1) + 4\bar{E}_r\right](P_g+\bar{E}_r/3)}
    \ . 
\end{align}
\end{subequations}
Among all our super-Eddington models, using the highest-resolution runs when available, we evaluate the quality metrics within the disk region that has reached inflow equilibrium.  The disk-averaged MRI quality factors range from approximately 10 to 40 in the vertical direction for $\big<Q_{\mathrm{MRI}}^z\big>_{\mathrm{disk}}$ and from 60 to 120 in the azimuthal direction for $\big<Q_{\mathrm{MRI}}^{\phi}\big>_{\mathrm{disk}}$, both satisfying the criteria provided by \citet{Hawley2011, Hawley2013}.  The disk-averaged thermal quality factor $\big<Q_{\mathrm{therm}}\big>_{\mathrm{disk}}$ falls in the range of 30 to 60, confirming that the thermal scale height is well resolved.  

\autoref{tab:quality} lists the disk-averaged quality factors measured at $10r_g$ for all models.  Values in parentheses indicate the corresponding measurements from lower-resolution runs, where applicable. 

\newpage
\subsection{Resolution Effect}
\label{sec:resolution_effect}

We perform a resolution study to evaluate the convergence of disk properties across different resolutions for all single-loop models.  Key measurements at both resolutions are presented earlier, including mass accretion rate (\autoref{fig:hst_compare} and \autoref{fig:mdot_compare}), magnetic flux (\autoref{fig:hst_compare}), stresses related to angular momentum transport (\autoref{fig:angmom1d}), and radiation luminosity (\autoref{fig:lum_compare}).  These quantities are generally consistent at both resolutions, with some deviations primarily arising from subtle differences in the accretion process.  In general, higher resolution yields a slightly higher accretion rate, which correspondingly reduces the mass-weighted magnetic flux and enhances the radiation luminosity.  

In \autoref{tab:quality}, we report the disk-averaged ratios of accretion rate, turbulent Reynolds stress, and Maxwell stress between the intermediate- and low-resolution runs at $10r_g$, where the cell size differs by approximately a factor of 2 in each dimension.  The overall trends in accretion rate across different resolutions broadly follow changes in the stresses responsible for outward angular momentum transport. 

The accretion process is primarily driven by Maxwell stress, which accounts for approximately 60--90\% of the outward angular momentum transport, as illustrated in \autoref{fig:angmom1d}.  The Maxwell stresses remain largely consistent across both resolutions, with only slight variations -- either higher or lower depending on the specific model.  This consistency ensures that accretion rates remain at similar levels and demonstrates convergence in resolving MRI, as already indicated by the high quality factors. 

The turbulent Reynolds stress, while subdominant, contributes around 10--40\% of the total outward angular momentum transport.  Its ratios between intermediate- and low-resolution runs exhibit more noticeable differences, indicating an additional contribution to the accretion process at higher resolution across all models.  This suggests that deviations in accretion rates might originate from differences in how effectively turbulence is resolved. 

\begin{figure*}
    \centering
    \includegraphics[width=\textwidth]{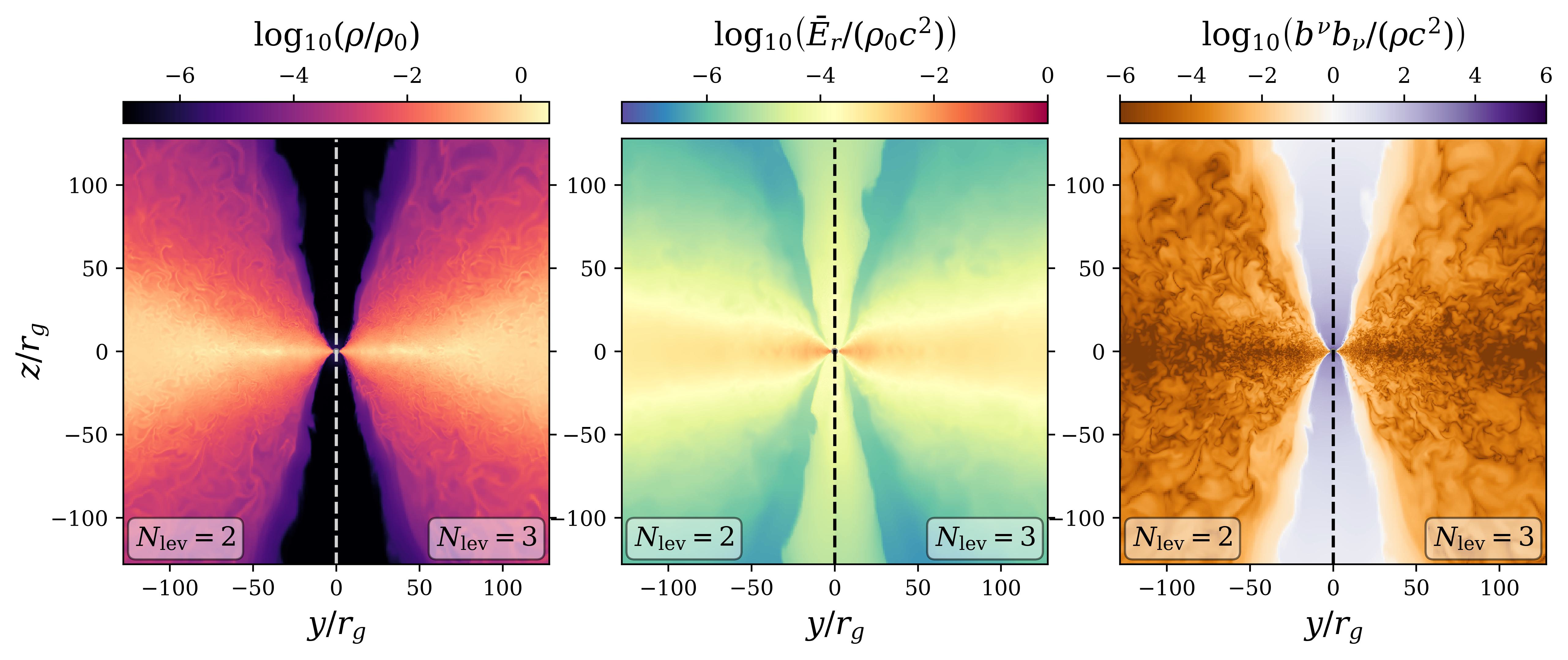}
    \caption{
    Comparison between models using different angular resolutions in radiation intensity.  From left to right, panels show side-by-side comparisons of density, fluid-frame radiation energy density, and magnetization at $t=30000r_g/c$, after steady states are achieved. In each panel, the left half adopts angular grid level-2 (42 angles), while the right half uses level-3 (92 angles), showing that the differences between the two models are nearly indistinguishable.  
    }
    \label{fig:nrad_compare}
\end{figure*}

Since the super-Eddington accreting system is radiation dominated, the turbulence length scale is set by the radiation viscosity, which is extremely small and nearly impossible to fully resolve in global simulations.  Fortunately, the accretion process is predominantly driven by Maxwell stress, which is reasonably well resolved.  As a result, the inability to fully capture the small-scale turbulence has only a modest impact on the overall quality of our simulations.  

In addition to the spatial resolution study, we perform an additional run for model E88-a3 with higher angular resolution in the radiation intensity.  Specifically, we increase the angular grid to level 3 (92 angles in total), and run the simulation to $t=30000r_g/c$.  This run begins with the same initial condition and reaches a steady state with results that are nearly indistinguishable, as demonstrated in \autoref{fig:nrad_compare}.  

It is worth noting that the angular resolution required to \lz{preserve} the necessary anisotropy in the radiation field can be significantly higher when the flow becomes highly relativistic (see equation~56 in \citealt{White2023}).  However, most of the disk and wind regions exhibit only moderate velocities, while relativistic flows typically occur in optically thin regions where gas and radiation are decoupled.  In such cases, radiation anisotropy has limited impact on the fluid dynamics.  This likely explains the close agreement between the standard and higher angular resolution runs.  Nevertheless, more definitive conclusions would require simulations at substantially higher angular resolution. 

\subsection{Comparison with Non-Radiative Runs}

\begin{figure}
    \centering
    \includegraphics[width=\columnwidth]{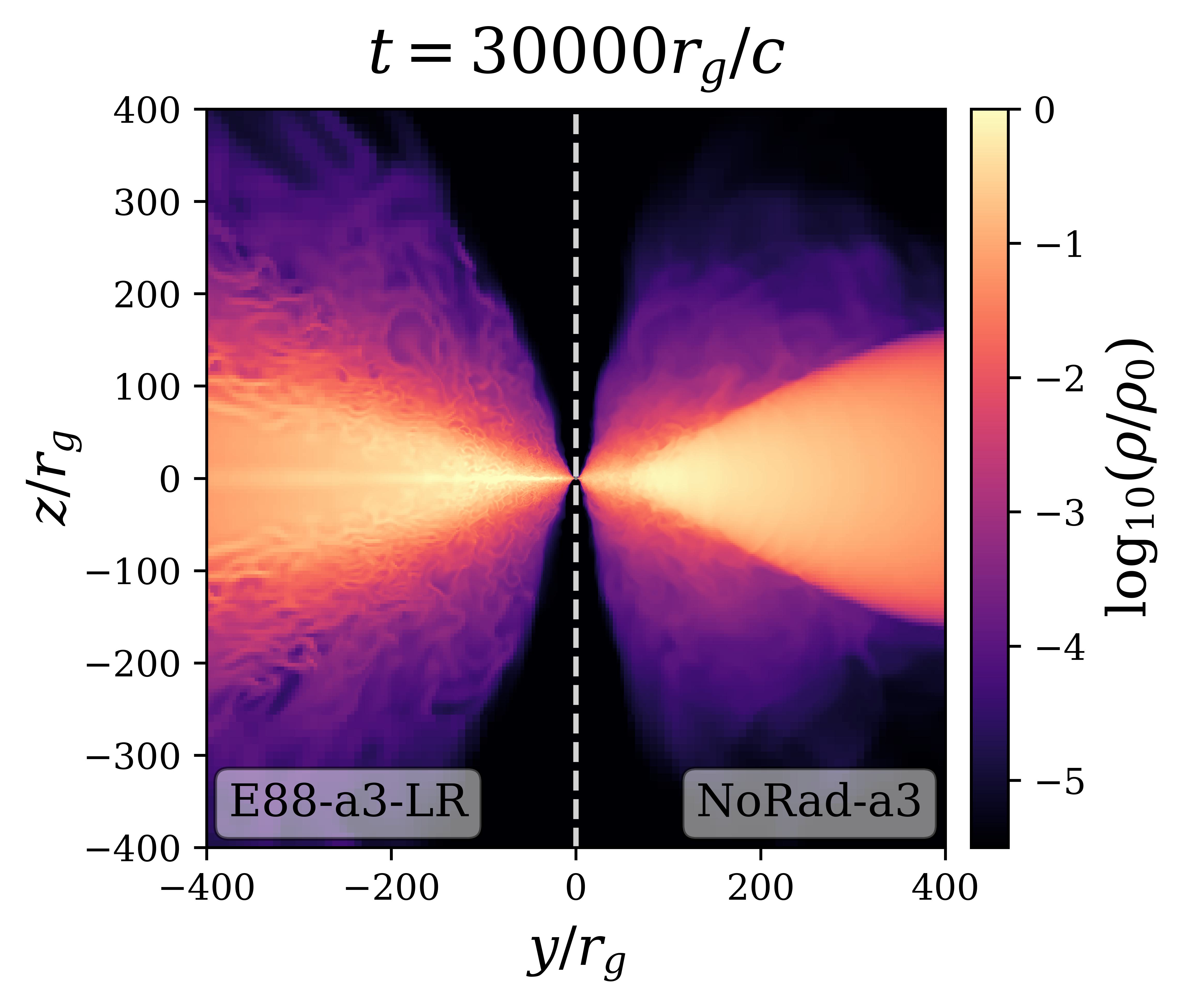}
    \caption{
    Side-by-side comparison of 2D density profiles between radiative (left half) and non-radiative (right half) models at $t=30000r_g/c$.  For the radiative case, we use the low-resolution model E88-a3-LR, which shares the same grid configuration as the non-radiative model NoRad-a3.  Although the overall disk morphology is similar, the radiative model exhibits significantly larger density fluctuations due to enhanced compressibility. 
    }
    \label{fig:den_norad_compare}
\end{figure}

Unlike the geometrically thin disks in the sub- and near-Eddington models presented in \citetalias{PaperI}, the disk structures in non-radiative runs are geometrically thick and closely resemble those in the super-Eddington models, despite fundamentally different dynamics.  \lz{Their radial density profiles are qualitatively similar, exhibiting nearly flat disk-averaged profiles and outward increasing midplane densities.}  As shown in \autoref{fig:den_norad_compare}, although the overall disk morphology is similar, the accretion flow in the radiative case appears significantly more inhomogeneous, as turbulence produces larger density fluctuations due to the enhanced compressibility of the radiation-dominated medium, consistent with previous findings \citep{Turner2003,Jiang2013b}. 

In \autoref{fig:hst_compare}, we compared the accretion rate and magnetic flux between radiative and non-radiative models.  For comparison, the accretion rates in non-radiative runs are scaled by the corresponding density units of each radiative model.  Their evolution histories are broadly similar, though radiative models generally show slightly higher mass accretion rates and lower magnetic flux. 

\begin{figure}
    \centering
    \includegraphics[width=\columnwidth]{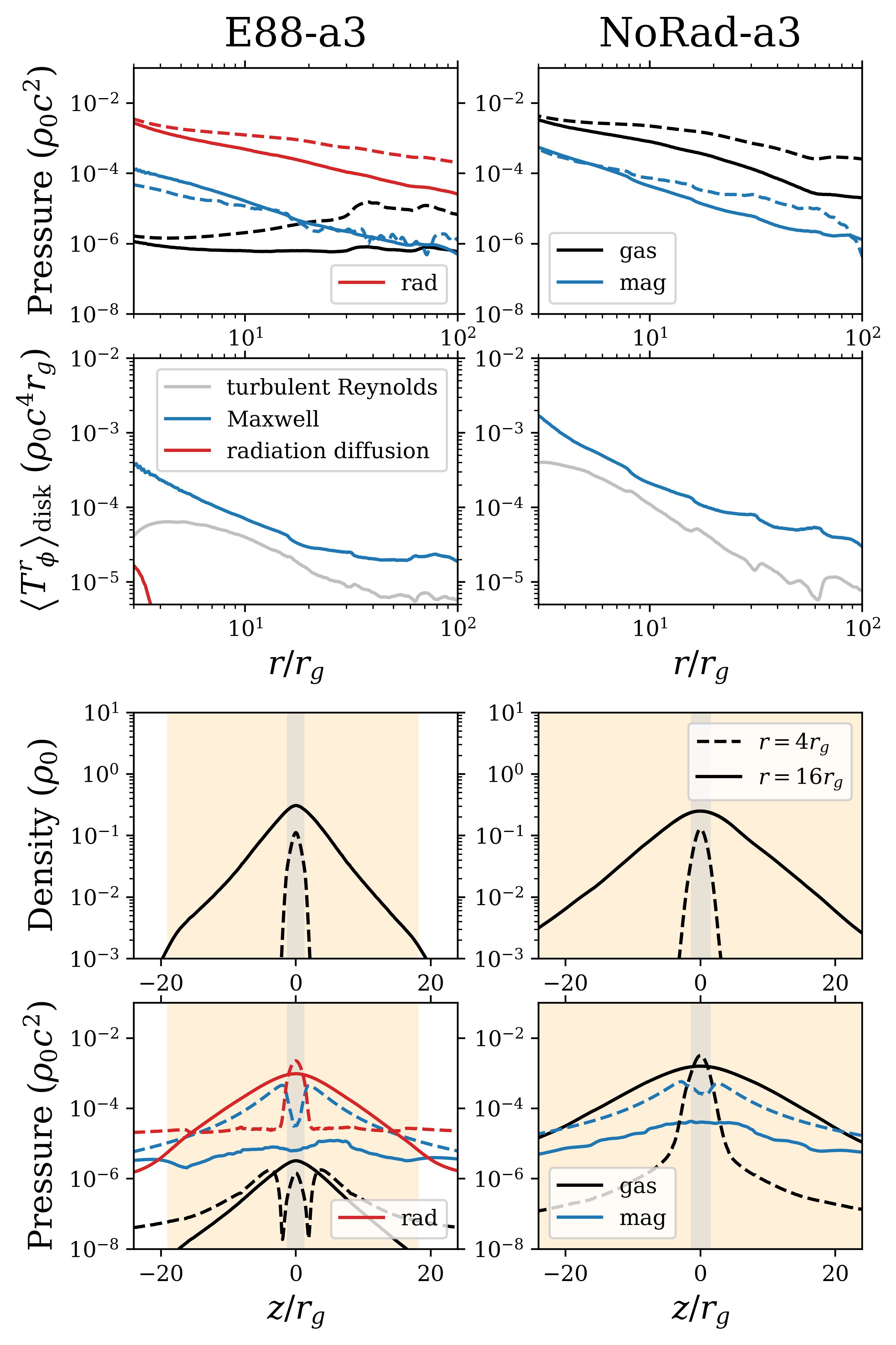}
    \caption{
    Comparison of 1D profiles between radiative (left column) and non-radiative (right column) models.  The first and second rows show radial profiles of pressure and angular momentum flux, respectively, measured as disk averages (solid) and at the midplane (dashed).  Pressure profiles include components from radiation (red), magnetic (blue), and gas (black).  Angular momentum fluxes consist of Maxwell stress (blue), turbulent Reynolds stress (gray), and radiation stress (red).  The last two rows compare the vertical profiles of density and pressure, using the same color scheme as the radial pressure plots.  Solid and dashed lines represent measurements at $16r_g$ and $4r_g$, respectively, with the disk regions highlighted in yellow and gray.
    }
    \label{fig:norad_compare}
\end{figure}

\autoref{fig:norad_compare} presents a side-by-side comparison of low-spin models E88-a3 and NoRad-a3, focusing on radial (pressure and angular momentum flux) and vertical (density and pressure) profiles.  In the radial profiles, solid lines represent disk-averaged quantities, while dashed lines indicate midplane values.  In the vertical profiles, solid and dashed lines correspond to measurements at $16r_g$ and $4r_g$, respectively, with the disk regions highlighted in yellow and gray. 

In both models, thermal pressure dominates over magnetic pressure (blue) and supports the disk against gravity.  In the non-radiative model, the thermal pressure arises purely from gas (black), whereas in the radiative case it is dominated by radiation (red).  Since super-Eddington flows are optically thick in most regions, radiation remains well coupled to the gas, and energy transport is dominated by advection (see Section~3.3 of \citetalias{PaperI}).  As a result, radiation behaves like a fluid, which effectively replaces the role of gas and results in similar disk morphologies in both models.  Near the photosphere where gas and radiation begin to decouple, the radiative model shows a more extended corona driven by radiation outflows compared to the non-radiative case, as shown in \autoref{fig:den_norad_compare}. 

The accretion processes in both models are also similar, with Maxwell stress (blue) dominating and turbulent Reynold stress (gray) playing a subdominant role.  The magnetic pressure shows similar trends in both radial and vertical profiles, with a slight dip near the midplane in the latter, likely due to buoyancy effects.  

\newpage
\subsection{Observational Applications}
\label{sec:observational_applications}

\begin{figure*}
    \centering
    \includegraphics[width=\textwidth]{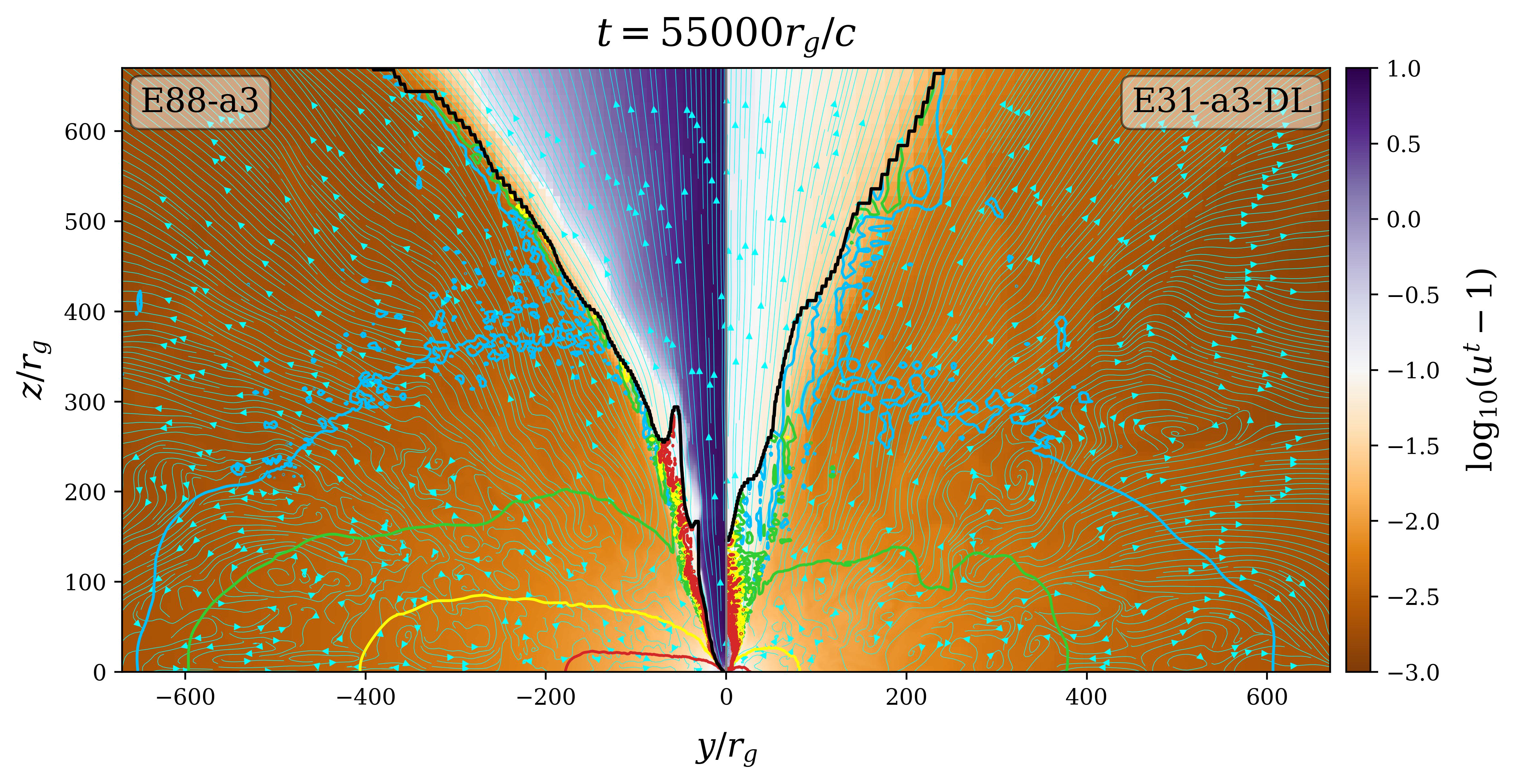}
    \caption{
    Snapshots at $t=55000r_g/c$ for two super-Eddington models: one with a strong jet (E88-a3, left) and one with a weak jet (E31-a3-DL, right).  The 2D profiles show flow speed, with the colormap indicating highly relativistic (purple), mildly relativistic (white), and non-relativistic (brown) regions.  The velocity field (cyan) is overlaid as streamlines, and gas temperatures are indicated by isothermal contours at logarithmic values of 7.5 (red), 7.2 (yellow), 6.8 (green), and 6.5 (blue).  Solid black lines denote the scattering photosphere.  When the strong jet evacuates the funnel region (left half), the inner disk becomes exposed and the scattering photosphere extends down to the horizon.  In contrast, the weak jet fails to clear the funnel (right half), allowing radiation-driven outflows to fill the region, resulting in the scattering photosphere lying above $\sim 150r_g$. 
    }
    \label{fig:glb_structure}
\end{figure*}

The observational implications of our numerical models across different regimes are discussed in detail in \citetalias{PaperI}.  Here, we provide additional insight by focusing on the physical pictures of super-Eddington accretion systems to constrain the observational signatures one should expect. 

The accretion disks in our super-Eddington models are broadly similar: radiation-dominated and geometrically thick, regardless of magnetic topology.  The thermally expanded flow forms a conical funnel that collimates energy output in the form of radiation or jet outflows.  However, the jet strength can significantly influence the observational properties of the system.  

\autoref{fig:glb_structure} presents snapshots of two super-Eddington models: one with a strong jet (E88-a3, left) and one with a weak jet (E31-a3-DL, right).  The 2D profiles display flow speed (colormap), overlaid with velocity streamlines (cyan) and temperature contours (red, yellow, green, and blue, from hot to cold).  The scattering photosphere is marked by black solid lines.  The colormap is chosen such that purple indicates highly relativistic speeds, white denotes mildly relativistic flow, and brown corresponds to non-relativistic regions. 

When a strong jet forms, it evacuates the funnel region, facilitating both radiation escape and relativistic jet streaming, along with a strong geometric beaming effect.  The cleared, optically thin funnel allows high-energy photons from the inner disk to escape directly, as indicated by the high-temperature region (red contours) just below the scattering photosphere (black solid line) in the left half of \autoref{fig:glb_structure}. 

When the jet is weak, radiation-driven outflows can fill the funnel region, suppressing jet propagation and dissipating jet power prematurely.  The resulting jet remains only mildly relativistic, magnetically driven near the black hole and radiation-driven at larger radii (for details see \autoref{sec:weak_jet}).  In this case, most high-energy photons are trapped by optically thick outflows, with only a tiny fraction expected to escape.  

Since jet properties are closely tied to the radiation output, they can serve as useful constraints for interpreting observations.  For example, super-Eddington accretion systems with strong jets are expected to exhibit a more steady radiation luminosity with low variability, while those with weak jets show greater variability due to photon trapping and enhanced magnetic turbulence, as also illustrated in Figure~10 of \citetalias{PaperI}. 

In addition, as shown in Figure~11 of \citetalias{PaperI}, radiation beaming effects are present in both near- and super-Eddington regimes.  When emission is assumed isotropic, these effects can introduce observational degeneracies when the apparent luminosity exceeds the Eddington limit.  This complicates identification of the underlying accretion regime, as both cases can produce super-Eddington flux despite having fundamentally different flow structures (see Figure~3 of \citetalias{PaperI}).  However, our numerical models offer a way to break this degeneracy by tracking changes in spectral hardness with luminosity.  

As illustrated in Figure~12 of \citetalias{PaperI}, when the accretion rate (or luminosity) decreases in the super-Eddington regime, the emerging spectrum becomes softer.  In contrast, as the system transitions from super- to near-Eddington accretion with further decrease in accretion rate, the spectrum gets harder due to the development of magnetically supported envelops above the thermal disk (see Section~3.5 of \citetalias{PaperI}). 

These theoretical behaviors of super-Eddington (and near-Eddington) accretion flows provide valuable insight into the observational properties of systems, such as ultraluminous X-ray sources (ULXs), little red dots (LRDs), tidal disruption events (TDEs), and stellar-mass black hole binaries (BHBs).  

For ULXs, a stable conical funnel structure cleared by strong jets provides a plausible physical environment for the high X-ray polarization observed in Cyg X-3 \citep{Veledina2024,Veledina2024b}. Moreover, the opposite trends in spectral hardness between the near- and super-Eddington accretion regimes may help explain the spectral pivoting observed in many ULX systems \citep{Kajava2009}.  When the system develops a weak jet, our model occupies a parameter regime similar to that of SS~443 \citep{Fabrika2004}, featuring a mildly relativistic jet that is partially driven by radiation pressure.

For LRDs, optically thick radiation-driven outflows can significantly obscure the system, potentially resulting in X-ray underdetection \citep[e.g.,][]{Greene2024,Wang2024} and an emergent spectrum with absorption features (e.g., a Balmer break, \citealt{Wang2024,Wang2024b}), even though the source remains bolometrically luminous due to super-Eddington accretion.  More detailed discussions of ULXs, SS~433, and LRDs can be found in \citetalias{PaperI}.

For transients sources, TDEs occur when a star passes too close to a supermassive black hole and is torn apart by tidal forces \citep{Rees1988}.  Roughly half of the stellar debris remains gravitationally bound and accretes onto the black hole at a super-Eddington rate during the early phase, with or without the presence of strong jets \citep[e.g.,][]{Burrows2011, Alexander2016}.  The geometry of the optically thick equatorial region (disk and wind) and the optically thin funnel region (jet) discussed above provides valuable insights into the relationship between X-ray and optical emission among TDE populations \citep[e.g., ][]{Roth2016,Dai2018,Huang2024}.  BHBs are potential sources of gravitational waves \citep{Abadie2011,Abbott2016}, with some thought to form in AGN disks and undergo super-Eddington accretion when observed as transient electromagnetic (EM) counterparts \citep{Stone2017,Rodriguez-Ramirez2025}.  Our super-Eddington models provide a physical framework for quantifying the properties of emergent radiation and radiation-driven outflows, including their dependence on jet formation.  In particular, these results place useful constraints on the radiation and outflow efficiencies in both TDE and BHB models. 

In future work, we plan to post-process our numerical models to convert this parameter survey into observable predictions, including spectral and polarization signatures, which will enable more direct comparisons with observational data.

\subsection{Comparison with the Slim Disk Model}

The slim disk model \citep{Abramowicz1988, Beloborodov1998, Sadowski2009, Abramowicz2013} is widely used to model super-Eddington accretion flows, as it incorporates radial energy advection that effectively captures the ``photon trapping'' effect within optically thick inflows near the black hole.  As shown in Figure~9 of \citetalias{PaperI}, despite differences in normalization, the slim disk model accurately reproduces the trend of radiation efficiency as a function of accretion rate across our numerical super-Eddington models. 

The construction of the slim disk model relies on two key assumptions: (1) the disk is geometrically thin, and (2) the effective viscosity (i.e., the $\alpha$ parameter) is constant throughout the disk.  

The slim geometry implies a simple vertical structure, allowing midplane profiles to be approximated by disk-averaged (or vertically integrated) quantities.  However, this assumption does not always hold in the presence of substantial outflows, as demonstrated by our numerical results (see \autoref{fig:hori_compare} and \autoref{fig:angmom1d}), where disk-averaged quantities can be significantly modified by outflows near the disk surface.  Moreover, the effective viscosity (or the pressure-scaled radial angular momentum flux) in our numerical models is generally not constant.  Instead, it follows a power-law with radius, as illustrated in the bottom row of \autoref{fig:angmom1d}, where $\alpha$ can vary by more than a factor of two across the inflow equilibrium region. 

When the outflow has limited impact, our numerical results show remarkable agreement with the slim disk solution, as seen in model E31-a3-DL.  In this case, the disk-averaged profiles closely follow the trends of the midplane profiles (see the middle column of \autoref{fig:hori_compare}), and the overall $\alpha$ value measured in the numerical model is consistent with the best-fit value inferred from the slim disk model (see the lower-middle panel of \autoref{fig:angmom1d}), despite the radial variation in $\alpha$. 

When the outflow is significant, as in models E88-a3 and E9-a3, clear discrepancies emerge between midplane and disk-averaged profiles, with increasingly pronounced differences in density and thermal pressure at larger radii due to outflow-driven advection (see the first and last columns in \autoref{fig:hori_compare}).  The $\alpha$ values measured in the simulations are generally higher than the best-fit values inferred from the slim disk model (see the lower-left and lower-right panels of \autoref{fig:hori_compare}).  Such differences can lead to substantial deviations from our numerical results, as the disk profiles inferred from the slim disk model can be highly sensitive to the choice of $\alpha$.  In such cases, outflow-driven effects must be taken into account, as explored by \citet{Poutanen2007}.

\subsection{Comparison with Previous Numerical Results}
\label{sec:compare_numerical_results}

As briefly discussed in \citetalias{PaperI}, direct comparisons with previous numerical models are challenging due to substantial differences in initial disk configurations and radiation transport treatments.  A comparison with our previous non-relativistic models \citep{Jiang2014a, Jiang2019b} using the full-transport method is discussed in \citetalias{PaperI}. Here, we provide a qualitative discussion of representative super-Eddington models that employ the M1 closure method within a general relativistic framework.  In the near future, we plan to perform simulations using the M1 approximation with the same initial conditions, which enables direct side-by-side comparisons with our full radiation transport models. 

Among general relativistic models that employ the M1 approximation, there are two main classes of solutions depending on whether the system evolves into the magnetically arrested disk (MAD) regime \citep{Tchekhovskoy2011}.  

When the system remains in the SANE regime, the accretion disk is generally geometrically and optically thick, with output luminosity typically beamed along the polar axis and characterized by very low radiation efficiency ($\lesssim$1\%; \citealt{McKinney2014, Sadowski2014, Sadowski2016b, Narayan2017, Utsumi2022}).  In these systems, radiation becomes trapped in the inner disk and is highly advective, with advection dominating over diffusion as the primary mechanism of thermal energy transport.  The total luminosity efficiency (including contributions from both radiation and kinetic outflows) remains at a few percent and increases with black hole spin, primarily due to enhanced jet power.  These results are broadly consistent with the physical picture found in our models.  

However, all of our super-Eddington simulations start from a relatively thin torus, with geometrically thick structures developing only after the system reaches a steady state.  In contrast, these M1-based models are initialized with a hot, geometrically thick disk, where the long thermal relaxation time can result in an outgoing radiation field dominated by cooling of the initial disk rather than by accretion-powered emission.  

Although \citet{Fragile2025} also reported a series of models initialized with super-Eddington accretion flows across a range of accretion rates, all of these settle into the near-Eddington regime through self-regulation by outflows.  The initial conditions are based on a generalized $\alpha$-prescribed analytical thin disk model.  As these systems evolve into the near-Eddington regime, the accretion flows remain geometrically thin and exhibit high radiation efficiency, ranging from 30--70\%.  We observe similar behavior in our near-Eddington models, which briefly enter a mildly super-Eddington phase before settling into the final steady state.  \lz{A detailed discussion of these results is beyond the scope of the present paper and will be presented in Paper III, which focuses on the near-Eddington regime. }

On the other hand, a geometrically thin MAD disk solution can typically be established when the system is initialized either with a single-loop magnetic field in a thick torus (e.g., \citealt{Narayan2017}) or with additional vertical magnetic flux in a thin torus (e.g., \citealt{McKinney2015,McKinney2017,Curd2023}).  These systems are generally much more radiation efficient ($\gtrsim$10\%) and in some cases achieve efficiencies as high as $\sim$80\% \citep{McKinney2017,Curd2023}, although the measured efficiency may be affected by the numerical floor imposed near the horizon \citep{McKinney2017}.  

With a spinning black hole, jet formation can help evacuate the polar funnel region, allowing radiation to stream out freely \citep{McKinney2015, Narayan2017}.  However, this alone does not fully account for the high radiation efficiency, as similar jet-clearing effects are also present in our SANE models.  This suggests that MAD accretion flows are intrinsically more dissipative, without accounting for numerical dissipation in the strongly magnetized plasma.  Nonetheless, our simulations exhibit a trend of increasing radiation efficiency as the funnel opens at lower accretion rates, consistent with the argument of \citet{McKinney2015} that a low-density channel can enhance radiation efficiency.  

In addition, an optically thick, magnetized wind can deplete the surface mass density, reduce the disk's optical depth, and advect radiation outward \citep{McKinney2015}.  This mechanism, however, is not unique to the MAD regime -- a non-MAD, magnetically dominated disk in \citet{Jiang2025} shows similar behavior. In some cases, the disk surface can even become optically thin, further enhancing radiative efficiency and altering the emergent spectrum \citep{Curd2023}.  These disks also exhibit strong variability in radiation luminosity, a feature uniquely associated with magnetic flux eruptions in the MAD regime.  This variability may serve as an observational diagnostic to distinguish MAD accretion flows from our SANE models. 

\section{Conclusions}
\label{sec:conclusions}

In \citetalias{PaperI}, we presented a parameter survey of black hole accretion in the radiation-dominated regime, covering a broad range of accretion rates, black hole spins, and magnetic field topologies, using full-transport radiation GRMHD simulations.  In this paper, we establish a unified analysis routine and provide a comprehensive analysis of the super-Eddington accretion models, including convergence studies in both spatial and angular resolution, along with two additional non-radiative models at different spins for comparison.  

Our main results for super-Eddington models are as follows: 

\begin{itemize}
\item Super-Eddington accretion develops geometrically thick disk structure supported by radiation pressure, regardless of magnetic field topologies.  Radiation in the inner disk is largely trapped within the optically thick inflow. 

\item Radiation generated by the accretion process drives significant outflows.  Along the disk surface, the unbound wind propagates outward at mildly relativistic speeds, undergoing gravitational deceleration and radiative cooling.  Near the polar regions, these optically thick outflows form conical funnels that (1) limit photon escape, resulting in low radiation efficiency, and (2) collimate the outgoing radiation, producing strong beaming effects. 

\item Radiation-dominated accretion flows are highly turbulent, with thermal energy transport dominated by radiation advection in the main body of the disk.  Radiation diffusion becomes important only near the photosphere. 

\item The accretion process is driven by outward angular momentum transport, primarily via Maxwell stress and secondarily via turbulent Reynolds stress, while radiation stress is nearly negligible.  In the main body of the disk, angular momentum is predominantly carried away by turbulence (through the turbulent components of Maxwell and Reynolds stresses), while near the disk surface, mean-field Maxwell stress becomes the dominant transport mechanism. 

\item A strong jet forms near a spinning black hole with sufficient poloidal magnetic fields, evacuating the funnel and allowing radiation to escape from the inner disk through strong geometric beaming.  In contrast, a weak jet, driven by both magnetic and radiative forces, fails to clear the funnel.  In this case, the inner disk is obscured by radiation-driven outflows, and the photosphere extends to a height of over a hundred gravitational radii above the horizon.  The power of strong jets is comparable to or exceeds the radiation luminosity, whereas that of weak jets is nearly negligible. 

\item In the plunging region, fluid profiles remain continuous across the ISCO, with angular momentum extraction dominated by Maxwell stress.  Spiral structures emerge due to local compressive effects and transport angular momentum outward.  These spiral waves exhibit a $90^{\circ}$ phase offset between density and velocity, similar to density waves. 
\end{itemize}

We also summarize the results relevant to numerical performance, broader applications, and model comparisons as follows:

\begin{itemize}
\item All simulations have quality factors that exceed the standard criteria for resolving both MRI and the thermal scale height.  However, the intermediate-resolution models show slightly higher accretion rates than their low-resolution counterparts, likely due to enhanced angular momentum transport from better-resolved small-scale turbulence.  Fully capturing such turbulence would require resolving down to the radiation viscous length scale, which is more feasible in local shearing-box calculations than in global simulations.  

\item The radial profiles of temperature, pressure, and stress, averaged over the disk closely follow power-law trends. The fitted power-law indices are reported in \autoref{tab:sim_overall} and Table~1 of \citetalias{PaperI}, providing useful inputs for developing or testing analytical models.

\item Our super-Eddington models, featuring both strong and weak jet formation, are applicable to a variety of astrophysical systems, including ULXs, LRDs, TDEs, and BHBs.  These observational applications are discussed in detail in \autoref{sec:observational_applications} and Section~4 of \citetalias{PaperI}. 

\item Despite fundamental differences in dynamics, super-Eddington models closely resemble their non-radiative counterparts in terms of disk morphology, pressure profiles, and angular momentum transport.  The key differences in radiative models arise from radiation dominance: (1) thermal pressure is primarily set by radiation rather than gas, (2) radiation-driven unbound outflows reach larger distances, and (3) the accretion flow is significantly more inhomogeneous due to enhanced compressibility. 

\item Our numerical results agree well with the slim disk model when outflows have limited influence on the disk structure.  However, when outflows become significant, their effects must be included to accurately capture the disk profiles. 

\item The physical picture emerging from our super-Eddington models is broadly consistent with previous non-relativistic models using full radiation transport, as well as with other relativistic SANE models using the M1 closure.  However, unlike earlier SANE simulations initiated with a geometrically thick torus, our models begin with a relatively thin torus, ensuring that the resulting radiation luminosity primarily reflects the accretion process rather than the initial conditions.  

\item None of our super-Eddington models reach the MAD regime.  In contrast, previous M1-based MAD simulations rely on specific numerical setups, such as injecting additional vertical magnetic flux or initializing with a geometrically thick single-loop field, to ensure sufficient magnetic flux in the accretion flow.  A comprehensive discussion of model comparisons can be found in \autoref{sec:compare_numerical_results} and Section~3.6 of \citetalias{PaperI}. 

\end{itemize}

\begin{acknowledgments}

We thank Omer Blaes, Sihao Cheng, Xiaoshan Huang, Bingjie Wang, \lz{and Alexander Dittmann} for helpful discussions, and \lz{Joseph Insley for assistance with the 3D visualization}.  This work was supported by the Schmidt Futures Fund, NASA TCAN grant 80NSSC21K0496, and the Simons Foundation.  The analysis made significant use of the following packages: NumPy \citep{Harris2020}, SciPy \citep{Virtanen2020}, and Matplotlib \citep{Hunter2007}.

An award for computer time was provided by the U.S. Department of Energy's (DOE) Innovative and Novel Computational Impact on Theory and Experiment (INCITE) Program. This research used supporting resources at the Argonne and the Oak Ridge Leadership Computing Facilities. The Argonne Leadership Computing Facility at Argonne National Laboratory is supported by the Office of Science of the U.S. DOE under Contract No. DE-AC02-06CH11357. The Oak Ridge Leadership Computing Facility at the Oak Ridge National Laboratory is supported by the Office of Science of the U.S. DOE under Contract No. DE-AC05-00OR22725. We thank Vassilios Mewes and Kyle Felker for support on these facilities.

Research presented in this article was supported by the Laboratory Directed Research and Development program of Los Alamos National Laboratory under project number 20220087DR.

This work has been assigned a document release number LA-UR-25-29188.

\end{acknowledgments}


\appendix

\section{Treatment of Compton Cooling}
\label{appendix:density_rescale}

Since radiation GRMHD is not scale-free, a density unit must be set to specify the accretion rate.  However, this becomes problematic at high accretion rates.  In low-density regions limited by the numerical floor, a large density unit can artificially enhance gas-radiation coupling and increase the optical depth.  This leads to several numerical issues, including: 1. artificial heat dissipation in jet and corona regions via the Compton process; 2. artificial Thomson scattering in vacuum regions, which prevents radiation from freely streaming;  3. spurious radiation generation in low-density regions, where the gas temperature is set by imposed density and pressure floors. 

In principle, these issues can be mitigated by lowering the density floor.  However, this would require an extremely small value to address all the problems above, which could in turn introduce more severe numerical issues in the highly magnetized funnel region.  Therefore, we apply a density reduction when computing the gas-radiation interaction coefficients (i.e. $\rho\kappa$) in low-density regions.  This approach allows gas-radiation coupling in low-density regions to be evaluated using a lower, more physically realistic density, thereby avoiding excessive coupling caused by the high density unit.  A key advantage of this method is that it does not alter the fluid density in the GRMHD equations nor affect the disk region where the density remains sufficiently high.  

We first define a truncation density $\rho_{\mathrm{trunc}}$, below which the density used to compute the radiation interaction coefficients is reduced towards a more physically realistic value (e.g., the Goldreich-Julian density), given by
\begin{equation}
    \rho_{\mathrm{trunc}} = \frac{\sigma_{\mathrm{m}}\tau_{\mathrm{trunc}}}{\kappa_T\Delta l}
    \ , \quad
    (\rho_{\mathrm{floor}} \le \rho_{\mathrm{trunc}} \le \rho_{\mathrm{trunc,max}})
\end{equation}
where $\Delta l$ is the cell size, and $\tau_{\mathrm{trunc}}$ is the maximum optical depth per cell that permits free-streaming radiation.  The truncation density is weighted by the magnetization factor, defined as $\sigma_{\mathrm{m}} = b^{\nu}b_{\nu}/(\rho c^2)$, allowing more aggressive density reduction in strongly magnetized regions.  To prevent over-reduction, it is constrained between the numerical floor $\rho_{\mathrm{floor}}$ and a physical upper limit $\rho_{\mathrm{trunc,max}}$, which represents the lowest reliable density in the funnel or corona region.  

The density rescaling is applied using a customized sigmoid function: 
\begin{subequations}    
\begin{align}
    \lg\rho_{\mathrm{op}} &= \lg\rho - \mathcal{C}_{\mathrm{reduce}}(\lg\rho_{\mathrm{floor}}-\lg\rho_{\mathrm{op,min}})
    \quad, 
    \\
    \mathcal{C}_{\mathrm{reduce}} &= 1 - \left\{  
    1+\exp\left[-\frac{1}{W}\left(\lg\rho - \lg\rho_{\mathrm{shift}} \right)\right]
    \right\}^{-1}
    \quad, 
    \\
    \lg\rho_{\mathrm{shift}} &= \lg\rho_{\mathrm{min}} + \frac{1}{2}\lg\left(\frac{\rho_{\mathrm{trunc}}}{\rho_{\mathrm{min}}}\right)
    \quad, 
    \\
    W &= \left. \frac{1}{2}\lg\left(\frac{\rho_{\mathrm{trunc}}}{\rho_{\mathrm{floor}}}\right) \right/ \log\left(\frac{1}{\varepsilon} - 1\right)
    \quad, 
\end{align}
\label{eq:density_rescale}
\end{subequations}
where $\rho_{\mathrm{op}}$ denotes the reduced density used to compute the radiation interaction coefficients.  We adopt the Goldreich-Julian density as the minimum physically allowed density $\rho_{\mathrm{op,min}}$ for gas-radiation coupling.  The small parameter $\varepsilon$, constrained to be less than 0.5, controls the width of the sigmoid transition. 

\begin{figure*}
    \centering
    \includegraphics[width=\textwidth]{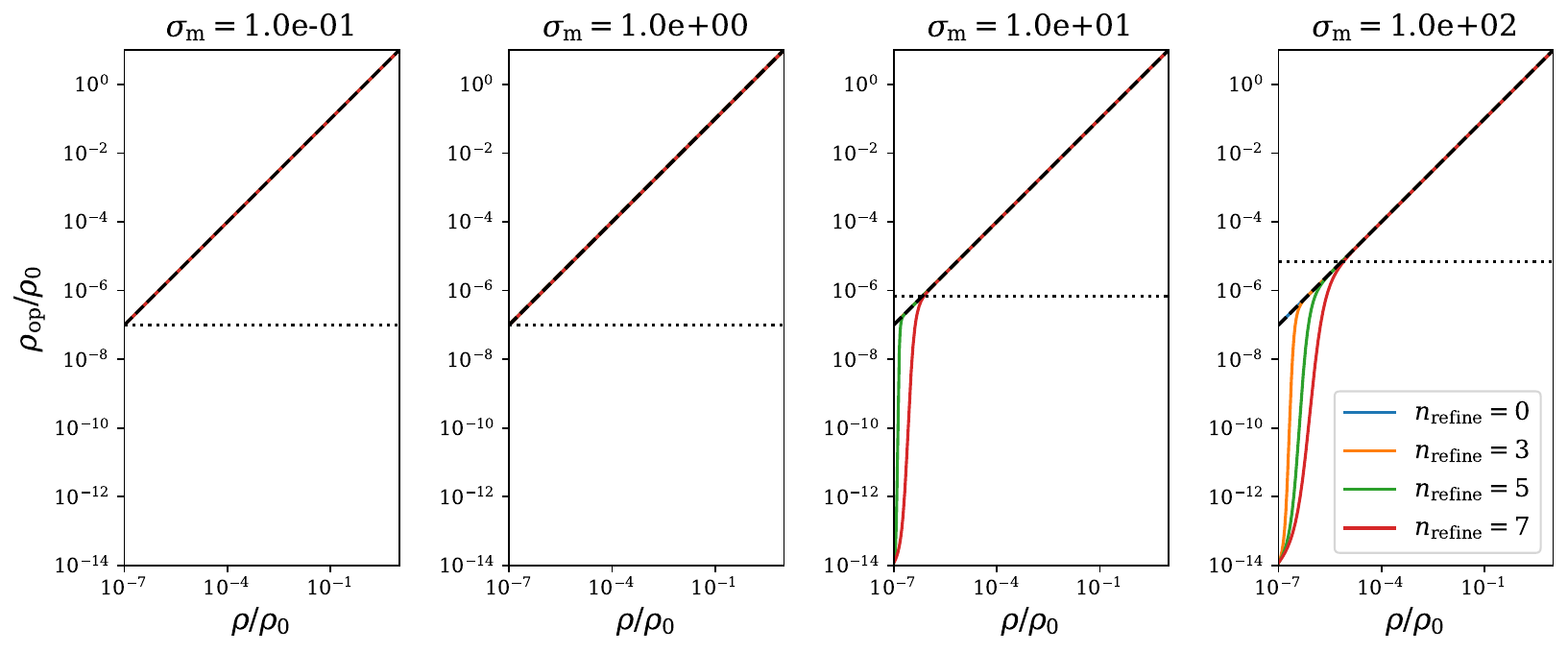}
    \caption{
    Density rescaling applied for gas-radiation coupling in model E150-a9.  The $x$-axis represents the input density, and the $y$-axis shows the rescaled output density.  Magnetization increases from left to right panels, as labeled by titles.  The reduction follows \autoref{eq:density_rescale} using parameters: $\tau_{\mathrm{trunc}}=$5e-4, $\rho_{\mathrm{trunc,max}}=$1e-7, $\rho_{\mathrm{trunc,max}}=$1e-4, $\rho_{\mathrm{op,min}}=$1e-14, and $\varepsilon=0.01$ (in simulation units).  Line colors indicate different levels of mesh refinement.  The horizontal dashed line marks the density threshold at which rescaling begins for the highest refinement level. 
   }
    \label{fig:density_rescale}
\end{figure*}

In \autoref{fig:density_rescale}, we use model E150-a9 to illustrate the density rescaling applied for gas-radiation coupling.  This rescaling is inactive in weakly magnetized or high-density regions, and therefore does not affect most of the disk body.  As magnetization increases, the density is progressively reduced in low-density regions to mitigate excessive gas-radiation coupling.  This approach is particularly effective at suppressing artificial heat dissipation via Compton process and enabling free-streaming radiation in the funnel region, where the density floor fails to reflect the true low-density environment. 

\begin{figure*}
    \centering
    \includegraphics[width=\textwidth]{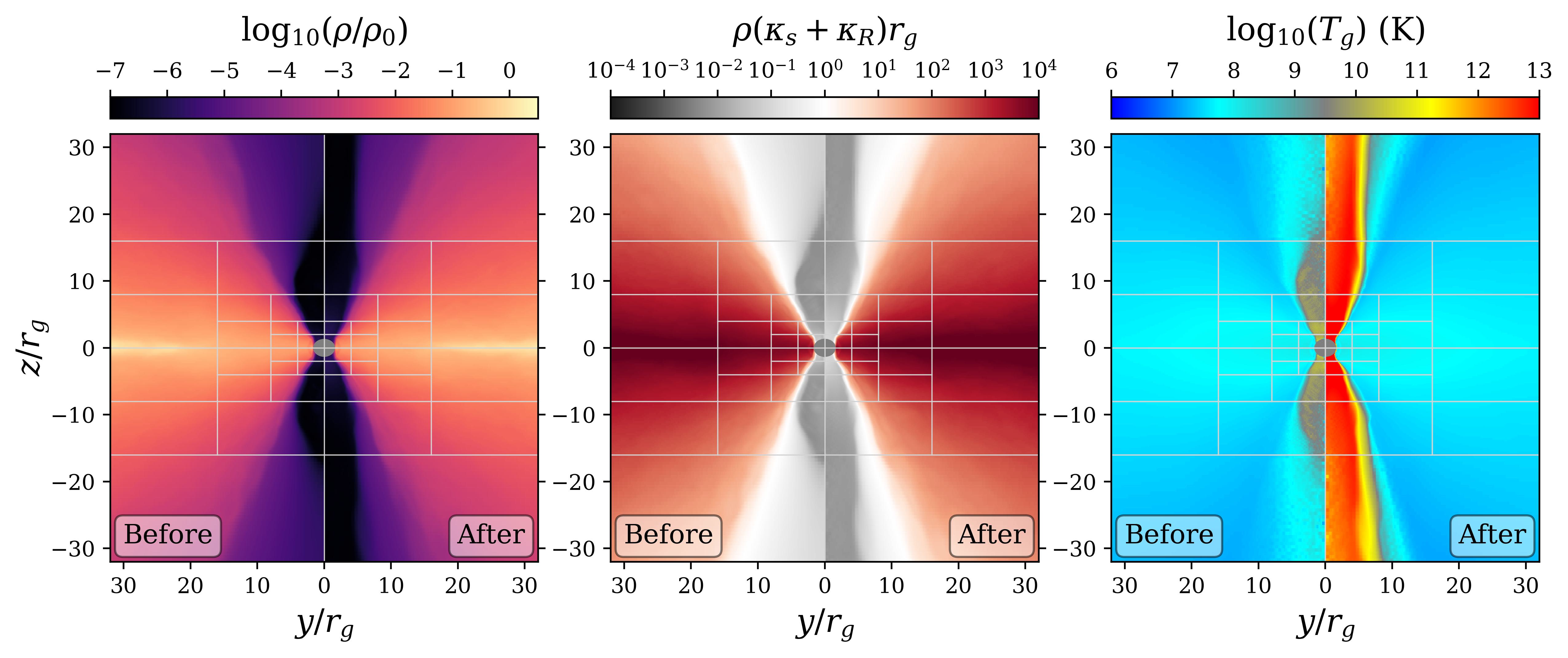}
    \caption{
    Comparison of models before and after applying density rescaling to the radiation source terms.  From left to right, panels show azimuthally averaged 2D profiles of gas density, flux-mean optical depth, and gas temperature at $t = 25000 r_g/c$.  Mesh blocks are overplotted as gray boxes to indicate levels of static mesh refinement.  In each panel, the left half shows results without density rescaling, while the right half shows results with rescaling applied to the gas-radiation coupling.      
    }
    \label{fig:compton_compare}
\end{figure*}

For example, \autoref{fig:compton_compare} presents a side-by-side comparison of models with and without density rescaling in the gas-radiation coupling.  This test is based on model E150-a9-LR, initialized from its starting condition.  From left to right, the panels show azimuthally averaged gas density, flux-mean optical depth per gravitational radius, and gas temperature at $t=25000 r_g/c$.  In each panel, the right half displays results with density rescaling applied, while the left half shows results without it.  The most significant difference appears in the funnel-region gas temperature: the model with density rescaling exhibits temperatures higher by two orders of magnitude near the jet core.  This difference arises because the gas-to-radiation temperature ratio is extremely large in the jet region, making Compton cooling highly sensitive to the local gas density.  However, the gas density in this region is artificially limited by the numerical floor, which fails to capture the true low-density environment.  When combined with a high-density unit, the floor imposes excessive Compton cooling.  Furthermore, the inflated density enhances scattering, which slows jet propagation and suppresses radiation free-streaming.  These effects are particularly pronounced in the lower-refined regions.

\section{Identification of the Jet Region}
\label{appendix:jet_id}

\begin{figure*}
    \centering
    \includegraphics[width=\textwidth]{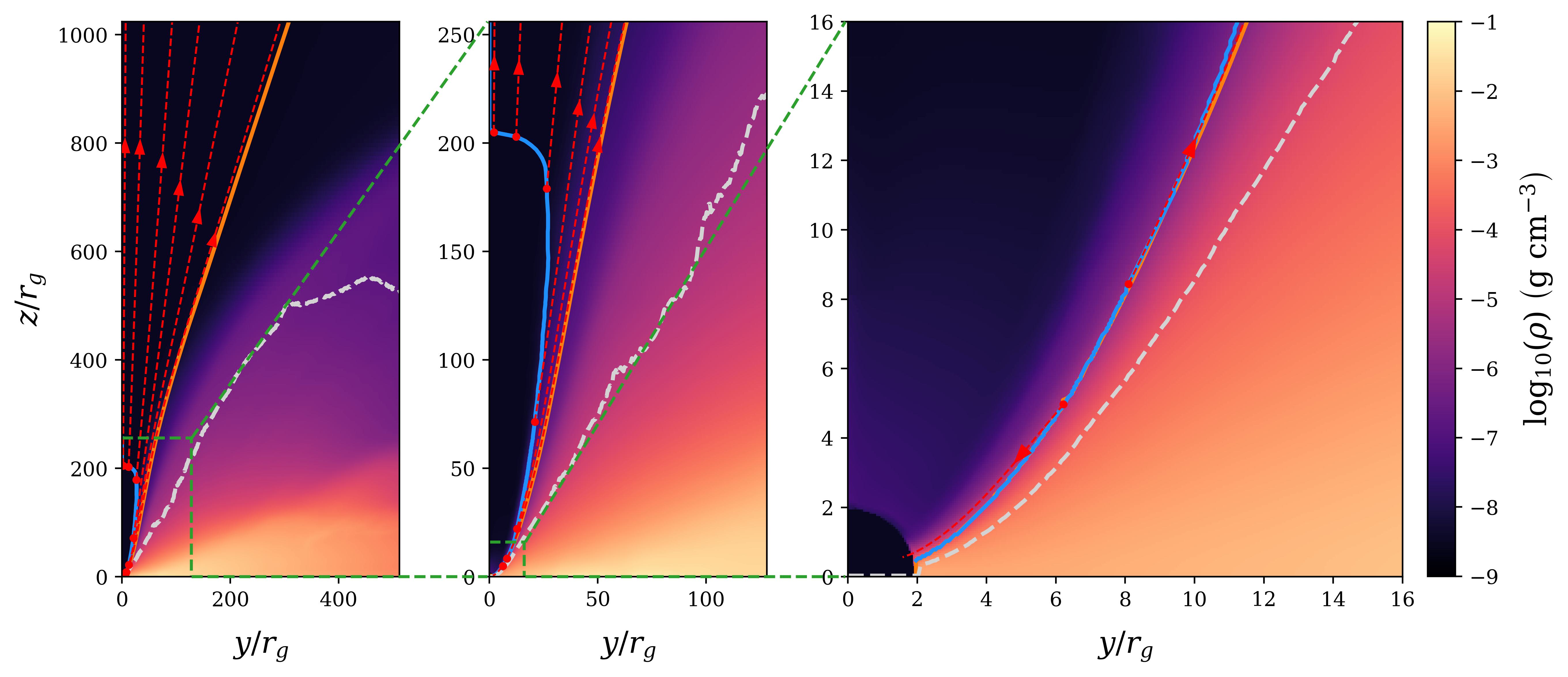}
    \caption{
    Identification of the relativistic jet region.  The panels zoom in on the jet region from left to right, with gray dashed lines indicating the disk surface.  The blue contour outlines the strongly magnetized region, where the magnetic energy exceeds fluid rest-mass energy.  Red streamlines trace the velocity field originating from the this region, and orange lines define the jet boundary based on the outermost streamlines. 
    }
    \label{fig:jet_id}
\end{figure*}

To identify the jet region, we begin with the strongly magnetized inner funnel, from which the relativistic jet is launched.  However, this magnetization criterion only captures the innermost part of the jet.  At larger radii, the jet loses magnetic confinement and transitions to free streaming.  To determine the full jet extent, we trace velocity streamlines outward from the strongly magnetized region, with the outermost streamlines defining the jet boundaries.  

An example of this identification process is shown in \autoref{fig:jet_id}.  The panels progressively zoom in on the relativistic jet from left to right.  The blue contour marks the surface where magnetic energy equals gas rest-mass energy.  Red streamlines trace the velocity field originating from this contour, while the solid orange lines denote the jet boundaries identified by this method.

\section{Computation of Effective Viscosity in General Relativity}
\label{appendix:viscosity}

\begin{figure*}
    \centering
    \includegraphics[width=0.88\textwidth]{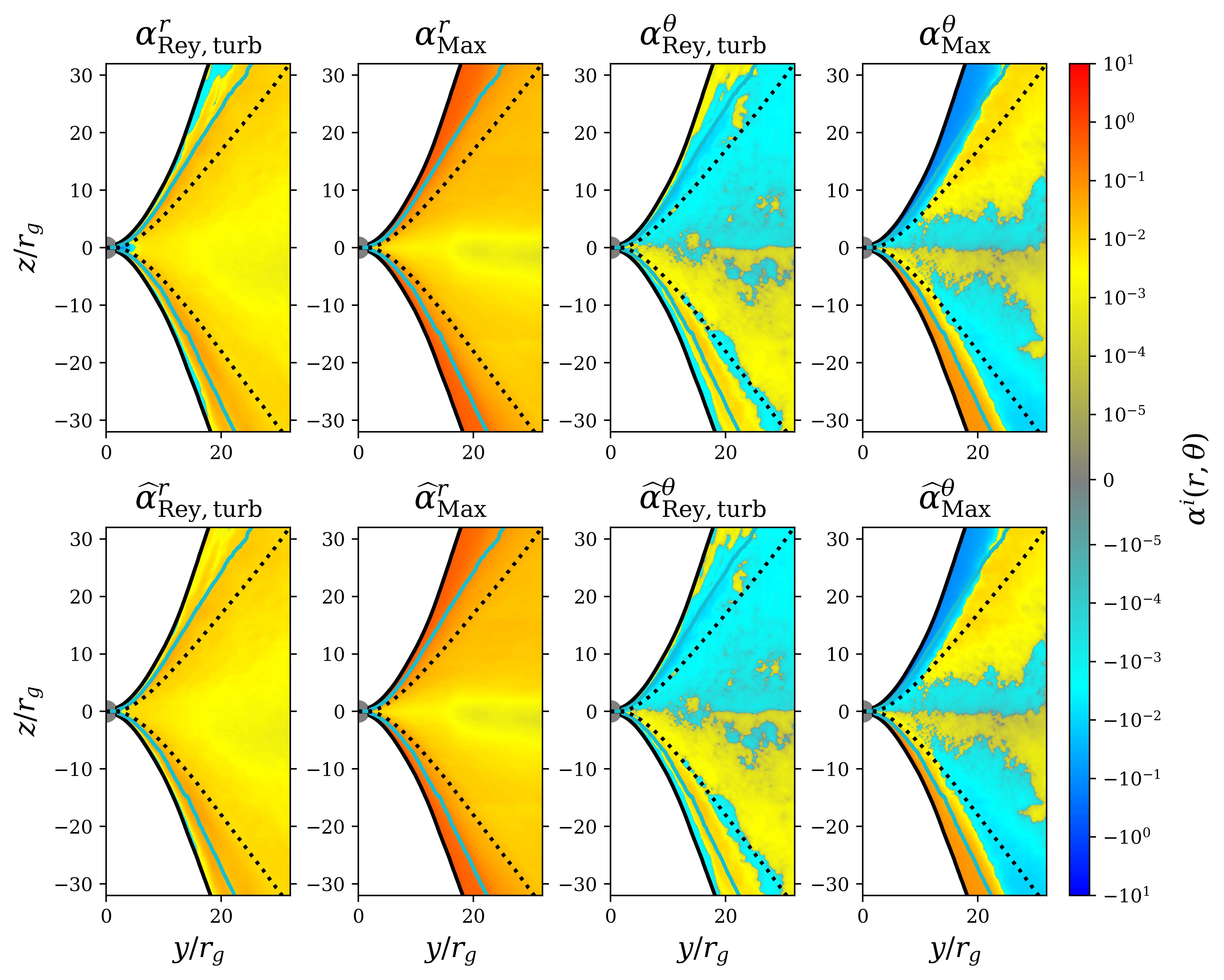}
    \caption{
    Consistency check between the effective viscosity measured in the co-rotating frame (top row) and approximated by coordinate-frame angular momentum flux scaled by total pressure (bottom row) for model E88-a3.  All quantities are averaged temporally and azimuthally.  Definitions of angular momentum flux are provided in \autoref{sec:angmom_transport}, and the method for measuring viscosity in the co-rotating tetrad frame is detailed in \autoref{appendix:viscosity}.  
    }
    \label{fig:alpha2d}
\end{figure*}

Effective viscosity is a local property of the accretion flow that characterizes angular momentum transport relative to the mean flow, and thus must be evaluated in a co-rotating frame.  This requires a well-defined mean flow, typically taken as the azimuthal average under the assumption of axisymmetry.  Although this assumption may break down in misaligned or intrinsically non-axisymmetric systems, our simulations focus on approximately axisymmetric configurations, allowing its use.  Following \citet{Krolik2005} (see also \citealt{Penna2013, White2019}), we compute the effective viscosity in Boyer-Lindquist coordinates, as summarized below.  We begin by converting the four-velocity from spherical Kerr-Schild to Boyer-Lindquist coordinates as follows: 
\begin{subequations}
\begin{align}
    u_{\mathrm{BL}}^{t} &= u^{t} - \frac{2Mr}{r^2-2Mr+a^2}u^{r}
    \ , 
    \\
    u_{\mathrm{BL}}^{r} &= u^{r}
    \ , 
    \\
    u_{\mathrm{BL}}^{\theta} &= u^{\theta}
    \ , 
    \\
    u_{\mathrm{BL}}^{\phi} &= u^{\phi} - \frac{a}{r^2-2Mr+a^2}u^{r}
    \ . 
\end{align}
\end{subequations} 
Given the mean flow defined as the azimuthal average, the components of the mean four-velocity in Boyer-Lindquist coordinates are
\begin{subequations}
\begin{align}
    \bar{u}_{\mathrm{BL}}^t &= 
    \frac{g_{t\phi}^{\mathrm{BL}}}{-g_{tt}^{\mathrm{BL}}}\bar{u}_{\mathrm{BL}}^{\phi}
    +
    \sqrt{
        \left(\frac{g_{t\phi}^{\mathrm{BL}}}{-g_{tt}^{\mathrm{BL}}}\bar{u}_{\mathrm{BL}}^{\phi}\right)^2 
        + \frac{g_{\phi\phi}^{\mathrm{BL}}}{-g_{tt}^{\mathrm{BL}}}\left(\bar{u}_{\mathrm{BL}}^{\phi}\right)^2 
        + \frac{1}{-g_{tt}^{\mathrm{BL}}}
    }
    \ , 
    \\
    \bar{u}^{r}_{\mathrm{BL}} &= 0
    \ ,
    \\
    \bar{u}^{\theta}_{\mathrm{BL}} &= 0
    \ ,
    \\
    \bar{u}^{\phi}_{\mathrm{BL}} &= \frac{\left< w_{\mathrm{disk}} u^{\phi}_{\mathrm{BL}} \right>_{\phi}}{\left<w_{\mathrm{disk}}\right>_{\phi}}
    \ ,
\end{align}
\end{subequations}
where the time component is determined by the normalization condition $g_{\mu\nu}^{\mathrm{BL}}\bar{u}_{\mathrm{BL}}^{\mu}\bar{u}_{\mathrm{BL}}^{\nu} = -1$, given the specified spatial components.  For reference, the relevant Boyer-Lindquist metric components are
\begin{subequations}    
\begin{align}
    g_{tt}^{\mathrm{BL}} &= -\left( 1 - \frac{2Mr}{r^2 + a^2\cos^2\theta} \right)
    \ , 
    \\
    g_{\phi\phi}^{\mathrm{BL}} &= \left( r^2 + a^2 + \frac{2Mra^2\sin^2\theta}{r^2 + a^2\cos^2\theta} \right) \sin^2\theta
    \ , 
    \\
    g_{t\phi}^{\mathrm{BL}} &= -\frac{2Mra\sin^2\theta}{r^2 + a^2\cos^2\theta}
    \ . 
\end{align}
\end{subequations}
With the mean-flow four-velocity defined above, we perform the frame transformation to the co-rotating tetrad frame using
\begin{subequations}
\begin{align}
    e_{\tilde{t}}^{\mu} &= \left( 
        \bar{u}_{\mathrm{BL}}^{t}, 
        \bar{u}_{\mathrm{BL}}^{r}, 
        \bar{u}_{\mathrm{BL}}^{\theta}, 
        \bar{u}_{\mathrm{BL}}^{\phi} \right)
    \ , 
    \\
    e_{\tilde{r}}^{\mu} &= \frac{s}{N_1}\left( 
        \bar{u}^{\mathrm{BL}}_{r}\bar{u}_{\mathrm{BL}}^{t}, 
        1+\bar{u}^{\mathrm{BL}}_{r}\bar{u}_{\mathrm{BL}}^{r}+\bar{u}^{\mathrm{BL}}_{\theta}\bar{u}_{\mathrm{BL}}^{\theta}, 
        0, 
        \bar{u}^{\mathrm{BL}}_{r}\bar{u}_{\mathrm{BL}}^{\phi} \right)
    \ , 
    \\
    e_{\tilde{\theta}}^{\mu} &= \frac{1}{N_2}\left( 
        \bar{u}^{\mathrm{BL}}_{\theta}\bar{u}_{\mathrm{BL}}^{t}, 
        \bar{u}^{\mathrm{BL}}_{\theta}\bar{u}_{\mathrm{BL}}^{r}, 
        1+\bar{u}^{\mathrm{BL}}_{\theta}\bar{u}_{\mathrm{BL}}^{\theta}, 
        \bar{u}^{\mathrm{BL}}_{\theta}\bar{u}_{\mathrm{BL}}^{\phi} \right)
    \ , 
    \\
    e_{\tilde{\phi}}^{\mu} &= \frac{1}{N_3}\left( 
        -\frac{\bar{u}^{\mathrm{BL}}_{\phi}}{\bar{u}^{\mathrm{BL}}_{t}}, 
        0, 
        0, 
        1 \right)
    \ , 
\end{align}
\end{subequations}
where the tilde denotes quantities in the co-rotating tetrad frame, and the normalization coefficients are defined as
\begin{subequations}
\begin{align}
    s &= -\frac{\bar{u}_{\mathrm{BL}}^{t}\bar{u}^{\mathrm{BL}}_{t} + \bar{u}_{\mathrm{BL}}^{\phi}\bar{u}^{\mathrm{BL}}_{\phi}}{\left| \bar{u}_{\mathrm{BL}}^{t}\bar{u}^{\mathrm{BL}}_{t} + \bar{u}_{\mathrm{BL}}^{\phi}\bar{u}^{\mathrm{BL}}_{\phi} \right|}
    \ , 
    \\
    N_1 &= \sqrt{
                \bar{u}^{\mathrm{BL}}_{r}\bar{u}^{\mathrm{BL}}_{r}\left(\bar{u}_{\mathrm{BL}}^{t}\bar{u}^{\mathrm{BL}}_{t} + \bar{u}_{\mathrm{BL}}^{\phi}\bar{u}^{\mathrm{BL}}_{\phi}\right) 
                + g^{\mathrm{BL}}_{rr}\left( \bar{u}_{\mathrm{BL}}^{t}\bar{u}^{\mathrm{BL}}_{t}+\bar{u}_{\mathrm{BL}}^{\phi}\bar{u}^{\mathrm{BL}}_{\phi} \right)^2
            }
    \ , 
    \\
    N_2 &= \sqrt{
                g^{\mathrm{BL}}_{\theta\theta}\left( 1+\bar{u}_{\mathrm{BL}}^{\theta}\bar{u}^{\mathrm{BL}}_{\theta} \right) 
            }
    \ , 
    \\
    N_3 &= \sqrt{
                g^{\mathrm{BL}}_{tt}\left( \frac{\bar{u}^{\mathrm{BL}}_{\phi}}{\bar{u}^{\mathrm{BL}}_{t}} \right)^2
                - 2g^{\mathrm{BL}}_{t\phi}\left( \frac{\bar{u}^{\mathrm{BL}}_{\phi}}{\bar{u}^{\mathrm{BL}}_{t}} \right) 
                + g^{\mathrm{BL}}_{\phi\phi}
            }
    \ . 
\end{align}
\end{subequations}
Therefore, the frame transformations of four-velocity, four-magnetic field, and radiation stress-energy tensor are given by
\begin{subequations}
\begin{align}
    u^{\tilde{\mu}} &= \eta^{\tilde{\mu}\tilde{\nu}} g_{\alpha\beta}^{\mathrm{BL}} e_{\tilde{\nu}}^{\alpha} u^{\beta}_{\mathrm{BL}}
    \ , 
    \\
    b^{\tilde{\mu}} &= \eta^{\tilde{\mu}\tilde{\nu}} g_{\alpha\beta}^{\mathrm{BL}} e_{\tilde{\nu}}^{\alpha} b^{\beta}_{\mathrm{BL}}
    \ , 
    \\
    R^{\tilde{\mu}\tilde{\nu}} &= \eta^{\tilde{\alpha}\tilde{\mu}}\eta^{\tilde{\beta}{\tilde{\nu}}}e_{\tilde{\alpha}}^{\sigma}e_{\tilde{\beta}}^{\lambda}R_{\sigma\lambda}^{\mathrm{BL}}
    \ , 
\end{align}
\end{subequations} 
where $\eta^{\tilde{\mu}\tilde{\nu}}=\mathrm{diag}(-1,1,1,1)$ is the Minkowski metric representing the locally flat spacetime.  With all relevant quantities defined in the co-rotating frame, the effective viscosity is measured as the local angular momentum flux scaled by the total pressure, given by
\begin{align}
    & \alpha_{\mathrm{X}}^{r}(r,\theta) = \left<\frac{T_{\mathrm{X}}^{\tilde{r}\tilde{\phi}}}{P_{\mathrm{tot}}}\right>_{\phi}
    \ , 
    \qquad\qquad
    \alpha_{\mathrm{X}}^{\theta}(r,\theta) = \left<\frac{T_{\mathrm{X}}^{\tilde{\theta}\tilde{\phi}}}{P_{\mathrm{tot}}}\right>_{\phi}
    \ , 
\end{align}
where the subscript `X' denotes any contributing component.  Recall that the viscosity can also be approximated directly from the coordinate-frame angular momentum flux, as given in equation~\autoref{eq:viscosity_approx}.  In \autoref{fig:alpha2d}, we compare the viscosity measured in the co-rotating frame (top panels) with that in the coordinate frame (bottom panels).  The results within the disk body (enclosed by the solid cyan lines) are nearly identical. 

\section{General Relativistic Four-Force}
\label{appendix:four_force}

The four-force equation can be derived by projecting the stress-energy equation onto a spacelike direction orthogonal to the local four-velocity $u^{\mu}$ \citep{Moller2015,White2020,Utsumi2022}.  Starting from the conservative form of the stress-energy equation, 
\begin{equation}
    \nabla_{\mu} T^{\mu}_{\ \nu} - G_{\nu} = 0
    \ ,
\end{equation}
we apply the projection $\mathcal{P}^{\alpha\nu}=g^{\alpha\nu}+u^{\alpha}u^{\nu}$, which decomposes the stress-energy equation into different components: 
\begin{equation}
    \underbrace{\mathcal{P}^{\alpha\nu}\nabla_{\mu}(w u^{\mu} u_{\nu})}_{\text{acceleration + frame}} 
    + \underbrace{\mathcal{P}^{\alpha\nu}\nabla_{\nu}P_g}_{\text{gas}} 
    + \underbrace{\mathcal{P}^{\alpha\nu}(\nabla_{\nu}P_m - \nabla_{\mu}b^{\mu}b_{\nu})}_{\text{magnetic}} 
    - \underbrace{\mathcal{P}^{\alpha\nu}G_{\nu}}_{\text{radiation}} = 0
    \ , 
\end{equation} 
where each term is labeled to indicate its physical origin.  The first term represents the four-acceleration and the frame force, and can be expressed in the form of the geodesic equation as: 
\begin{align}
    \mathcal{P}^{\alpha\nu}\nabla_{\mu}(w u^{\mu} u_{\nu}) &= w\left(\frac{du^{\alpha}}{d\tau} + \Gamma^{\alpha}_{\mu\nu}u^{\mu}u^{\nu}\right)
    \ . 
\end{align}
The radiation source term can be directly computed as: 
\begin{equation}
    \mathcal{P}^{\alpha\nu}G_{\nu} = \mathcal{P}^{\alpha\nu}\left(\bar{\chi}_R + \bar{\chi}_T\right) u^{\sigma}R_{\nu\sigma}
    \ . 
\end{equation}
This leads to the final form of the equation, which can be interpreted as a geodesic equation sourced by various four-forces:
\begin{equation}
    \frac{du^{\alpha}}{d\tau} = 
    \underbrace{- \Gamma^{\alpha}_{\mu\nu}u^{\mu}u^{\nu}}_{\text{frame}}
    \underbrace{- \frac{1}{w}\mathcal{P}^{\alpha\nu}\nabla_{\nu}P_g}_{\text{gas pressure}}
    \underbrace{- \frac{1}{w}\mathcal{P}^{\alpha\nu}\nabla_{\nu}P_m}_{\text{magnetic pressure}}
    \underbrace{+ \frac{1}{w}\mathcal{P}^{\alpha\nu}\nabla_{\mu}b^{\mu}b_{\nu}}_{\text{magnetic tension}}
    \underbrace{+ \frac{1}{w}\mathcal{P}^{\alpha\nu}\left(\bar{\chi}_R + \bar{\chi}_T\right) u^{\sigma}R_{\nu\sigma}}_{\text{radiation}}
    \ . 
\end{equation}
Each labeled four-forces reduces to its Newtonian counterpart in the appropriate limit.  This formulation is particularly useful, as it is equivalent to the stress-energy equation solved in the simulations, providing clear insight into the influence of individual physical components on the flow evolution. It also connects naturally to Newton's second law, facilitating physical interpretation.

However, it is important to recognize that both the frame force and magnetic tension force can depend on the choice of metric, leading to different physical interpretations.  For example, in the spherical Kerr-Schild (SKS) metric, the frame force includes an additional centrifugal component due to the rotating frame, which is absent in the Cartesian Kerr-Schild (CKS) metric. Similarly, the magneto-centrifugal component of the magnetic tension force appears in the partial derivative terms of the four-magnetic field in the CKS metric, but is absorbed into the affine connection terms in the SKS metric.  Therefore, these subtleties must be carefully considered when interpreting physical processes within a specific coordinate framework.

\section{Energy Flux Decomposition in Disk, Wind, and Jet Regions}
\label{appendix:energy_decomposition}

To analyze the energetics in the steady state, we need to examine the energy flow across each region (i.e. disk, wind, and jet) and identify the dominant heating and cooling mechanisms.  Accordingly, we categorize the energy transport rates into the following components: fluid rest-mass advection, gas thermal energy advection, magnetic advection, radiation advection, kinetic energy advection, magnetic convection, and total radiation energy transport, defined sequentially as follows: 
\begin{subequations}    
\begin{align}
    \dot{E}_{\mathrm{mass}}^{(\mathrm{zone})} &= -\int_{\mathrm{zone}} \left< \rho u^r u_t\right>_{t} \sqrt{-g} d\theta d\phi 
    \ ,
    \label{eq:edot_mass}
    \\
    \dot{E}_{\mathrm{egas}}^{(\mathrm{zone})} &= -\frac{\gamma}{\gamma-1} \int_{\mathrm{zone}} \left<P_g u^r u_t \right>_{t} \sqrt{-g} d\theta d\phi    
    \ ,
    \label{eq:edot_egas}
    \\
    \dot{E}_{\mathrm{emag}}^{(\mathrm{zone})} &= -\int_{\mathrm{zone}} \left< b^{\nu}b_{\nu}u^r u_t \right>_{t} \sqrt{-g} d\theta d\phi
    \ ,
    \label{eq:edot_emag}
    \\
    \dot{E}_{\mathrm{erad}}^{(\mathrm{zone})} &= -\frac{4}{3} \int_{\mathrm{zone}} \left< \bar{E}_r u^r u_t \right>_{t} \sqrt{-g} d\theta d\phi    
    \ ,
    \label{eq:edot_erad}
    \\
    \dot{E}_{\mathrm{kin}}^{(\mathrm{zone})} &= \int_{\mathrm{zone}} \left< -\rho u^r\left(u_t+\sqrt{-g_{tt}}\right)\right>_{t} \sqrt{-g} d\theta d\phi
    \ ,
    \label{eq:edot_kin}
    \\
    \dot{E}_{\mathrm{Max}}^{(\mathrm{zone})} &= \int_{\mathrm{zone}} \left< b^r b_t \right>_{t} \sqrt{-g} d\theta d\phi    
    \ ,
    \label{eq:edot_max}
    \\
    \dot{E}_{\mathrm{rad}}^{(\mathrm{zone})} &= -\int_{\mathrm{zone}} \left<R^r_{\ t}\right>_{t} \sqrt{-g} d\theta d\phi    
    \ ,
    \label{eq:edot_rad}
\end{align}
\end{subequations}
where `zone' refers to the integration domain defined as the disk, wind or jet region.  The radiation energy density $\bar{E}_r$ is computed by integrating the radiation intensity $\bar{I}$ over the solid angle $\bar{\Omega}$ in the fluid frame (denoted by bars).  The radiation stress-energy tensor $R^{\alpha}_{\ \beta}$ is obtained through a second-order angular integration of the tetrad-frame intensity $\hat{I}$.  Each energy transport rate is calculated by performing a spherical integration at a specified radius within the corresponding zone.  Note that the rest-mass advection term~\autoref{eq:edot_mass} includes contributions from rest-mass energy, gravitational binding energy, and kinetic energy.  

In the disk region, the total energy transport rate is decomposed into the following components: 
\begin{subequations}   
\begin{equation}
    \dot{E}^{(\mathrm{disk})} = \dot{E}_{\mathrm{acc}}^{(\mathrm{disk})} + \dot{E}_{\mathrm{Max}}^{(\mathrm{disk})} + \left(\dot{E}_{\mathrm{rad}}^{(\mathrm{disk})} - \dot{E}_{\mathrm{erad}}^{(\mathrm{disk})}\right)
    \ ,
\end{equation}
where $\dot{E}_{\mathrm{acc}}^{(\mathrm{disk})}$ represents the total energy input from the accretion process, defined as: 
\begin{equation}
    \dot{E}_{\mathrm{acc}}^{(\mathrm{disk})} = \dot{E}_{\mathrm{mass}}^{(\mathrm{disk})} + \dot{E}_{\mathrm{egas}}^{(\mathrm{disk})} + \dot{E}_{\mathrm{emag}}^{(\mathrm{disk})} + \dot{E}_{\mathrm{erad}}^{(\mathrm{disk})}
    \ .
\end{equation}
\end{subequations}
Within the disk, since radiation and fluid are well-coupled, radiation advection is included in the total enthalpy.  This contribution is then subtracted from the total radiation energy flux to isolate the term that corresponds to radiation diffusion in the classical limit. 

In the wind and jet regions, where radiation and gas are largely decoupled, radiation energy transport is treated separately. To evaluate the energy carried by the outflow, the kinetic energy component is explicitly extracted, while the advection and convection of the magnetic field are combined to compute the Poynting flux. Accordingly, the total energy transport rate is partitioned as follows:
\begin{subequations}
\begin{align}
    &\dot{E}^{(\mathrm{zone})} = \left(\dot{E}_{\mathrm{mass}}^{(\mathrm{zone})}
        - \dot{E}_{\mathrm{kin}}^{(\mathrm{zone})}\right) 
        + \dot{E}_{\mathrm{kin}}^{(\mathrm{zone})}
        + \dot{E}_{\mathrm{egas}}^{(\mathrm{zone})} 
        + \dot{E}_{\mathrm{mag}}^{(\mathrm{zone})} 
        + \dot{E}_{\mathrm{rad}}^{(\mathrm{zone})}
        \ , 
    \\
    &\dot{E}_{\mathrm{mag}}^{(\mathrm{zone})} = \dot{E}_{\mathrm{emag}}^{(\mathrm{zone})} + \dot{E}_{\mathrm{Max}}^{(\mathrm{zone})}
    \ .
\end{align}
\end{subequations}

\bibliography{edd_paper_ii}
\bibliographystyle{aasjournal}
\end{CJK*}
\end{document}